\documentclass[preprint]{aastex}
\usepackage{graphicx}
\usepackage{natbib}
\usepackage{amsmath,amsthm,amssymb}
\usepackage{color}
\usepackage{ulem}
\usepackage[T1]{fontenc}
\newcommand{\gps}{\ensuremath{g_{\rm P1}}}
\newcommand{\rps}{\ensuremath{r_{\rm P1}}}
\newcommand{\ips}{\ensuremath{i_{\rm P1}}}
\newcommand{\zps}{\ensuremath{z_{\rm P1}}}
\newcommand{\yps}{\ensuremath{y_{\rm P1}}}

\newcommand{\PS}{\protect \hbox {Pan-STARRS1}}
\newcommand{\WISE}{{\it WISE}}
\newcommand{\um}{$\mu$m}
\newcommand{\mjup}{$M_{\mathrm{Jup}}$}           
\newcommand{\lsun}{$L_{\odot}$}                   
\newcommand{\hto}{H$_2$O}
\newcommand{\htoa}{H$_2$O--1}
\newcommand{\htob}{H$_2$O--2}
\newcommand{\htod}{H$_2$OD}
\newcommand{\fehz}{FeH$_{\rm z}$}
\newcommand{\voz}{VO$_{\rm z}$}
\newcommand{\kij}{K~I$_{\rm J}$}
\newcommand{\iz}{\ips$-$\zps}
\newcommand{\iy}{\ips$-$\yps}
\newcommand{\zy}{\zps$-$\yps}
\newcommand{\ywa}{\yps$-W1$}
\newcommand{\wawb}{$W1-W2$}
\newcommand{\wbwc}{$W2-W3$}
\newcommand{\mytilde}{\raise.17ex\hbox{$\scriptstyle\mathtt{\sim}$}}
\newcommand{\fldg}{\mbox{\textsc{fld-g}}}
\newcommand{\intg}{\mbox{\textsc{int-g}}}
\newcommand{\vlg}{\mbox{\textsc{vl-g}}}

\shorttitle{Pan-STARRS + WISE L/T Transition Dwarfs}
\shortauthors{Best, W. M. J. et al}

\slugcomment{\large Published in ApJ, 2015 December 1}

\begin{document}

\title{A Search for L/T Transition Dwarfs With \PS\ and \WISE. II. \\
  L/T Transition Atmospheres and Young Discoveries}
\author{William M. J. Best\altaffilmark{1,6}, 
  Michael C. Liu\altaffilmark{1,6},
  Eugene A. Magnier\altaffilmark{1}, 
  Niall R. Deacon\altaffilmark{2}, 
  Kimberly M. Aller\altaffilmark{1,6}, 
  Joshua Redstone\altaffilmark{3},
W. S. Burgett\altaffilmark{4}, 
K. C. Chambers\altaffilmark{1}, 
P. Draper\altaffilmark{5}, 
H. Flewelling\altaffilmark{1}, 
K. W. Hodapp\altaffilmark{1}, 
N. Kaiser\altaffilmark{1}, 
N. Metcalfe\altaffilmark{5}, 
J. L. Tonry\altaffilmark{1}, 
R. J. Wainscoat\altaffilmark{1}, 
C. Waters\altaffilmark{1} 
}

\altaffiltext{1}{Institute for Astronomy, University of Hawaii, 2680 Woodlawn
  Drive, Honolulu, HI 96822, USA; wbest@ifa.hawaii.edu}
\altaffiltext{2}{Centre for Astrophysics Research, University of Hertfordshire, College Lane Campus, Hatfield AL10 9AB, UK}
\altaffiltext{3}{Equatine Labs, 89 Antrim Street, \#2, Cambridge, MA 02139, USA}
\altaffiltext{4}{GMTO Corporation, 251 S. Lake Ave., Suite 300, Pasadena, CA 91101, USA}
\altaffiltext{5}{Department of Physics, Durham University, South Road, Durham DH1 3LE, UK}
\altaffiltext{6}{Visiting Astronomer at the Infrared Telescope Facility, which
  is operated by the University of Hawaii under Cooperative Agreement
  no. NNX-08AE38A with the National Aeronautics and Space Administration,
  Science Mission Directorate, Planetary Astronomy Program.}


\begin{abstract}

  The evolution of brown dwarfs from L to T spectral types is one of the least
  understood aspects of the ultracool population, partly for lack of a large,
  well-defined, and well-characterized sample in the L/T transition.  To improve
  the existing census, we have searched $\approx$28,000~deg$^2$ using the \PS\
  and \textit{Wide-field Infrared Survey Explorer} surveys for L/T transition
  dwarfs within 25~pc.  We present 130 ultracool dwarf discoveries with
  estimated distances $\approx9-130$~pc, including 21 that were independently
  discovered by other authors and 3 that were previously identified as
  photometric candidates.  Seventy-nine of our objects have near-IR spectral
  types of L6--T4.5, the most L/T transition dwarfs from any search to date, and
  we have increased the census of L9--T1.5 objects within 25~pc by over 50\%.
  The color distribution of our discoveries provides further evidence for the
  ``L/T gap,'' a deficit of objects with $(J-K)_{\rm MKO}\approx0.0$--0.5~mag in
  the L/T transition, and thus reinforces the idea that the transition from
  cloudy to clear photospheres occurs rapidly.  Among our discoveries are 31
  candidate binaries based on their low-resolution spectral features.  Two of
  these candidates are common proper motion companions to nearby main sequence
  stars; if confirmed as binaries, these would be rare benchmark systems with
  the potential to stringently test ultracool evolutionary models.  Our search
  also serendipitously identified 23 late-M and L dwarfs with
  spectroscopic signs of low gravity implying youth, including 10 with
  \vlg\ or \intg\ gravity classifications and another 13 with indications of low
  gravity whose spectral types or modest spectral signal-to-noise ratio do not
  allow us to assign formal classifications.  Finally, we identify 10
  candidate members of nearby young moving groups (YMG) with spectral types
  L7--T4.5, including three showing spectroscopic signs of low gravity.  If
  confirmed, any of these would be among the coolest known YMG members and would
  help to determine the effective temperature at which young brown dwarfs cross
  the L/T transition.

\end{abstract}


\section{Introduction}
\label{intro}
Over the past 20~years some 1,500 brown dwarfs have been discovered in the
field, yet fundamental questions about their formation, evolution, and
atmospheres remain. Without sustained hydrogen fusion in their cores, brown
dwarfs cool continuously, creating an observational degeneracy between their
masses, ages and luminosities. Their photospheres are dominated by molecules and
dusty condensates, and undergo significant chemical changes as they cool
\citep[e.g.,][]{Burrows:2001iv}.  The relationship between the observable
properties (fluxes and spectra) and the underlying physical properties (masses,
ages, metallicities, and gravities) of ultracool dwarfs is therefore complex and
challenging to disentangle, and evolutionary trends are difficult to identify.

This is particularly true in the L/T transition (spectral types
$\approx\!$~L6--T4.5), where spectral features undergo significant changes and
near-infrared colors become bluer by $\approx$2 magnitudes over a narrow range
of effective temperature
\citep[$T_{\rm eff}\approx1400-1200$~K;][]{Golimowski:2004en,Stephens:2009cc}.
These changes are thought to arise from the depletion of thick condensate clouds
as brown dwarfs cool
\citep[e.g.,][]{Allard:2001fh,Burrows:2006ia,Saumon:2008im}.  Several scenarios
have been proposed wherein condensate clouds thin gradually, rain out suddenly,
or break up
\citep[e.g.,][]{Ackerman:2001gk,Knapp:2004ji,Tsuji:2005cd,Marley:2010kx}.  The
process is still not well understood, however, and state-of-the-art evolutionary
and atmospheric models typically yield inaccurate luminosities and inconsistent
temperatures for L/T objects with dynamical masses and/or age determinations
\citep[e.g.,][]{Dupuy:2009ga,Liu:2010cw,Dupuy:2014iz}.  Color-magnitude diagrams
with accurate luminosities are still rather sparsely populated in the L/T
transition \citep{Dupuy:2012bp}, hindering our ability to test the models.

A large and well-defined sample is a necessary starting point, but L/T
transition dwarfs are known to be more elusive than those with higher and lower
effective temperatures.  At optical wavelengths, L/T transition dwarfs are
faint.  In the near-infrared, where they are brightest, their colors make them
difficult to distinguish from low-mass stars \citep[e.g.,][]{Reid:2008fz}.  The
most productive previous searches so far each focused on $\lesssim$10\% of the
sky: \citet{Chiu:2006jd} used the Sloan Digital Sky Survey
\citep[SDSS;][]{York:2000gn} to find 46 L6--T4.5 dwarfs over
$\approx$3,500~deg$^2$, and \citet{Marocco:2015iz} found 48 L6--T4.5 dwarfs in
$\approx$4,000~deg$^2$ by cross-matching the UKIRT Infrared Deep Sky Survey
\citep[UKIDSS;][]{Lawrence:2007hu} Large Area Survey with SDSS.  What has been
missing is an all-sky search specifically targeting nearby, bright L/T
transition dwarfs.

To address this deficiency, we have conducted an extensive search with these key
features: (1) We used the new \PS\ Survey \citep[PS1;][]{Kaiser:2010gr}
cross-matched with the Wide-Field Infrared Survey Explorer
\citep[\WISE;][]{Wright:2010in} All-sky Release, thereby exploiting the combined
broad wavelength coverage of these optical and mid-infrared surveys; (2) we
searched $\approx$28,000~deg$^2$, nearly the full area of the PS1 $3\pi$ survey;
and (3) we searched to within $3^{\circ}$ of the Galactic plane, whereas most
previous searches stopped at $b=10^{\circ}$ or $b=15^{\circ}$
\citep[e.g.,][]{Cruz:2003fi,Scholz:2011gs}.  In \citet[][hereinafter Paper
I]{Best:2013bp}, we presented seven initial discoveries from our search, all
bright L/T transition dwarfs within 15~pc.  In this paper, we present the
complete results of our search, including 79 total L/T transition dwarfs and 23
young or potentially young late-M and L dwarfs.

We describe our search in Section~\ref{method} and our observations in
Section~\ref{obser}.  In Section~\ref{results} we present the results of our
search, including descriptions of interesting individual objects.  In
Section~\ref{lttrans.atmos} we discuss implications of our discoveries for
evolutionary models of the L/T transition.  We discuss our young discoveries in
more detail in Section~\ref{young} and summarize our findings in
Section~\ref{summary}.

\section{Search Method}
\label{method}

\subsection{Input Catalogs}
\label{catalogs}
The PS1 $3\pi$ survey (K. C. Chambers et al., in preparation) has obtained an
average of $\approx$12 epochs of imaging in five optical bands
($g_{\rm P1}, r_{\rm P1}, i_{\rm P1}, z_{\rm P1}, y_{\rm P1}$) with a 1.8-meter
wide-field telescope on Haleakala, Maui, covering the entire sky north of
$-30^{\circ}$~declination.  Images were processed nightly through the Image
Processing Pipeline
\citep[IPP;][]{Magnier:2006uj,Magnier:2007wn,Magnier:2008jf}, with photometry on
the AB magnitude scale \citep{Tonry:2012gq}.  Imaging began in May 2010 and
concluded in March 2014.  We conducted our search using PS1/IPP Processing
Version 1 photometry, and constructed object names according to the PS1
  convention using object coordinates as of January 2012.  The \WISE\ All-sky
Source Catalog \citep{Cutri:2012wm} comprises data taken between January and
August 2010 in four mid-infrared bands: $W1$~($3.6\,\mu$m), $W2$~($4.5\,\mu$m),
$W3$~($12\,\mu$m), and $W4$~($22\,\mu$m).

\subsection{Search Parameters}
\label{parameters}
Our search is described in detail in Paper I. Briefly, we merged all PS1
detections through January 2012 with the \WISE\ All-sky catalog using a $3.0''$
matching radius.  We removed objects within $3^\circ$ of the Galactic plane and
in the heavily reddened areas of the sky defined by \citet{Cruz:2003fi}, except
for objects in these regions for which PS1 reported a proper motion with
S/N~$>3$.  We searched between $\delta=-30^{\circ}$ (the southern limit of PS1)
and $\delta=+70^{\circ}$ (the northern limit of the NASA Infrared Telescope
Facility (IRTF), which we used for spectroscopic follow-up).  We identified
candidate L/T dwarfs using a suite of quality and color cuts applied to our
merged PS1+\WISE\ database.  After visually screening these candidates using
images from PS1, \WISE, and the Two Micron All Sky Survey
\citep[2MASS;][]{Skrutskie:2006hl}, we obtained near-infrared photometry from
2MASS, UKIDSS, and our own observations (Section~\ref{obser.nearIR}), and used
this to apply a final screening based on colors and magnitudes.  We summarize
our photometric criteria here:

\begin{enumerate}
\item Detected in at least two separate \yps\ frames with $\mathrm{S/N}>5$ in each.
\item Good quality photometry in \yps, no saturated objects or cosmic rays.
\item No more than one total detection in either \gps\ or \rps.
\item \iz\ $\ge1.8$~mag (only applied when the \ips\ and \zps\ photometry for an
  object met the same quality standards required for \yps).
\item \iy\ $\ge2.8$~mag (only applied when \ips\ photometry met the same quality
  standards required for \yps).
\item \zy\ $\ge0.6$~mag (only applied when \zps\ photometry met the same quality
  standards required for \yps).
\item $W1$ and $W2$ detections have $\mathrm{S/N}>2$ (for most candidates, PS1
  establishes the sensitivity limit).
\item $W1$ and $W2$ detections are point sources, not saturated, and unlikely to
  be variable.
\item \ywa\ $\ge3.0$~mag.
\item \wawb\ $\ge0.4$~mag.
\item \wbwc\ $\le2.5$~mag.
\item $y_{\rm P1}-J_{2MASS}\ge1.8$~mag or $y_{\rm P1}-J_{MKO}\ge1.9$~mag.
\end{enumerate}

We then obtained and classified near-IR spectra for 142 candidates using
standard procedures described in Section~\ref{obser}.  In
Table~\ref{tbl.ps1.wise} we present the PS1 and \WISE\ photometry for the
objects we observed spectroscopically, and Table~\ref{tbl.nir} shows their
near-infrared photometry.  We did not re-observe objects also found by other
concurrent PS1 searches for ultracool dwarfs (M. C. Liu et al., in preparation).

\subsection{A \WISE\ Photometric Criterion for L/T Transition Dwarfs Within 25~pc}
\label{WISE.25pc}
Prior to obtaining spectra for our candidates, we used photometry to estimate
distances.  In Paper I, we noted that \yps\ absolute magnitudes are roughly flat
across the L/T transition, and we identified $y_{\rm P1}=19.3$~mag as a limit
for single objects expected to lie within 25~pc.  We therefore used
$y_{\rm P1}<19.3$~mag to prioritize candidates for spectroscopic observations
(though in the end, we did observe a few objects with $y_{\rm P1}>19.3$~mag.)
However, some of our first spectroscopic confirmations proved to be L/T
transition dwarfs with spectrophotometric distances of $30-35$~pc and earlier L
dwarfs at greater distances, so we sought a better criterion than the \yps\
cutoff.

\begin{figure}
\begin{center}   
  \includegraphics[width=1.00\columnwidth, trim = 20mm 0 5mm 0]{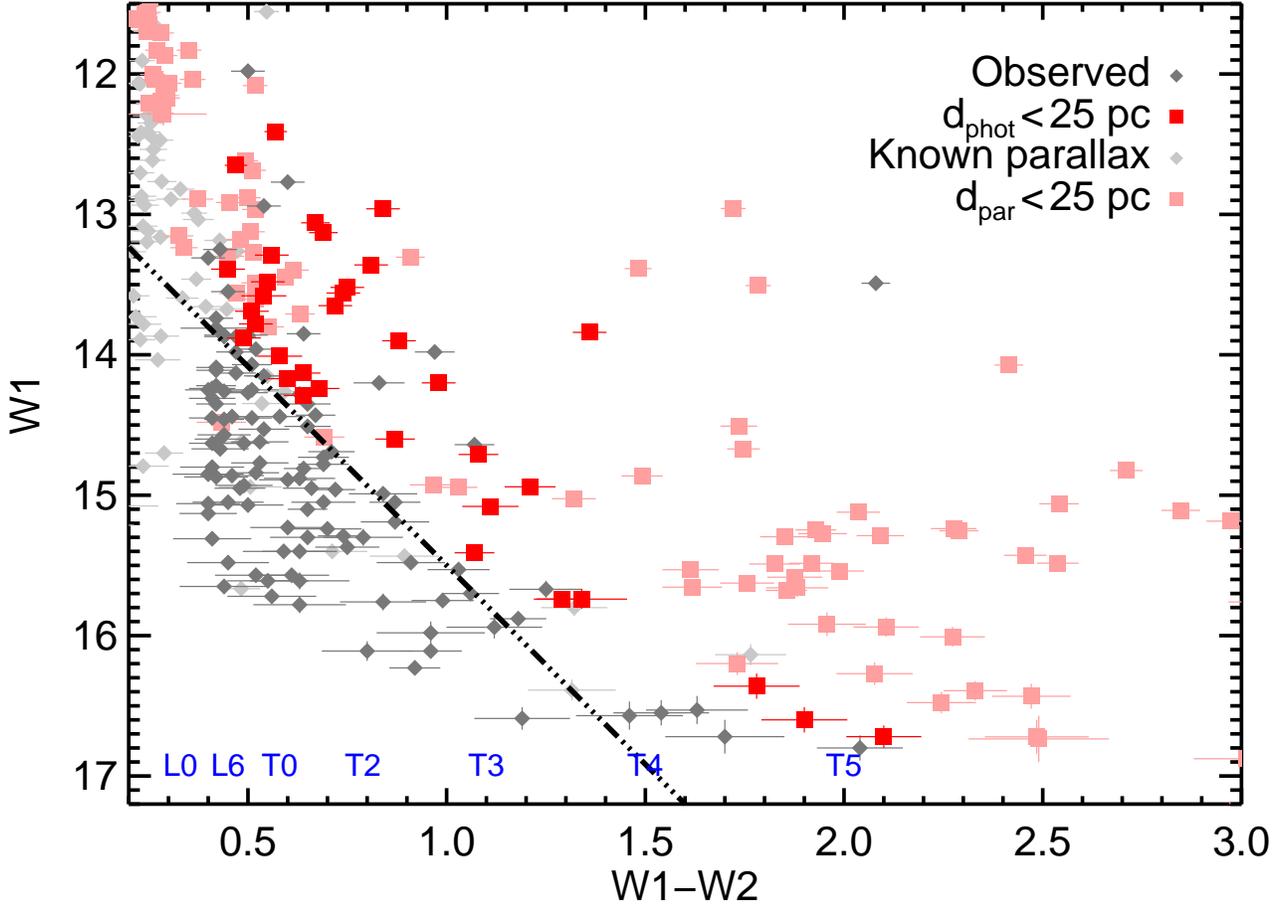}
  \caption{$W1$ vs. \wawb\ diagram for known ultracool dwarfs. Objects with
    known parallaxes \citep{Dupuy:2012bp} and within 25~pc are shown as pink
    squares; those beyond 25~pc are light gray diamonds. Our new discoveries
    with photometric distances (Section~\ref{results}) less than than 25~pc are
    plotted with red squares, and those with $d_{\rm phot}>25$~pc are dark gray
    diamonds.  Approximate spectral types for \wawb\ colors are indicated in
    blue along the bottom.  The region above and to the right of the black
    dashed-dotted line, $W1\le2.833\times (W1-W2) + 12.667$~mag, includes 100\% of
    our L/T transition discoveries with $d_{\rm phot}<25$~pc but only 33\%
    beyond 25~pc.}
  \label{fig.w1.w1w2.25pc}
\end{center}
\end{figure}

We examined the relationships between colors and magnitudes in the PS1, 2MASS,
and \WISE\ bands and the distances to ultracool dwarfs with known parallaxes
from \citet{Dupuy:2012bp}\footnote{An updated list can be found in the Database
  of Ultracool Parallaxes maintained by Trent Dupuy at
  http://www.as.utexas.edu/~tdupuy/plx/Database\_of\_Ultracool\_Parallaxes.html.
  Here we used the version posted on 2013-09-09.}.  We identified an inequality
in the $W1$ vs. $W1-W2$ color-magnitude diagram that selects L/T transition
dwarfs with $d<25$~pc:
\begin{equation}
\label{eq.25pc}
W1\le2.833\times (W1-W2) + 12.667\ {\rm mag}
\end{equation}
This inequality excludes nearly all ultracool dwarfs with trigonometric
distances beyond 25~pc for {\protect \hbox
  {$0.5\lesssim W1-W2\lesssim1.2$}}~mag, equivalent to spectral types
$\approx{\rm L8-T3.5}$ (Figure~\ref{fig.w1.w1w2.25pc}).  For earlier and later
spectral types, there is still contamination from distant objects, but the
relationship helps.

Once we identified this inequality, we used it instead of $y_{\rm P1}<19.3$~mag
to prioritize candidates for spectroscopic follow-up.  This increased our rate
of success at confirming late-L and T dwarfs within 25~pc, but also meant that
our final sample of 142 candidates was heterogeneously selected.  If we had used
the $W1$ vs. $W1-W2$ inequality from the beginning of the search, we would have
observed almost none of our discoveries with spectral types earlier than
$\approx$L7.

\section{Observations}
\label{obser}

\subsection{Near-infrared Photometry}
\label{obser.nearIR}
Following our initial PS1+\WISE\ database search, our candidates all had
red-optical (\yps, possibly \ips\ and \zps) and mid-infrared ($W1$ and $W2$,
possibly $W3$) photometry.  Our red-optical and mid-IR photometry were drawn
from single sources, so we sought a similarly homogenous set of near-IR
photometry.  The only near-IR survey covering our entire search area is 2MASS,
but most of our candidates were too faint to have been well detected (S/N~$>10$)
by 2MASS, and $\approx$30\% were not detected at all. Thus, we obtained
additional near-IR photometry in order to further vet our candidates prior to
spectroscopic observations.

We therefore searched the UKIDSS Data Release 9 \citep[DR9;][]{Lawrence:2013wf}
and VISTA Hemisphere Survey \citep[VHS;][]{Cross:2012jz} catalogs for JHK
photometry of our candidates on the Mauna Kea Observatories (MKO) filter system
\citep{Simons:2002hh,Tokunaga:2002ex}.  For objects not found in either survey,
we obtained follow-up images using WFCAM \citep{Casali:2007ep} on the 3.8-meter
United Kingdom InfraRed Telescope (UKIRT) as part of the UKIRT Service
Programme.  Observations took place on multiple nights spanning 2010 September
to 2013 December.  We obtained J band images for all observed targets, as well
as H and K bands when time constraints permitted.  Integrations were 5~sec
$\times$ 5~dithers in J and H bands and 10~sec $\times$ 5~dithers in K band,
sufficient to reach S/N$>$20 in most cases.  Data were reduced and calibrated at
the Cambridge Astronomical Survey Unit using the WFCAM survey pipeline
\citep{Irwin:2004ej,Hodgkin:2009jr}.

For objects for which we did not obtain both H and K band images, we used our
near-IR spectra (Section~\ref{obser.spec}) to synthesize photometry in the
missing band(s), using our measured $J$ magnitudes to flux-calibrate the
synthetic magnitudes.  For nine candidates with existing 2MASS photometry for
which we did not obtain UKIRT photometry, we synthesized MKO $JHK$ photometry
from the near-IR spectra using the corresponding 2MASS magnitudes to calibrate
each synthetic magnitude.  All observed and synthetic magnitudes are included in
Table~\ref{tbl.nir}.  Altogether we have MKO system $JHK$ photometry for all but
one of our 142 candidates.

\subsection{Near-infrared Spectroscopy and Spectral Typing}
\label{obser.spec}
We obtained low-resolution near-IR spectra for our candidates between 2012 July
and 2014 January using the NASA Infrared Telescope Facility (IRTF).  We used the
facility spectrograph SpeX \citep{Rayner:2003hf} in prism mode with the $0.5''$
($R\approx120$) and $0.8''$ ($R\approx75$) slits.
We re-observed eight targets between 2015 January and June with the
  $0.5''$ slit to obtain higher S/N and assess possible low-gravity spectral
  signatures (Section~\ref{results.gravity}).
Details of our observations
are listed in Table~\ref{tbl.obslog}.  Contemporaneously with each science
target, we observed a nearby A0V star for telluric calibration.  All spectra
were reduced in the standard way using versions 3.4 and 4.0 of the
Spextool software package \citep{Vacca:2003fw,Cushing:2004bq}.  We aimed for S/N
$\gtrsim30$, sufficient for accurate spectral typing based on overall morphology
(i.e., visual comparison of JHK bands).

Spectral typing of our observed objects was performed by visually comparing our
spectra to the near-infrared M and L~dwarf standards of
\citet{Kirkpatrick:2010dc} and T~dwarf standards of \citet{Burgasser:2006cf},
substituting the T3 standard SDSS~J1206+2813 suggested by \citet{Liu:2010cw}.
When assigning M and L types we followed the procedure of
\citet{Kirkpatrick:2010dc}, first comparing only the $0.9-1.4\ \mu$m portions of
the spectra to evaluate the goodness of fit, and subsequently determining if the
object's $1.4-2.4\ \mu$m flux was unusually red or blue for its spectral type.
For T dwarfs, we compared our spectra to the standards over the entire
$0.9-2.4\ \mu$m window simultaneously.  Our visual typing was able to identify
spectra whose features clearly placed them in between consecutive spectral
standards, so we assume a default uncertainty of $\pm0.5$ sub-types.  In cases
of larger uncertainty, we use ``:'' (e.g., spectral type L6:) to indicate an
uncertainty of $\pm1$ sub-type, and ``::'' to indicate larger uncertainties.

We also determined spectral types for our discoveries using two index-based
systems, which enable objective spectral typing based on specific spectral
features.  \citet[][hereinafter AL13]{Allers:2013hk} developed a system of
near-IR indices that are sensitive to spectral type while insensitive to
differences in gravity.  At least one index is defined for each spectral
sub-type spanning M4--L7, so we calculated these indices for our discoveries in
that range (M7--L7).  Following \citet{Aller:2015ti}, we determined the spectral
type uncertainties derived from each index by performing Monte Carlo simulations
for each object to propagate the measurement errors of our reduced spectra and
the rms uncertainties on index-spectral type conversions into the index
calculations.  We calculated a final index-based spectral type for each object
equal to the weighted average of spectral types from all Monte Carlo runs,
excluding those that fell outside the valid range for each index.

The second system of near-IR indices we used is that of \citet[][hereinafter
B06]{Burgasser:2006cf}, assigning spectral types based on each index using the
polynomial fits from \citet{Burgasser:2007fl}.  The indices were originally
designed to classify T dwarfs, but the polynomials for three of the five indices
are valid for L dwarfs as well.  We calculated a final B06 spectral type for
each object as the mean of the individual index-based types, following the
approach of \citet{Burgasser:2007fl}.  For uncertainties, we use the standard
deviations of the individual index-based types, which are typically $1.0-2.0$
subtypes for L~dwarfs and $0.5-1.5$ subtypes for T~dwarfs.

Neither of these index-based systems covers the full spectral type range of our
objects, whereas visual typing was performed for every object.  Therefore, we
adopt the visually-assigned types as final spectral types for our discoveries.

\section{Results}
\label{results}

\subsection{Ultracool Discoveries}
Our search found 130 ultracool dwarfs, comprising 92\% of our 142 spectroscopic
targets.  Of these, 106 are completely new discoveries.  Twenty-one were
independently discovered and published by other teams during our search, and 3
are previously identified photometric candidates for which we present the first
spectroscopic confirmation.  The SpeX prism spectra for 122 of our
ultracool discoveries are presented in Figure~\ref{fig.all.stack}, and their
spectral types are listed in Table~\ref{tbl.spt.kin}.  The remaining eight
discoveries are candidate members of the Scorpius-Centaurus Association and the
Taurus star-forming region, and their spectra will be presented in a future
paper (W. M. J. Best et al., in preparation).  We include these eight objects in
the summary figures of this paper in order to accurately characterize the
overall results of our search.  The objects previously published by other teams
are listed in Table~\ref{tbl.scooped}.  Seven of our discoveries, all with
photometric distances less than 15~pc, were initially presented in Paper I.  We
also identified 12 non-ultracool objects including a carbon star, an emission
line galaxy, and background stars, which are detailed in
Table~\ref{tbl.interlopers}.

\begin{figure}
\begin{center}   
  \includegraphics[width=0.75\columnwidth, trim = 20mm 0 5mm 0]{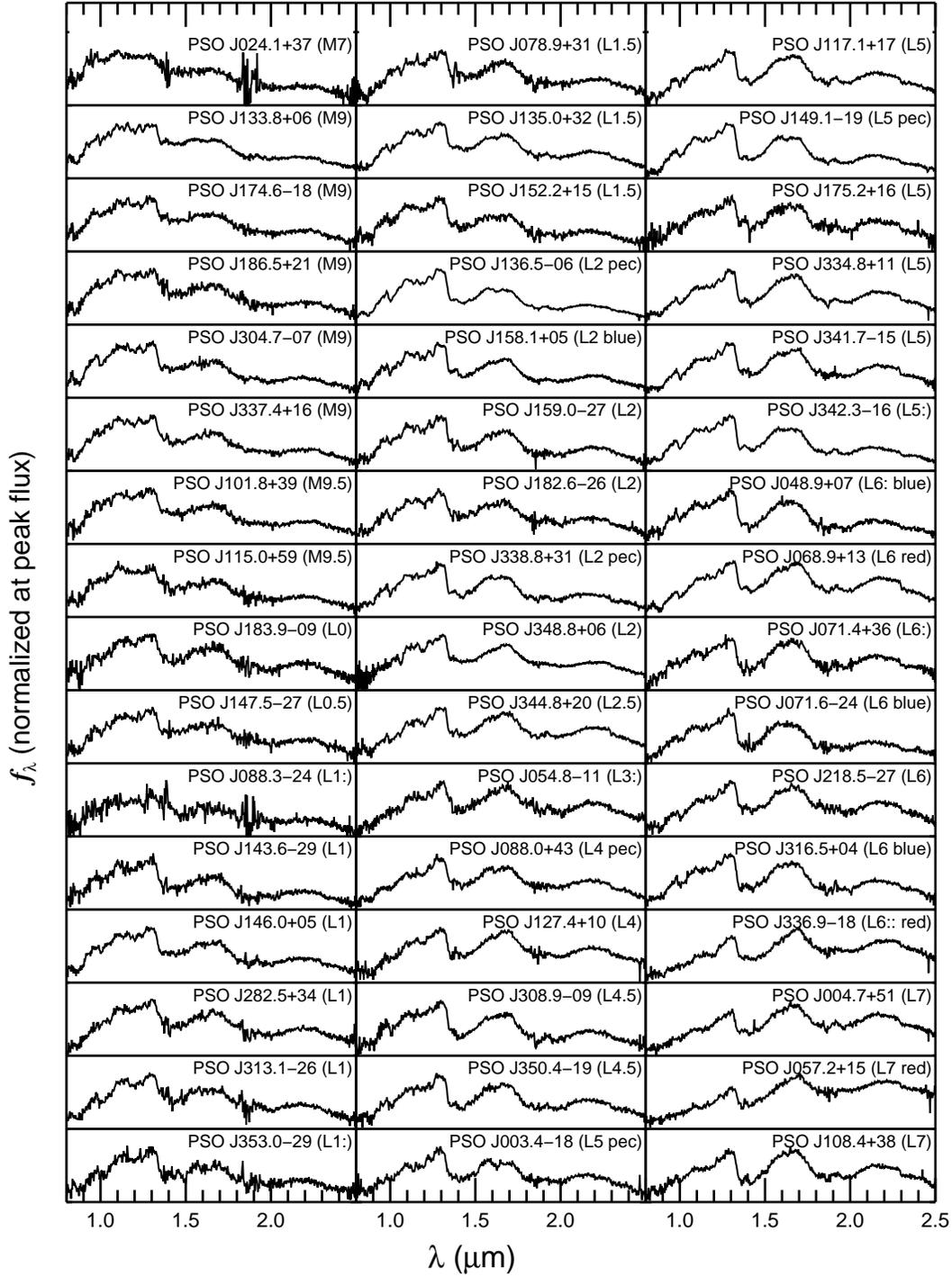}
  \caption{SpeX prism spectra for our discoveries, normalized at
    the peak flux value for each spectrum and arranged from earliest to latest spectral
    type.  Spectra were typed by visual comparison with
    the near-infrared standards defined by \citet{Burgasser:2006cf} and
    \citet{Kirkpatrick:2010dc}.}
  \figurenum{fig.all.stack.1}
\end{center}
\end{figure}

\begin{figure}
\begin{center}
  \includegraphics[width=0.75\columnwidth, trim = 20mm 0 5mm 0]{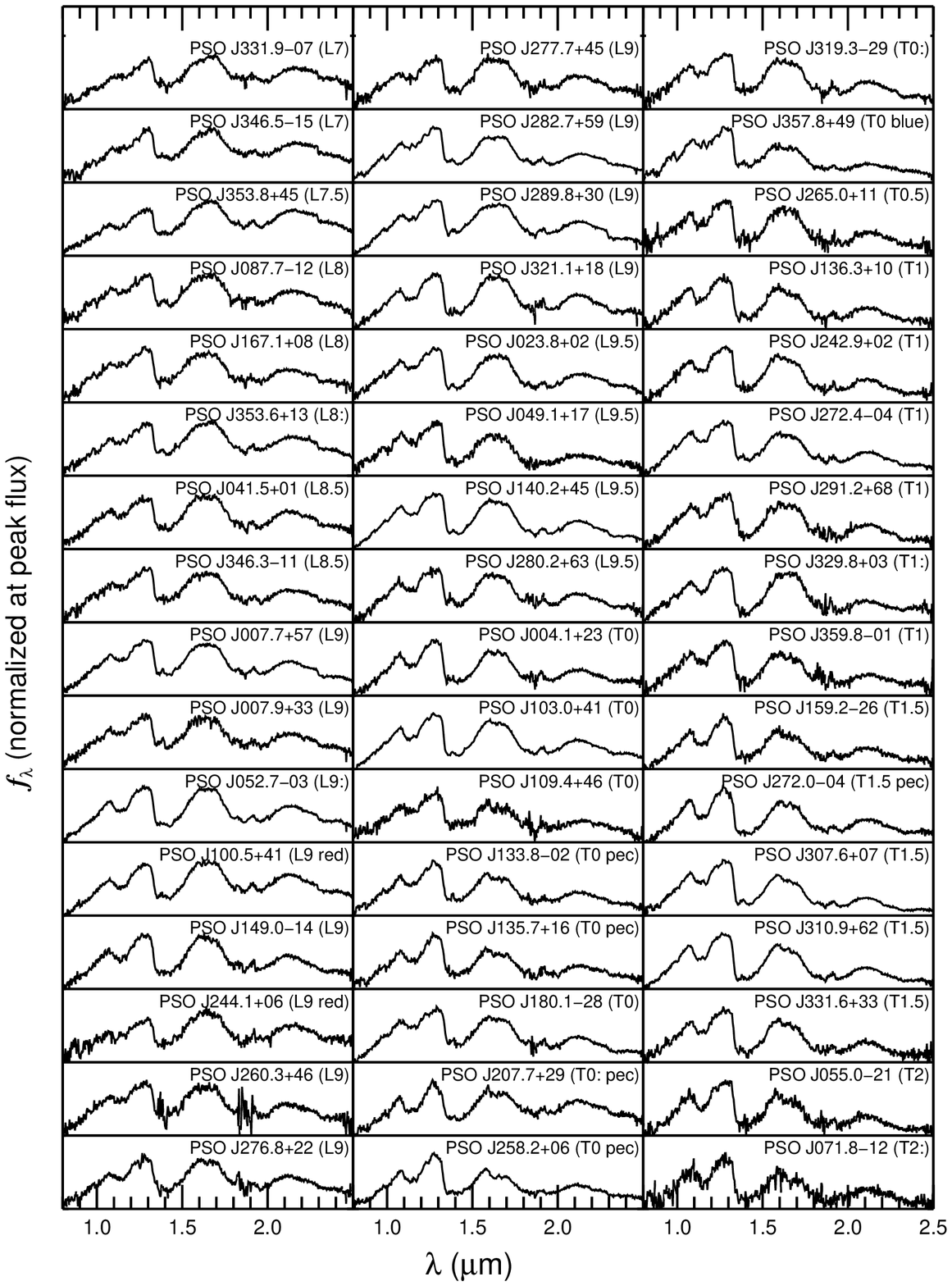}
  \caption{continued.}
  \figurenum{fig.all.stack.2}
\end{center}
\end{figure}

\begin{figure}
\begin{center}
  \includegraphics[width=0.75\columnwidth, trim = 20mm 0 5mm 0]{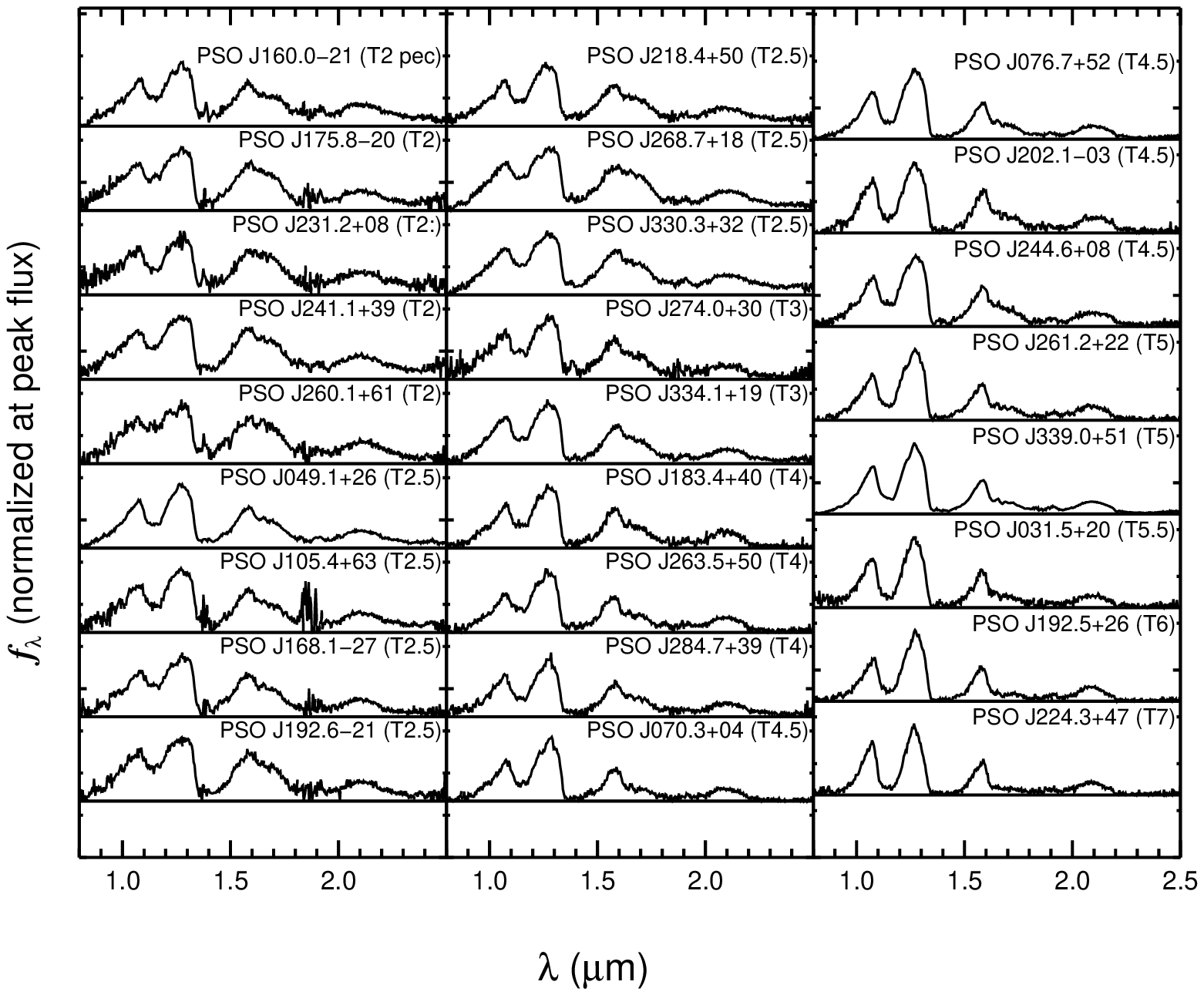}
  \caption{continued.}
  \label{fig.all.stack}
\end{center}
\end{figure}

\begin{figure}
\begin{center}   
  \includegraphics[width=1.00\columnwidth, trim = 20mm 0 5mm 0]{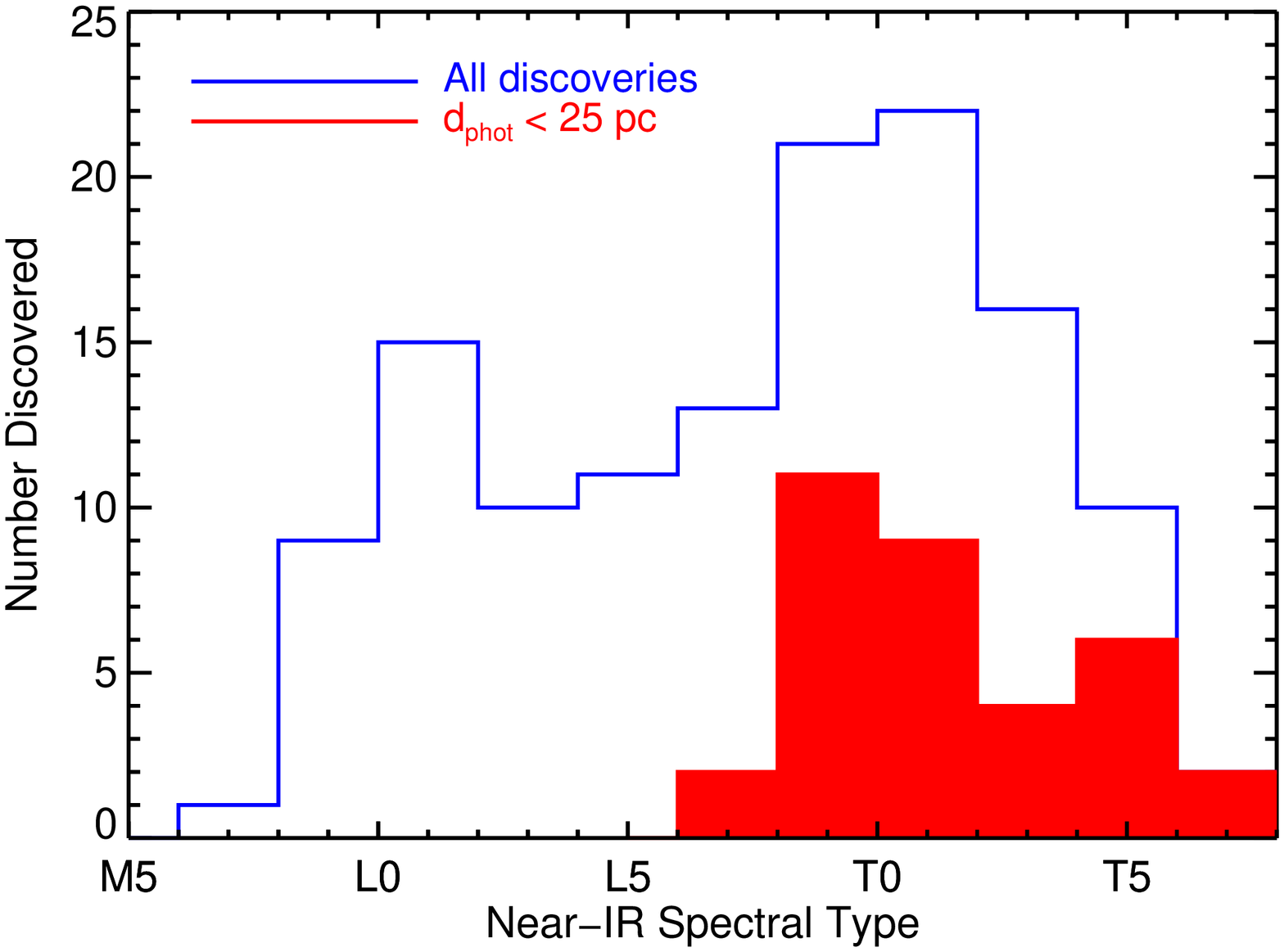}
  \caption{The spectral type distribution of our ultracool discoveries (blue
    open histogram), highlighting objects with W2 spectrophotometric distances
    less than 25~pc (solid red).  We identified 79 objects with
    spectral types L6--T4.5, including 30 within 25~pc.}
\label{fig.spt.hist}
\end{center}
\end{figure}

\begin{figure}
\begin{center}   
  \includegraphics[width=1.00\columnwidth, trim = 20mm 0 5mm 0]{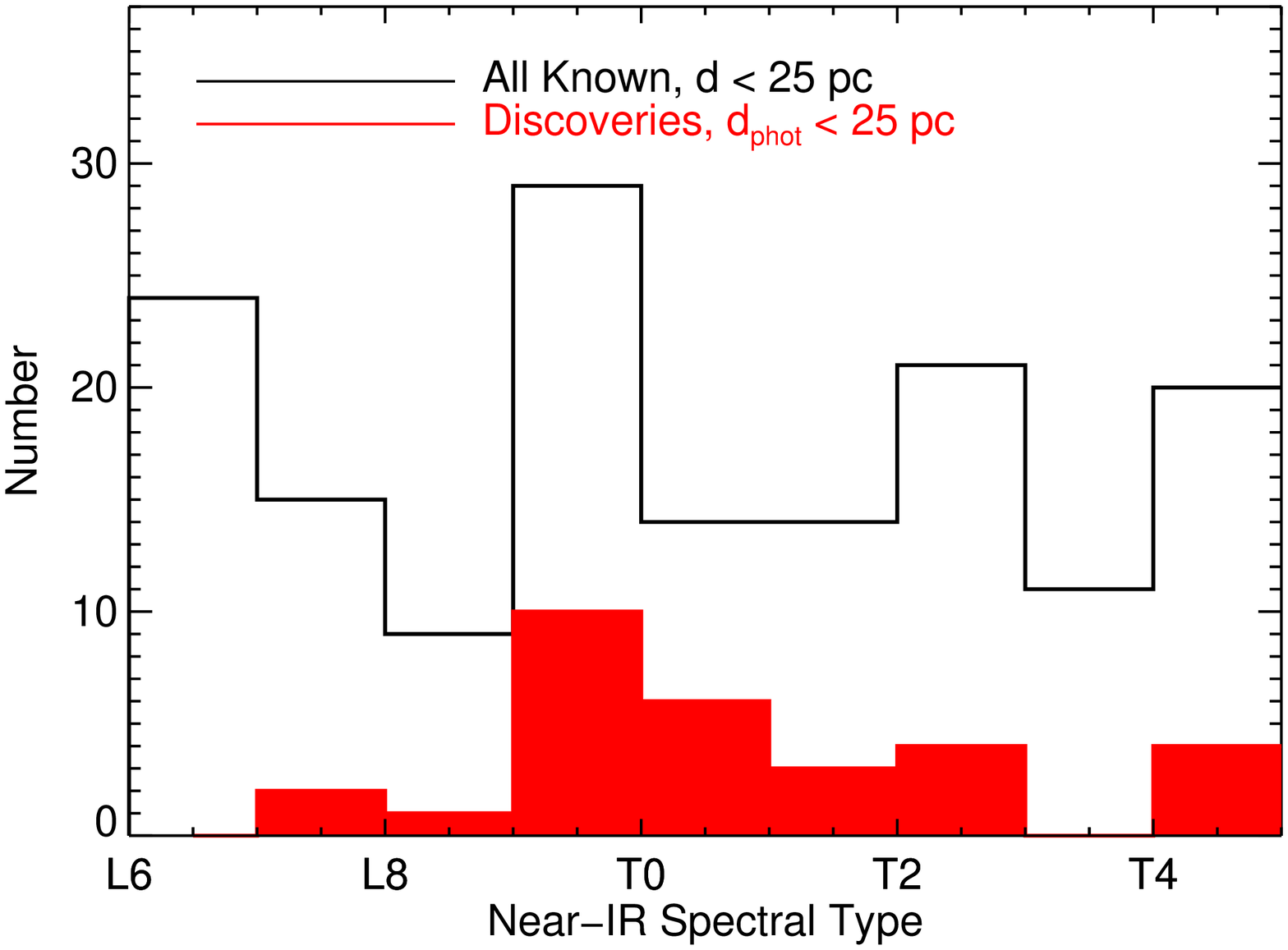}
  \caption{The near-infrared spectral type distribution of all known L/T
    transition dwarfs within 25~pc including our discoveries and previously
    published objects (black open histogram), compared with just our discoveries
    within 25~pc (solid red).  In the middle of the L/T transition (L9--T1.5),
    our discoveries increase the census by over 50\%.}
\label{fig.spt.hist.lttrans}
\end{center}
\end{figure}

Figure~\ref{fig.spt.hist} shows the spectral type distribution of our ultracool
discoveries.  These include 79 L6--T4.5 dwarfs ($\approx$55\% of our sample),
giving us the largest number of L/T transition dwarfs identified by any search
to date.  Figure~\ref{fig.spt.hist.lttrans} compares the spectral type
distributions within 25~pc of our 30 L/T transition discoveries and all known
L/T transition dwarfs.  Note that some previously published L/T transition
dwarfs have spectral types based on optical spectra, while others have near-IR
spectral types.  For a fair comparison with our near-IR discoveries, we can only
use near-IR spectral types for known objects because optical and near-IR types
may not be the same for a given object \citep[e.g.,][]{Kirkpatrick:2010dc}, and
the optical spectral standards for L dwarfs do not include type L9
\citep{Kirkpatrick:1999ev}.  To obtain as complete a sample as possible of
near-IR spectral types for known L/T transition dwarfs, we searched the
literature and identified eight brown dwarfs within 25~pc with optical spectral
types $\ge$L4 but no near-IR spectral types.  Two of these have spectra in the
SpeX Prism Library\footnote{http://pono.ucsd.edu/\mytilde
  adam/browndwarfs/spexprism} which we used to determine near-IR spectral types
(Table~\ref{tbl.new.nir.spt}) following the visual method described in
Section~\ref{obser.spec}.  Two more have optical spectral types L6 and L7,
respectively, and we adopt these as the near-IR types for use in
Figure~\ref{fig.spt.hist.lttrans}.  The remaining four all have optical spectral
types of L5, so we do not include them in the known L/T transition sample.
Figure~\ref{fig.spt.hist.lttrans} shows that our contribution is most
significant for spectral types L9--T1.5, a range of particular interest for
studies focused on photometric variability induced by clouds clearing in
photospheres \citep{Radigan:2014dj}.  Our 19 L9--T1.5 discoveries have increased
the 25~pc census by over 50\%.

We note also that the $W1$ vs. $W1-W2$ inequality we identified in
Section~\ref{WISE.25pc} (Figure~\ref{fig.w1.w1w2.25pc}) preserves all of our L/T
transition discoveries within 25~pc while excluding two-thirds of those farther
away, and also excludes almost all of our discoveries having earlier spectral
types (which all lie beyond 25~pc).

Eleven of our discoveries have spectral features we deemed unusual enough to
assign the object's spectral type a ``peculiar'' designation.  All of these
objects were identified as candidate unresolved binaries, and we discuss them in
Section~\ref{results.binaries}.

\subsection{Spectral Indices and Spectral Types}
\label{results.indices}
In Section~\ref{obser.spec}, we described three methods we used to determine
spectral types for our discoveries: visual comparison with field standards, and
two index-based methods which applied to limited spectral type ranges.  Because
visual typing was the only method used for all objects, we adopted those types
as the final spectral types for our discoveries.  Here we describe the results
of the index-based methods and compare our visual and index-based spectral
types.

\subsubsection{Allers \& Liu (2013) Indices}
\label{results.indices.allers}
\begin{figure}
\begin{center}   
  \includegraphics[width=0.99\columnwidth]{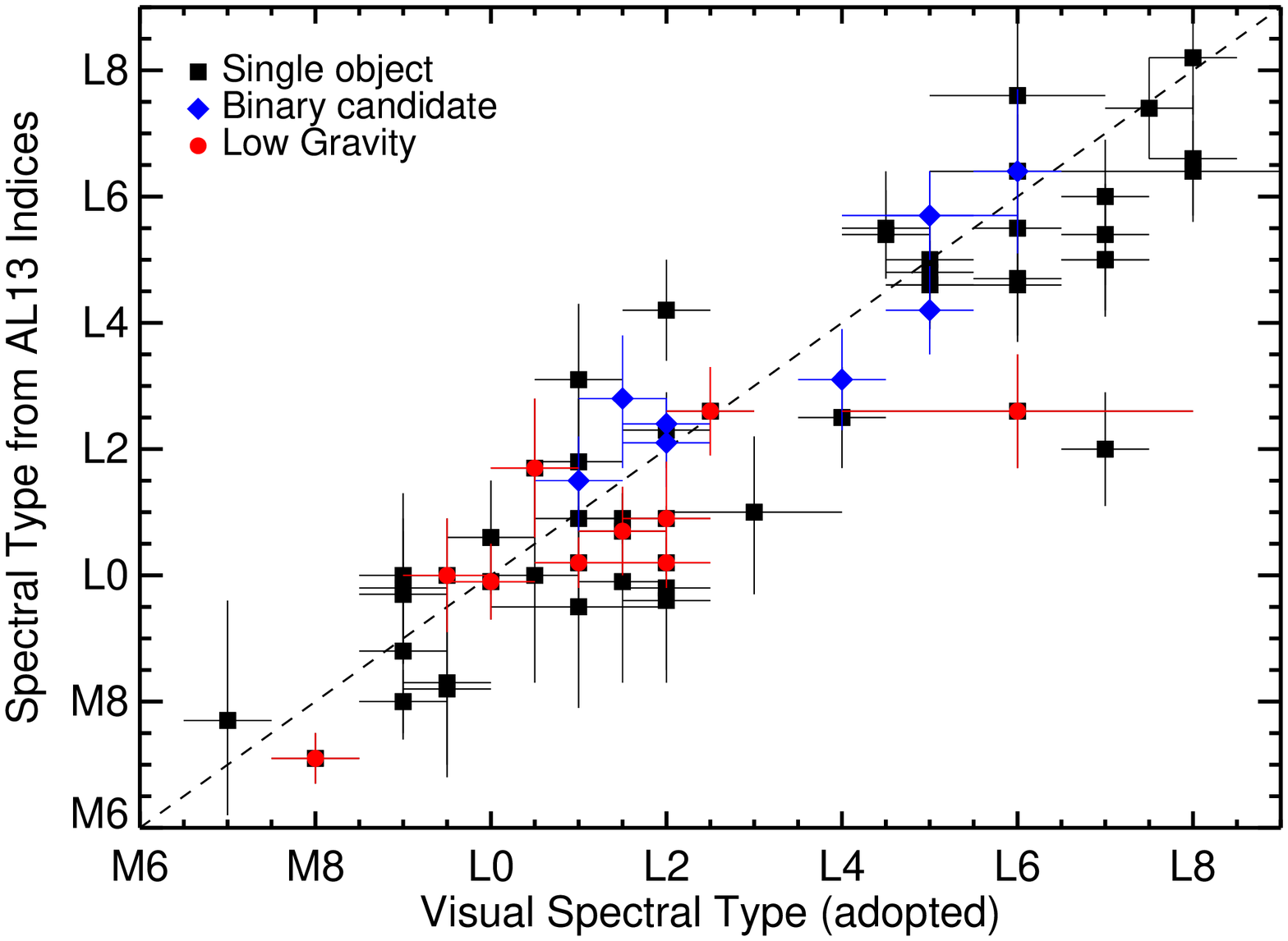}
  \caption{Comparison of our visual spectral types with spectral types
    calculated using the indices of \citet{Allers:2013hk}.  Single objects are
    marked with black squares, binary candidates
    (Section~\ref{results.binaries}) with blue diamonds, and objects having
    \intg\ or \vlg\ gravity classifications
    (Section~\ref{results.gravity}) with red circles.  The dashed black line
    indicates equal spectral types.  There is an overall tendency for our visual
    spectral types to be $\approx$1 sub-type later than the index-based types,
    but no other trend is apparent in the typing of binary candidates or
    low-gravity objects.}
\label{fig.spt.al13.comp}
\end{center}
\end{figure}

Spectral types determined using the AL13 indices are presented in
Table~\ref{tbl.indices.allers}, along with our visual spectral types for these
objects.  Figure~\ref{fig.spt.al13.comp} compares our visual and index-derived
spectral types.  The final index spectral types are mostly consistent with our
adopted visual spectral types, agreeing within the joint $1\sigma$ uncertainties
in 49 out of 60 cases.  Of the remaining 11 objects, only one (PSO~J057.2+15,
discussed below) has an index-derived spectral type more than $2\sigma$
different from the visual type, and none are candidate binaries
(Section~\ref{results.binaries}).  Figure~\ref{fig.spt.al13.comp} shows an
apparent tendency for visual spectral types to be slightly later
($\approx\!0.5-1$ subtypes) than the AL13-based types, but the bias is within
the uncertainties of both typing methods, and does not appear correlated with
low-gravity objects or with possible binaries.  Overall, our results generally
support the effectiveness and insensitivity to gravity of the AL13
low-resolution spectral type indices.

We now discuss the objects whose index-derived spectral types are different from
our adopted visual types by more than $1\sigma$.  The overall tendency here is
for the AL13 index-based spectral types to be earlier than our visual types for
unusually red objects.

{\it PSO~J004.7+51} (visual L7, index L$5.4\pm0.8$) --- The index-based spectral
type is determined by only one index (\htod) and is too late-type for the other
indices to apply.  The spectrum is redder overall than the L7 standard, and this
has probably affected the \htod\ index which measures the depth of the
$\approx$1.9~\um\ water absorption band.

{\it PSO~J057.2+15} (visual L7 red, index L$2.0\pm0.9$) --- This late-L dwarf is
very red. As with PSO~J004.7+51, the index-based spectral type is determined by
only the \htod\ index, and is $3.6\sigma$ different (joint uncertainties) from
the visual L7 type. The $\approx$1.9~\um\ water absorption band measured by the
\htod\ index is significantly shallower than for the L7 standard.

{\it PSO~J068.9+13 (Hya12)} (visual L6 red, index L$4.6\pm0.8$) --- Another
unusually red object, cool enough that only the \htod\ index is available to
determine the spectral type. Visually, the J band is an excellent match to the
L6 standard.  The object was first identified photometrically by
\citet{Hogan:2008ha} and confirmed by \citet{Lodieu:2014jo} as an L3.5 dwarf
based on its optical spectrum. \citet{Lodieu:2014jo} classify Hya12 as a
candidate member of the Hyades based on it sky location, proper motion, and
photometric distance. We tentatively assign this object a gravity classification
of \intg\ (Section~\ref{results.gravity}), which would imply a younger
age than the Hyades. A higher-S/N spectrum would confirm the youth of
PSO~J068.9+13, and parallax and radial velocity measurements are needed to
assess its membership in the Hyades.

{\it PSO~J127.4+10} (visual L4, index L$2.5\pm0.8$) --- Also a redder object,
with three indices contributing to the final index type, but with a good visual
J band match to the L4 standard.  We tentatively give this object a gravity
classification of \vlg\ (Section~\ref{results.gravity}), noting the
triangular H band profile, but the lower S/N of the spectrum precludes a firm
gravity determination.

{\it PSO~J143.6$-$29} (visual L1, index L$3.1^{+1.2}_{-1.3}$) --- The S/N of
this spectrum limits our ability to declare a firm spectral type and increases
the uncertainties in the index-based type, but the J band matches the L1
standard quite well. This object is also discussed in
Section~\ref{results.indices.burg}.

{\it PSO~J159.0$-$27} (visual L2 blue, index L$4.2\pm0.8$) --- The object is
atypically blue for an L2, which may increase the depth of the water absorption
bands used by the indices to determine the spectral type.  Visually it is a good
match in J band to the L2 standard and shows signs of low gravity.  We
tentatively assign this object a gravity classification of \intg\
(Section~\ref{results.gravity}).

{\it PSO~J218.5$-$27} (visual L6, index L$3.9\pm0.8$) --- The modest S/N of this
spectrum affects the indices (and results in the noise spikes in the H band
peak).   The J band spectrum is a good match to the L6 standard.

{\it PSO~J331.9$-$07} (visual L7, index L$5.0\pm0.8$) --- This object is an
excellent visual match to the L7 standard, and the index-based spectral type is
determined by only one index (\htod).

{\it PSO~J336.9-18} (visual L6::~red, index L$2.1\pm0.7$) --- This extremely red
L dwarf has a \vlg\ gravity class (Section~\ref{results.gravity}).
Visually, it is not a clear J band match to any field standard, but the shape of
the J band peak and the depth of the $\approx$1.4~\um\ water absorption band are
decent matches to the L6 standard.  The index-based type depends on two
indices. This object demonstrates the difficulty in assigning spectral types to
unsually red L dwarfs.

{\it PSO~J346.5$-$15} (visual L7, index L$5.0\pm0.9$) --- This object is a good
visual match to the L7 standard, albeit slightly blue, and the index-based
spectral type is determined by only one index (\htod) in the fairly noisy
spectrum.

{\it PSO~J348.8+06} (visual L2, index L$0.2\pm0.4$) --- We classify this object
as \vlg\ (Section~\ref{results.gravity}), as it shows many signs of low
gravity in its spectra.  The J band spectrum is a good match to the L2 standard.

\subsubsection{Burgasser et al. (2006) Indices}
\label{results.indices.burg}
\begin{figure}
\begin{center}   
  \includegraphics[width=0.99\columnwidth]{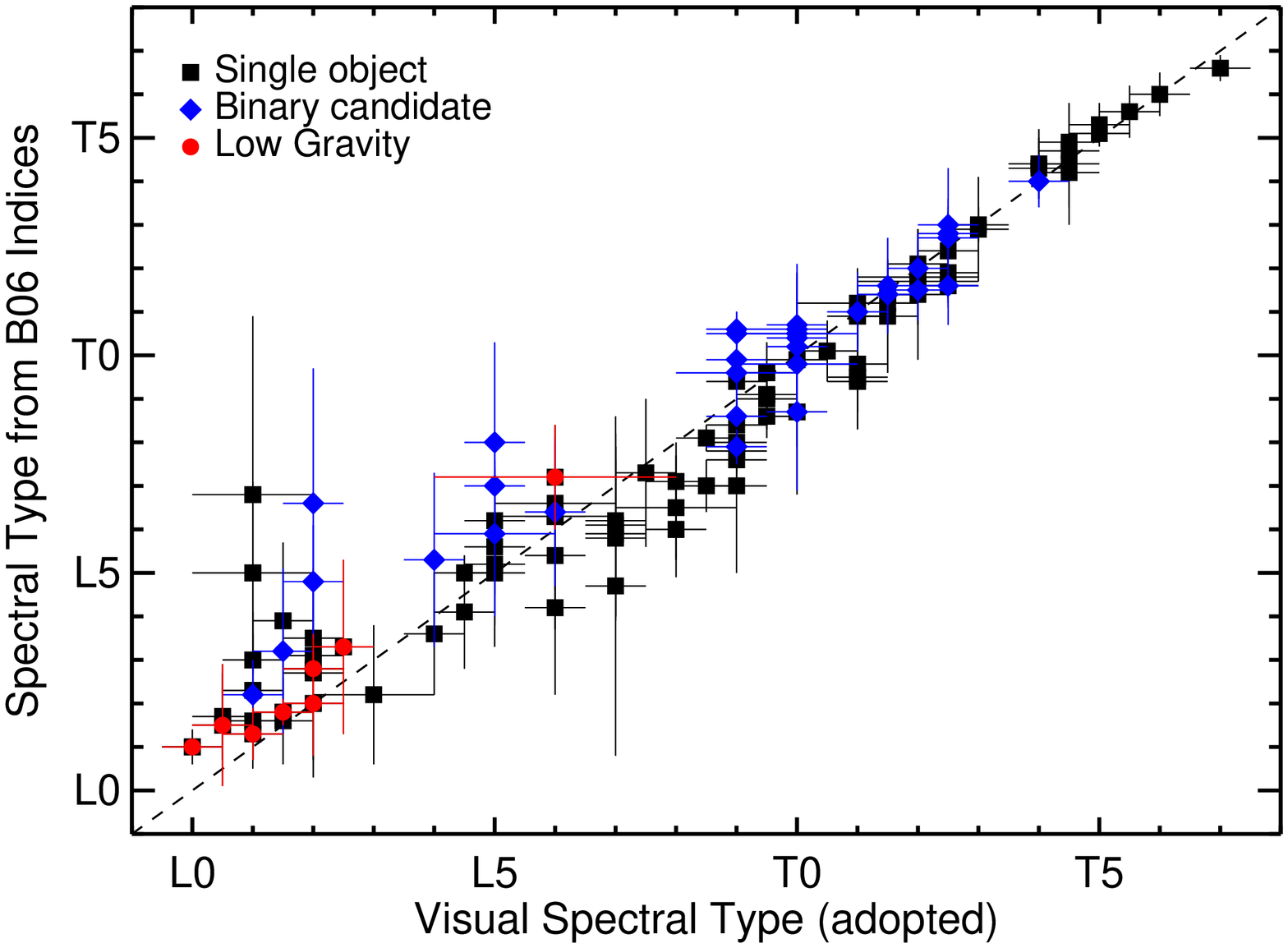}
  \caption{Comparison of our visual spectral types with spectral types
    calculated using the indices of \citet{Burgasser:2006cf}, using the same
    symbols as Figure~\ref{fig.spt.al13.comp}.  Compared to the index-based
    spectral types, our visual spectral types are $\approx$1 sub-type earlier
    for early-L dwarfs, $\approx$1 sub-type later for late-L dwarfs, and in good
    agreement for T dwarfs.  No other trend is apparent in the typing of
    low-gravity objects, but the objects with the largest discrepancy in types
    tend to be binary candidates.  The two objects with visual L1 types and
    index-based types $\ge$L5 have spectra with S/N$<20$, so their index-based
    types are not reliable.}
\label{fig.spt.burg.comp}
\end{center}
\end{figure}

We show spectral types calculated using the B06 indices for all of our L and T
dwarf discoveries in Table~\ref{tbl.indices.burg}, along with the adopted visual
spectral types.  Figure~\ref{fig.spt.burg.comp} compares our visual and B06
index-based spectral types.  The spectral types from the two methods agree very
well for T dwarfs (none differ by more than $1\sigma$).  17 of the 70 L dwarfs
have $>$1$\sigma$ differences in spectral type, even though the uncertainties on
the L dwarf spectral types are larger.  This larger scatter in L dwarf types is
consistent with the smaller number of indices used as well as the wider variety
of spectral features and colors seen in L dwarfs than in T dwarfs
\citep[e.g.,][]{Kirkpatrick:2010dc}, and was previously noted by
\citet{Burgasser:2010df} who compared literature and B06 spectral types.  The
visual types appear to skew $\approx$1 sub-type earlier than the B06 types for
early-L dwarfs and $\approx$1 sub-type later for late-L dwarfs.  We see no
significant correlation between low gravity and the differences in visual and
B06 spectral types.  The uncertainties in the index-based types are typically
larger than the rms uncertainties of the \citet{Burgasser:2007fl} polynomials
but do not appear to be correlated with the S/N of our spectra.  Overall, our
results strongly support the effectiveness of the B06 indices for T dwarf
classification, but B06 index-based spectral types for L dwarfs may differ
visually determined ones by $\approx0.5-1.0$ subtypes.

Below we comment on objects whose B06 index-derived spectral types 
differ from our adopted visual types by more than $1\sigma$.

{\it PSO~J003.4$-$18 (2MASS~J0013$-$1816)} (visual L5 pec, index L$8.0\pm2.3$)
--- We identify this object as a strong binary candidate in
Section~\ref{results.binaries.strong}.  Our visual L5 type was determined by the
J band shape, but H and K bands have features typical of cooler dwarfs.

{\it PSO~J007.9+33} (visual L9, index L$7.6\pm0.4$) --- We find a good fit to
this object's J band profile with the L9 standard, but PSO~J007.9+33 has
slightly shallower water absorption bands at $\approx$1.15~\um\ and
$\approx$1.4~\um\ which are suggestive of an earlier spectral type.

{\it PSO~J087.7$-$12} (visual L8, index L$6.0\pm1.1$) --- Low S/N likely affects
the indices for this spectrum, which is a good visual fit to the L8 standard.

{\it PSO~J088.3$-$24} (visual L1:, index L$6.8\pm4.1$) --- This spectrum has
only S/N $\approx10$.  The overall early-L morphology is apparent, but more
accurate typing by any method will require a higher S/N spectrum.

{\it PSO~J136.5-06} (visual L2 pec, index $L6.6\pm3.1$) --- This strong binary
candidate (Section~\ref{results.binaries.strong}) shows multiple signs of an L+T
blend, and consequently the individual indices gives spectral types ranging from
L4.2 to T1.2.

{\it PSO~J143.6$-$29} (visual L1, index L$3.0\pm0.6$) --- This is the only
object among our discoveries whose visual spectral type disagrees by more than
$1\sigma$ with spectral types derived from both the AL13
(Section~\ref{results.indices.allers}) and B10 indices.  The index-based
classifications are L3.1 and L3.0, in close agreement, but this spectrum is
noisy enough to make those types unreliable.  We see a good J band match to the
L1 standard.

{\it PSO~J149.0$-$14} (visual L9, index T$0.5\pm0.5$) --- This medium-ranked
binary candidate (Section~\ref{results.binaries.medium}) shows an overall L9
morphology, but there are subtle signs of methane absorption that shift the
index-based type to a T dwarf.

{\it PSO~J149.1$-$19} (visual L5, index L$7.0\pm0.6$) --- One of our strongest
binary candidates (Section~\ref{results.binaries.strong}), this object's
spectrum shows several L+T dwarf blend features along with a good J band fit to
the L5 standard.

{\it PSO~J158.1+05} (visual L2~blue, index L$3.5\pm0.3$) --- The J-band spectrum
is a good fit to the L2 standard, but the overall bluer spectral slope includes
deeper water bands that point to a later index-based spectral type.

{\it PSO~J183.9$-$09} (visual L0, index L$1.0\pm0.4$) --- The spectral types differ by
only $1.1\sigma$, which is actually surprising given the very low S/N of this
early-L spectrum.


{\it PSO~J244.1+06} (visual L9 red, index L$7.8\pm0.5$) --- The spectral types differ by
only $1.2\sigma$.  This minor discrepancy may be due to the modest S/N of the spectrum
and/or the object's unusually red color.

{\it PSO~J282.7+59 (WISE~J1851+5935)} (visual L9, index L$7.9\pm0.4$) --- This
weak binary candidate (Section~\ref{results.binaries.weak}) was discussed in
detail in Paper I, and was assigned a spectral type of L9 pec by
\citet{Thompson:2013kv}.  The object's blue color may contribute to the slightly
earlier index-based spectral type.

{\it PSO~J321.1+18} (visual L9, index T$0.6\pm0.4$) --- This weak binary
candidate (Section~\ref{results.binaries.weak}) features water absorption bands
and an H band peak more similar to an early-T dwarf, which explains the T0.6
index-based type.

{\it PSO~J338.8+31} (visual L2 pec, index L$4.8\pm1.3$) --- The deeper water
absorption bands of this strong binary candidate
(Section~\ref{results.binaries.strong}) lead to a later index-based spectral
type.

{\it PSO~J346.3$-$11} (visual L8.5, index L$7.0\pm0.6$) --- The J band shape is
a clear fit to the L8 and L9 standards.  Surprisingly, this object was assigned
an earlier spectral type by the indices despite the depth of the water
absorption bands and the slightly bluer color.

{\it PSO~J353.0$-$29} (visual L1:, index L$5.0\pm2.1$) --- This spectrum has
only S/N $\approx15$.  The overall early-L morphology is apparent, but more
accurate typing by any method will require a higher S/N spectrum.

\subsection{Colors and PS1 Photometry}
\begin{figure}
\begin{center}   
  \includegraphics[width=1.00\columnwidth, trim = 20mm 0 9mm 0]{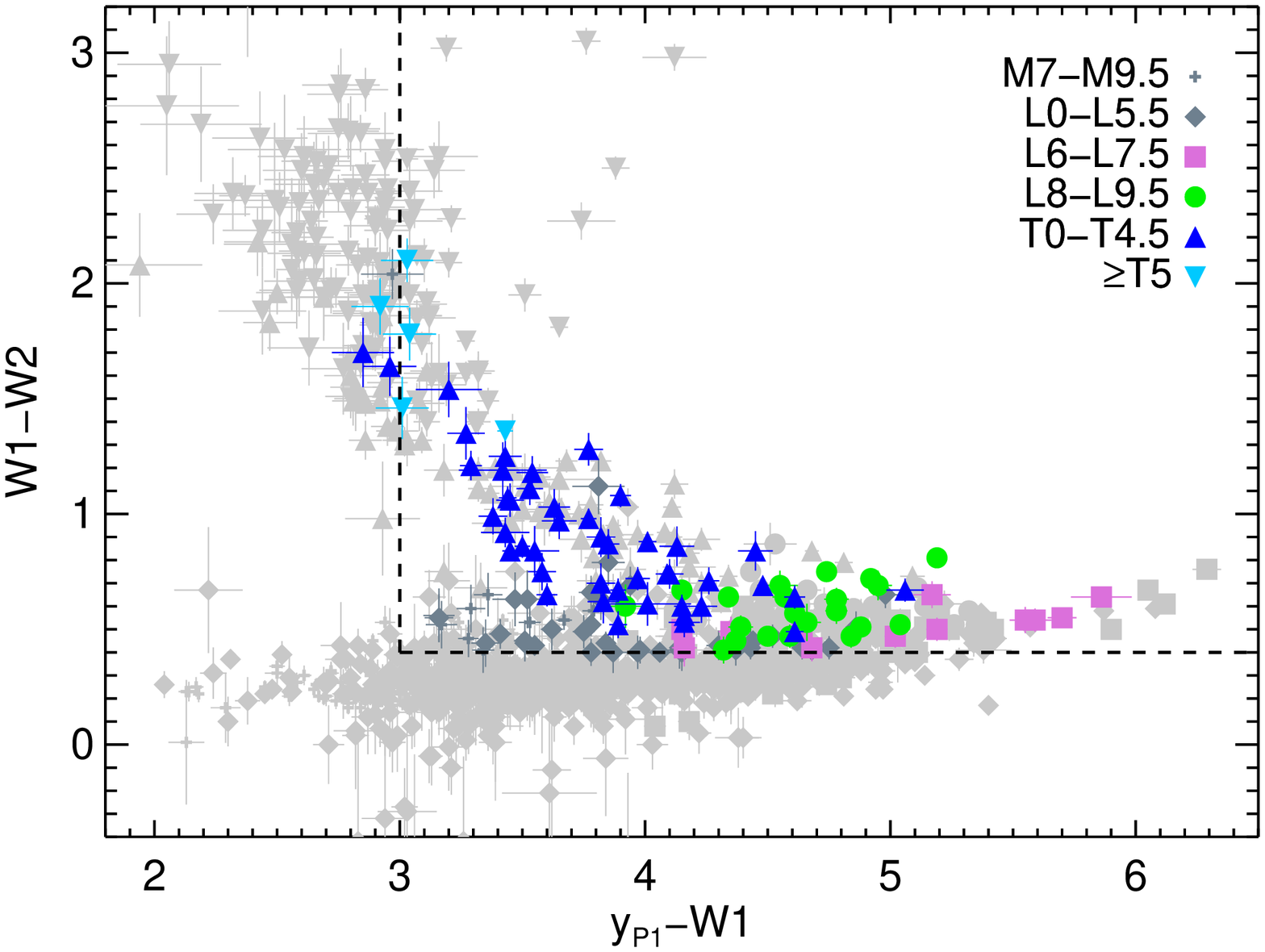}
  \caption{$W1-W2$ vs. $y_{\rm P1}-W1$ diagram showing our discoveries in dark
    gray and colors for spectral type bins (see legend at upper right), and
    previously known ultracool dwarfs in light gray using the same symbols as
    for our discoveries.  We selected objects above and to the right of the
    dashed lines using \yps\ photometry from 2012 January; this plot shows \yps\
    values as of 2015 March.  Only 4 of our 130 discoveries would have been
    excluded from our search using the newer \yps\ photometry.}
\label{fig.w1w2.yw1}
\end{center}
\end{figure}

\begin{figure}
\begin{center}   
  \includegraphics[width=1.00\columnwidth, trim = 20mm 0 5mm 0]{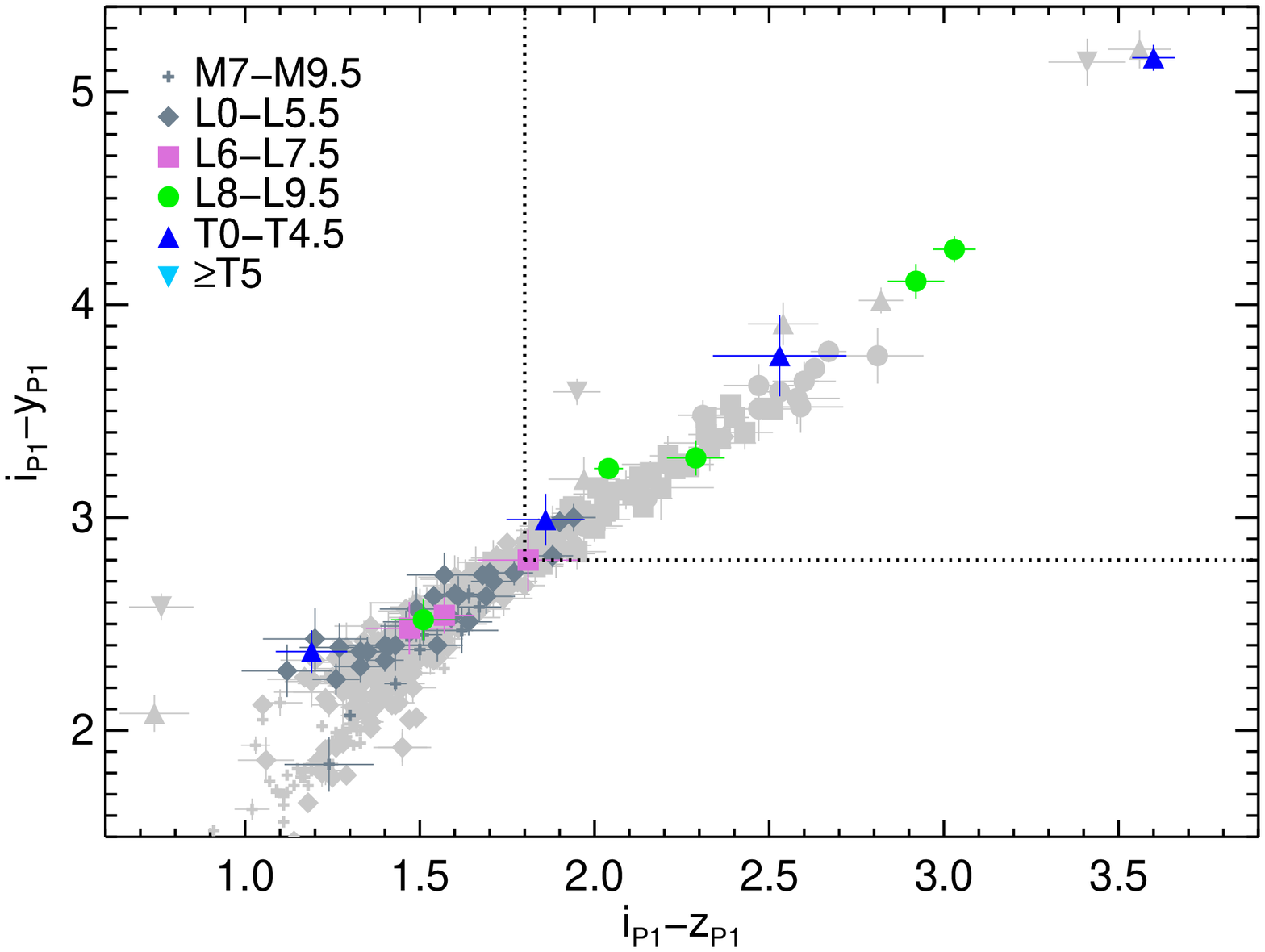}
  \caption{\iy\ vs. \iz\ diagram for our discoveries and known ultracool dwarfs,
    using PS1 photometry from 2015 March and the same symbols as in
    Figure~\ref{fig.w1w2.yw1}.  The dotted black lines indicate the color cuts
    used in our search; we selected objects above and to the right of the dotted
    lines, but only enforced each cut for objects having $\sigma<0.2$~mag and
    at least two detections in both \ips\ and \zps\ in the 2012 January epoch of PS1
    photometry.  Most of our discoveries with spectral types less than L6 would
    have been culled from our search using the most recent PS1 photometry, which has
    many more detections in \ips\ for our objects.  This would have resulted in
    a significantly higher fraction ($\approx$80\%) of L/T transition discoveries,
    but far fewer discoveries of young objects.}
\label{fig.iy.iz}
\end{center}
\end{figure}

\begin{figure}
\begin{center}   
  \includegraphics[width=1.00\columnwidth, trim = 20mm 0 5mm 0]{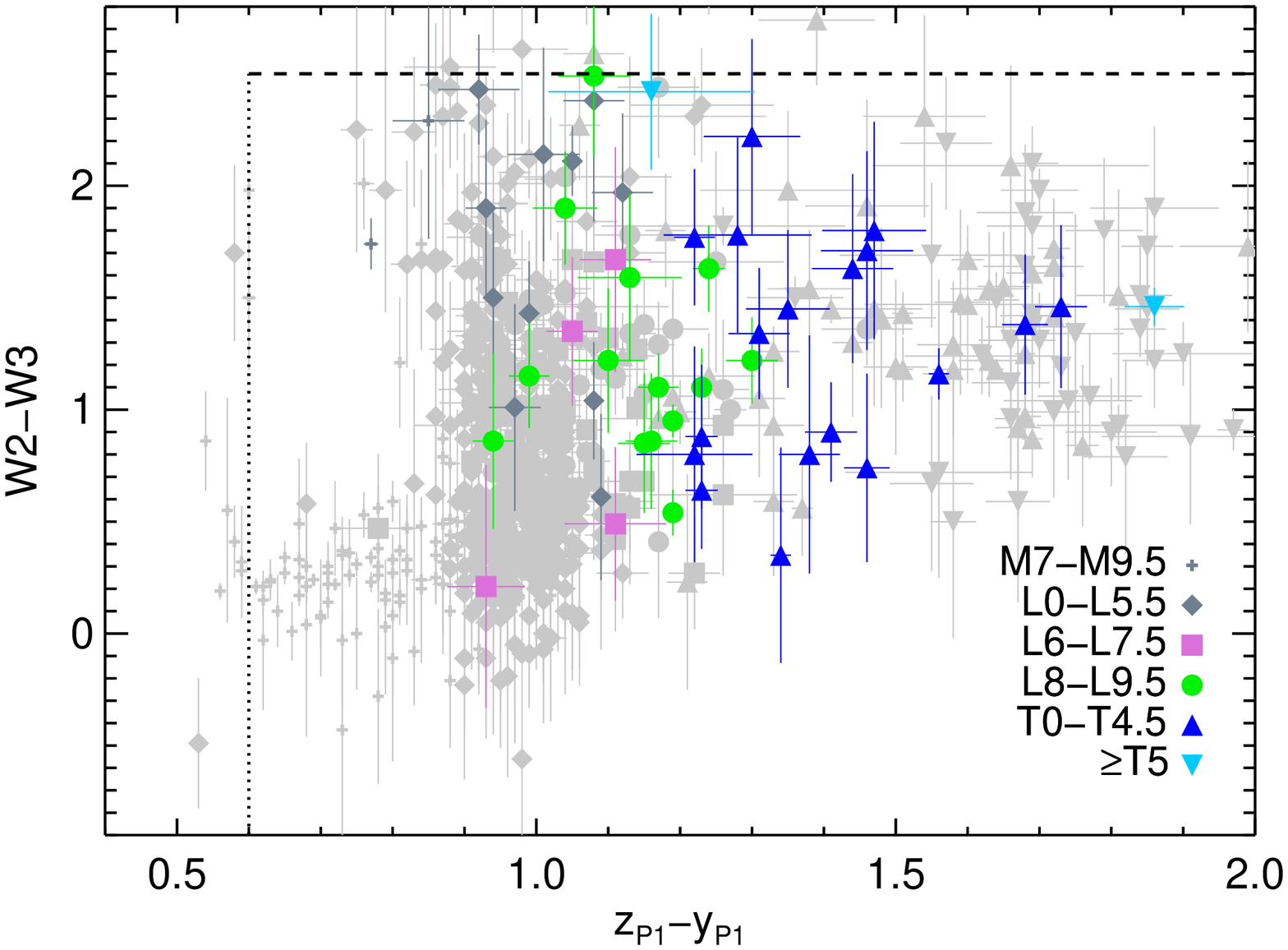}
  \caption{ \wbwc\ vs. \zy\ diagram for our discoveries and known ultracool
    dwarfs, using PS1 photometry from 2015 March and the same symbols as in
    Figure~\ref{fig.w1w2.yw1}.  The vertical dotted line indicates our \zy\ cut,
    which we applied only to objects with $\sigma_z<0.2$~mag and having at least
    two \zps\ detections in the 2012 January epoch of PS1 photometry.  The
    horizontal dashed line represents our \wbwc\ cut, which we applied to all
    objects in our search in order to exclude galaxies.  We selected objects
    below and to the right of these lines.}
\label{fig.w2w3.zy}
\end{center}
\end{figure}

\begin{figure}
\begin{center}   
  \includegraphics[width=1.00\columnwidth, trim = 10mm 0 5mm 0]{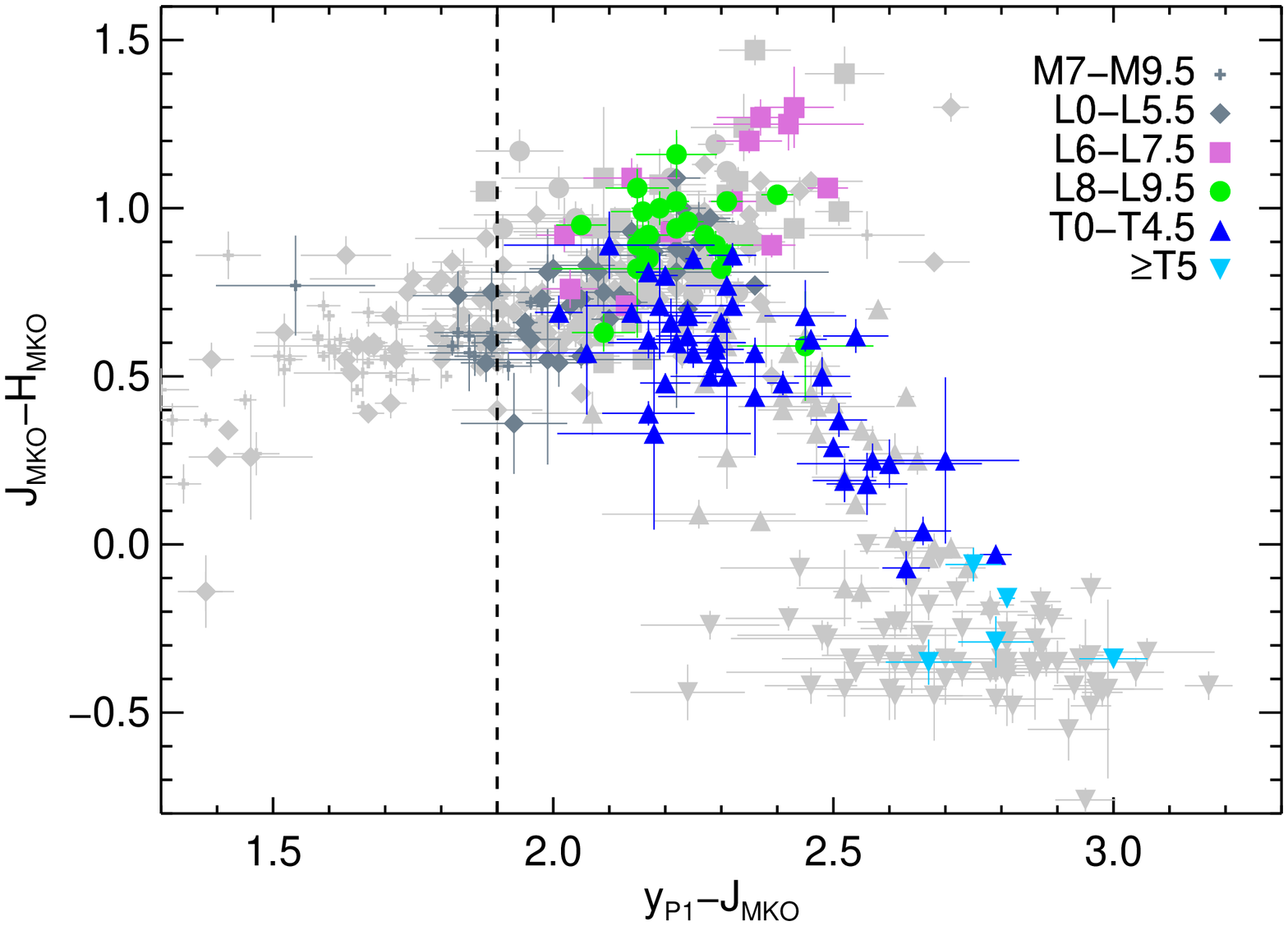}
  \caption{$J_{\rm MKO}-H_{\rm MKO}$ vs. $y_{\rm P1}-J_{\rm MKO}$ diagram for
    our discoveries and known ultracool dwarfs, using \yps\ photometry from
    2015 March and the same symbols as in Figure~\ref{fig.w1w2.yw1}.
    We selected objects to the right of the dashed line using \yps\ photometry
    from 2012 January.  (We did not use $J-H$ color to screen targets in our
    search, but it has been used in many previous near-IR searches for T
    dwarfs.)  The updated PS1 photometry has shifted eleven late-M and early-L
    dwarfs outside of our $y_{\rm P1} -J_{\rm MKO}\ge1.9$~mag cut.}
\label{fig.JH.yJ}
\end{center}
\end{figure}

\begin{figure}
\begin{center}   
  \includegraphics[width=1.00\columnwidth, trim = 10mm 0 5mm 0]{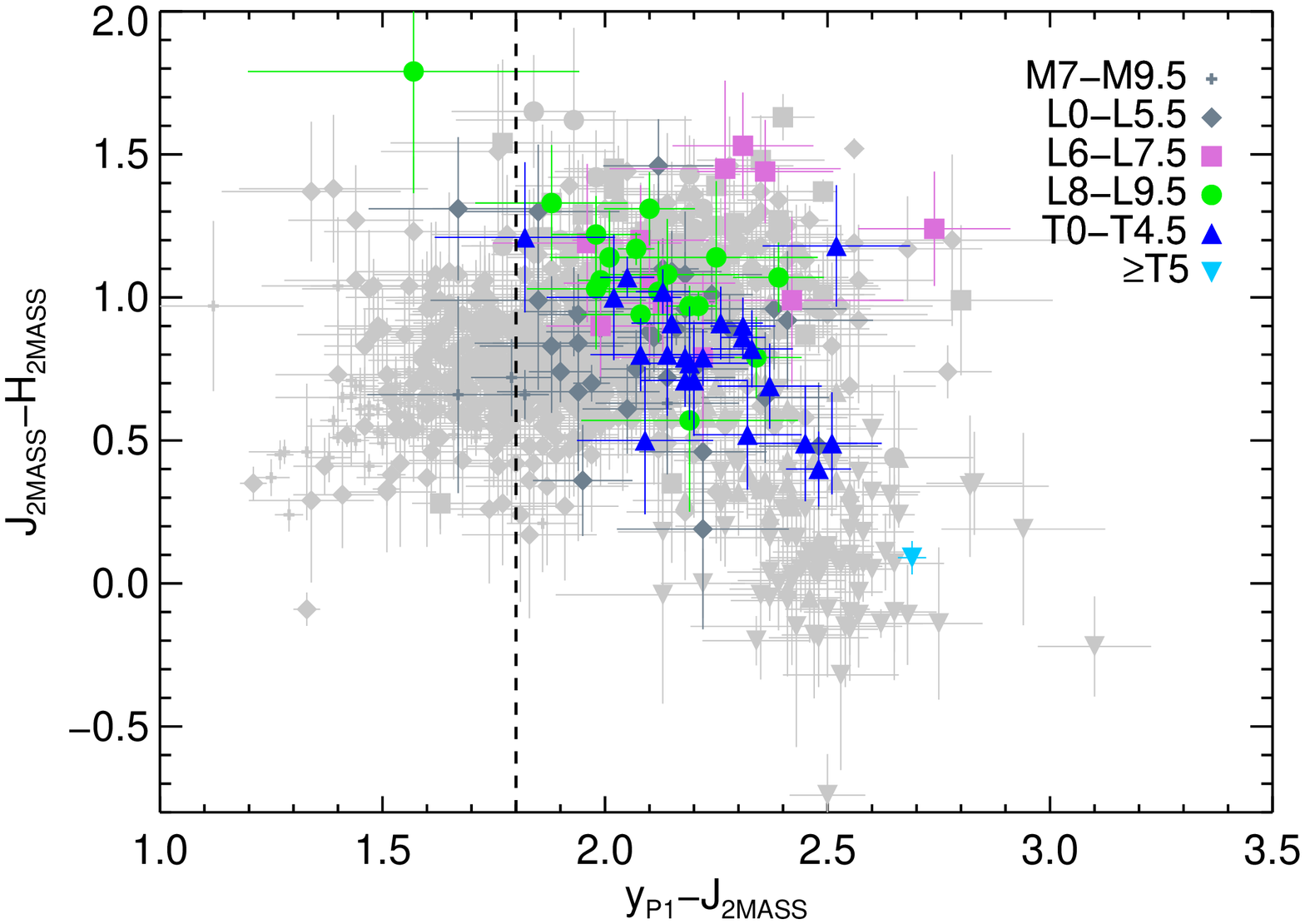}
  \caption{Same as Figure~\ref{fig.JH.yJ}, but using 2MASS photometry for
    J and H bands instead of MKO.  The updated PS1 photometry has revised the
    unusually blue $y_{\rm P1} -J_{\rm 2MASS}$ colors of discoveries we
    presented in Paper I, bringing them in line with other field objects.}
\label{fig.J2H2.yJ2}
\end{center}
\end{figure}

Figures~\ref{fig.w1w2.yw1}--\ref{fig.J2H2.yJ2} show the colors of our
discoveries and previously known ultracool dwarfs that we used to identify the
candidate L/T transition dwarfs in our search (Section~\ref{method}).  Those
color criteria were designed using PS1 photometry from January 2012 (Processing
Version 1).  Since then, ongoing PS1 observations and image processing have
produced more detections and improved measurements.  We have chosen to use PS1
data from March 2015 (Processing Version 2) to make
Figures~\ref{fig.w1w2.yw1}--\ref{fig.J2H2.yJ2} because of the improved
photometric precision and the increased number of detections, particularly
valuable in \ips.  WISE photometry is from the All-sky release
\citep{Cutri:2012wm}.

These figures demonstrate the success of our color criteria.  In particular,
Figure~\ref{fig.w1w2.yw1} shows the two colors at the core of our screening
process, \ywa\ and \wawb.  Our \ywa\ $\ge3.0$~mag cut is very effective,
removing only a few T dwarfs at the cool end of the L/T transition (spectral
type $\approx$ T4--T5).  Our \wawb\ $\ge0.4$~mag cut is similarly effective,
excluding some L6--L7 dwarfs but also culling many more earlier-type objects.
We also note that four of our discoveries now have \ywa\ $<3.0$~mag with the
updated PS1 photometry; these objects have spectral types M7, T4, T4.5, and
T5.5.

The updated PS1 photometry includes \ips\ detections of 50 of our spectroscopic
targets (compared with only 3 detections from the January 2012 PS1 photometry),
nearly all of which have spectral types earlier than L6.  The new \iy\ and \iz\
colors (Figure~\ref{fig.iy.iz}) would actually have culled most of these
SpT~$<$~L6 objects from our candidate list, significantly increasing the
efficiency of our search for L/T transition dwarfs (from 55\% to $\approx$80\%)
but also eliminating most of our young discoveries
(Section~\ref{results.gravity}).  \ips\ detections of T dwarfs remain rare
($\approx$10 in Processing Version 2), as these objects are optically extremely
faint.  Figure~\ref{fig.w2w3.zy} shows the usefulness of \zy\ for separating M
dwarfs from L and T dwarfs, and similarly for $y_{\rm P1} -J_{\rm MKO}$ in
Figure~\ref{fig.JH.yJ}.  The new \yps\ photometry would also have rejected 11 of
our late-M and early-L dwarf discoveries which now have
$y_{\rm P1} -J_{\rm MKO}<1.9$.

In Paper I, we reported unusually blue $y_{\rm P1} -J_{\rm 2MASS}$ colors for
six of our bright nearby discoveries.  We note that this color deviation has now
disappeared.  The updated \yps\ photometry for these objects is slightly
fainter, which brings the $y_{\rm P1} -J_{\rm 2MASS}$ colors of these
discoveries into the locus of other field objects (Figure~\ref{fig.J2H2.yJ2}).

\subsection{Low-Gravity Objects}
\label{results.gravity}
Signatures of low gravity in the spectra of ultracool dwarfs are a reflection of
the extended radii of young objects that are still contracting.  AL13 identified
a set of near-IR spectral indices at low ($R\approx100$) and intermediate
($R\approx1000$) spectral resolution to assess the surface gravity of M4--L7
dwarfs, and thereby to identify ultracool dwarfs younger than $\approx$200 Myr.
Briefly, the low-resolution indices measure the depths of the \fehz\ (0.99~\um),
\voz\ (1.06~\um), and \kij\ (1.24~\um) absorption features relative to the
continuum, as well as the shape of the H band continuum over $1.47-1.67$~\um.
Based on these indices, an object is assigned a score of 0, 1, or 2, which
correspond to classes of field gravity (\fldg, ages $\gtrsim200$~Myr),
intermediate gravity (\intg, ages $\approx50-200$~Myr), and very low
gravity (\vlg, ages $\approx10-30$~Myr), respectively.  (Note that the
age calibration of the these gravity classes is only notional, and more work is
needed in this area.)  The median value of the index scores is the final gravity
score for the object.

We calculated low-resolution indices and gravity scores for our M and L dwarfs
(through L7) using the approach described in \citet{Aller:2015ti}, performing
Monte Carlo simulations for each object to propagate the measurement errors of
our reduced spectra into the index calculations.  Most of our spectra have
$R\approx75$, so the indices were computed using only a few resolution elements.
We found that spectra with a mean S/N $\lesssim30$ measured over the interval
$1.20-1.31$~\um\ (encompassing the bulk of the J-band flux for L dwarfs) produced
gravity scores with uncertainties too large to contain useful information, and
we discarded the scores for those objects.  We visually inspected the remaining
(higher S/N) spectra to confirm the gravity class, and flagged those with low
enough S/N that we could not confirm the gravity class by eye.

\begin{figure}
\begin{center}   
  \includegraphics[width=0.99\columnwidth, trim = 0 95mm 0 0]{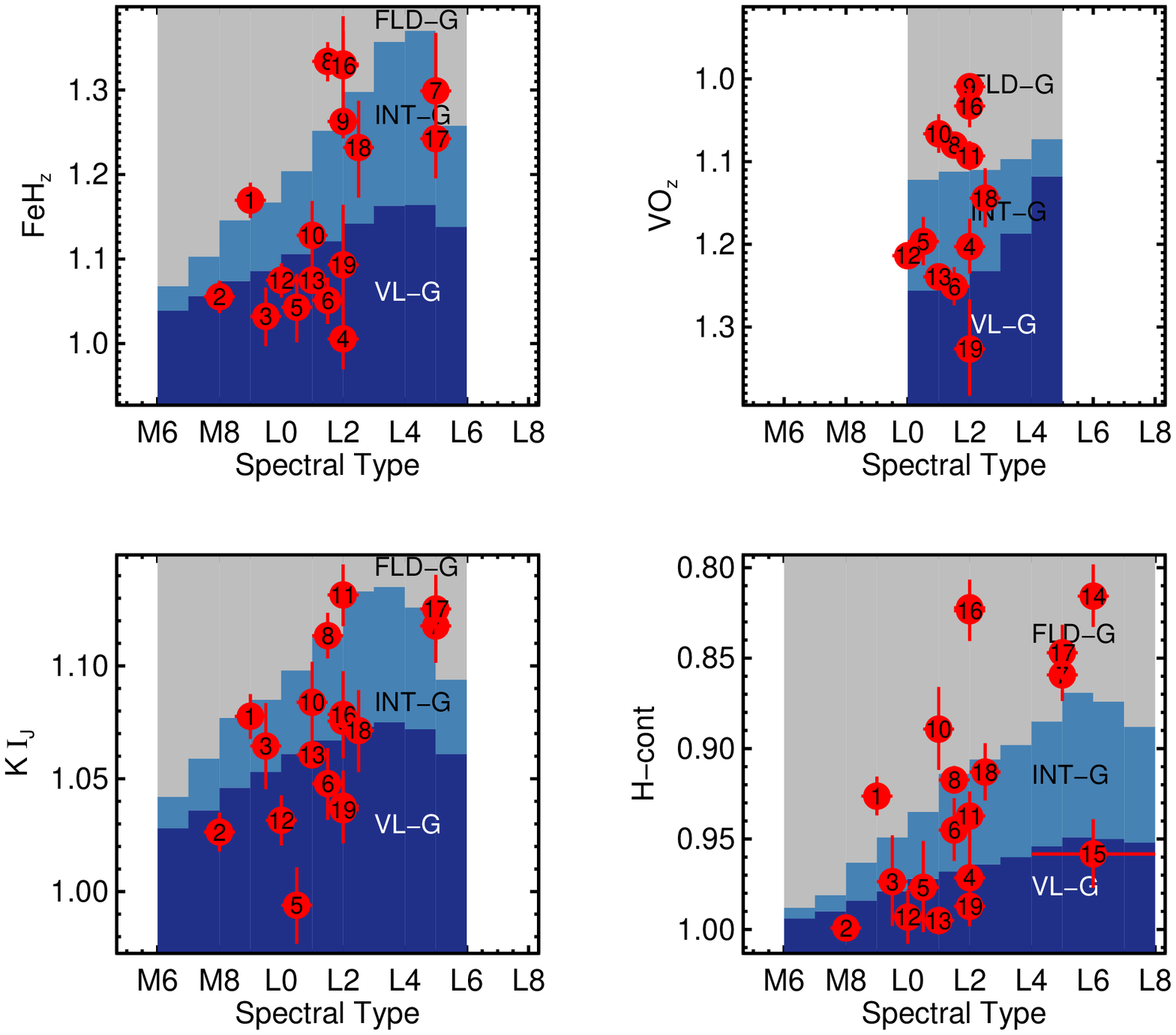}
  \caption{Values of the low-resolution gravity-sensitive spectral indices from
    \citet{Allers:2013hk} for the objects whose gravity classes we confirm
    visually, including 9 \fldg, 1 \intg, and 9 \vlg\
    objects.  The index values are plotted in red, with values for the same
    object in different plots labelled with the same number.  The gray, slate,
    and dark blue bars represent the ranges of index values corresponding to the
    \fldg, \intg, and \vlg\ gravity classes,
    respectively, and indicate the spectral types for which each index is valid
    for gravity classification. Given that our search targeted field L/T
    transition dwarfs and not young M and L dwarfs, discovering this many
    objects with low-gravity spectral signatures was unexpected.}
\label{fig.lego.al13.def}
\end{center}
\end{figure}

\begin{figure}
\begin{center}   
  \epsscale{1.8}
  \plottwo{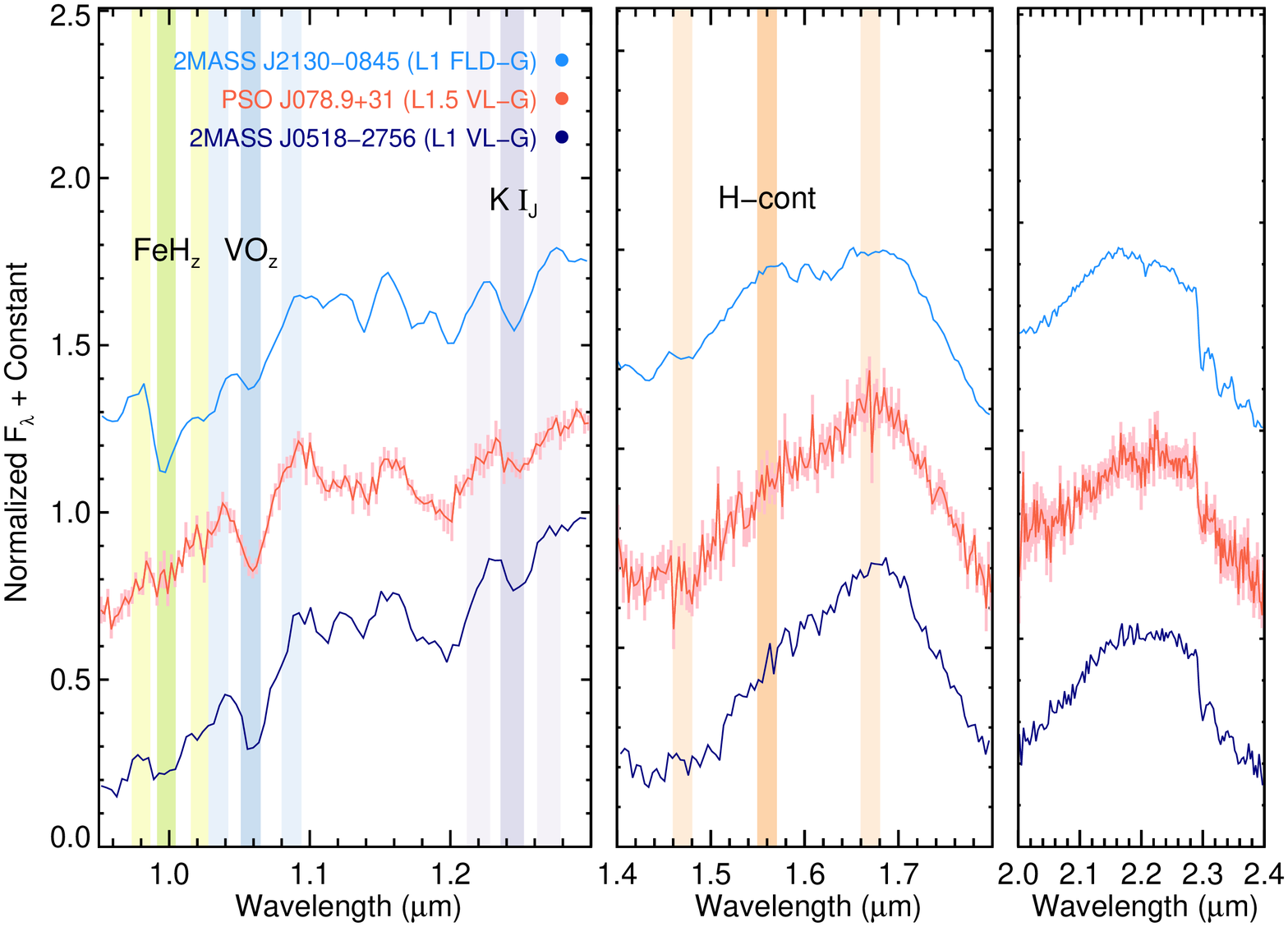}{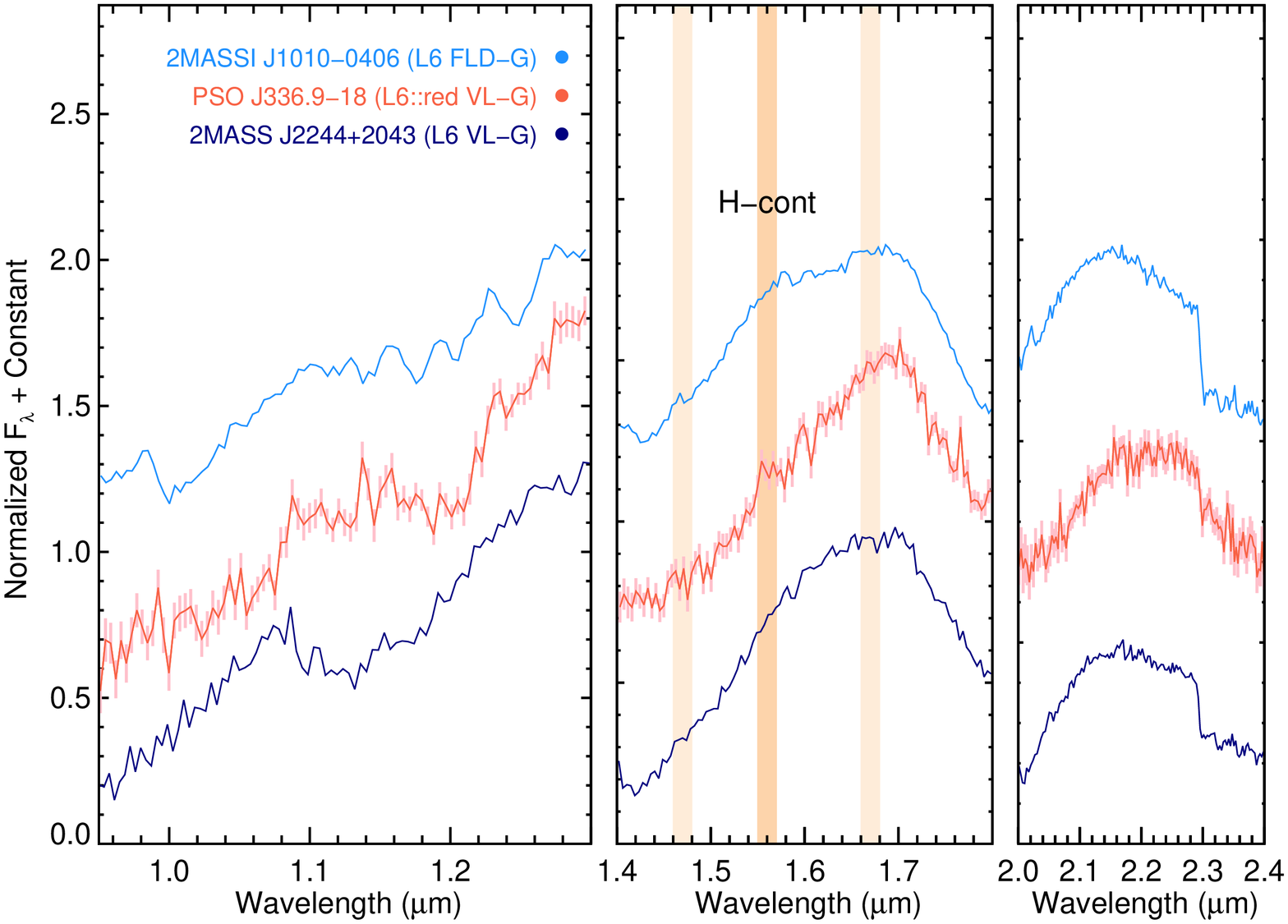}
  \caption{Plots showing our four newly identified field \intg\ and \vlg\
    objects (middle, with error bars) compared with field standards (top) from
    \citet{Kirkpatrick:2010dc} and \vlg\ standards (bottom) from AL13 of the
    same spectral type (within 0.5 subtypes). The vertical colored bars show the
    spectral regions used to calculate the indicated indices, for visual
    comparison. Each plot shows only the indices that are valid for the object's
    spectral type.}
  \figurenum{fig.gravplots.def.1}
\end{center}
\end{figure}

\begin{figure}
\begin{center}
  \epsscale{1.8}
  \plottwo{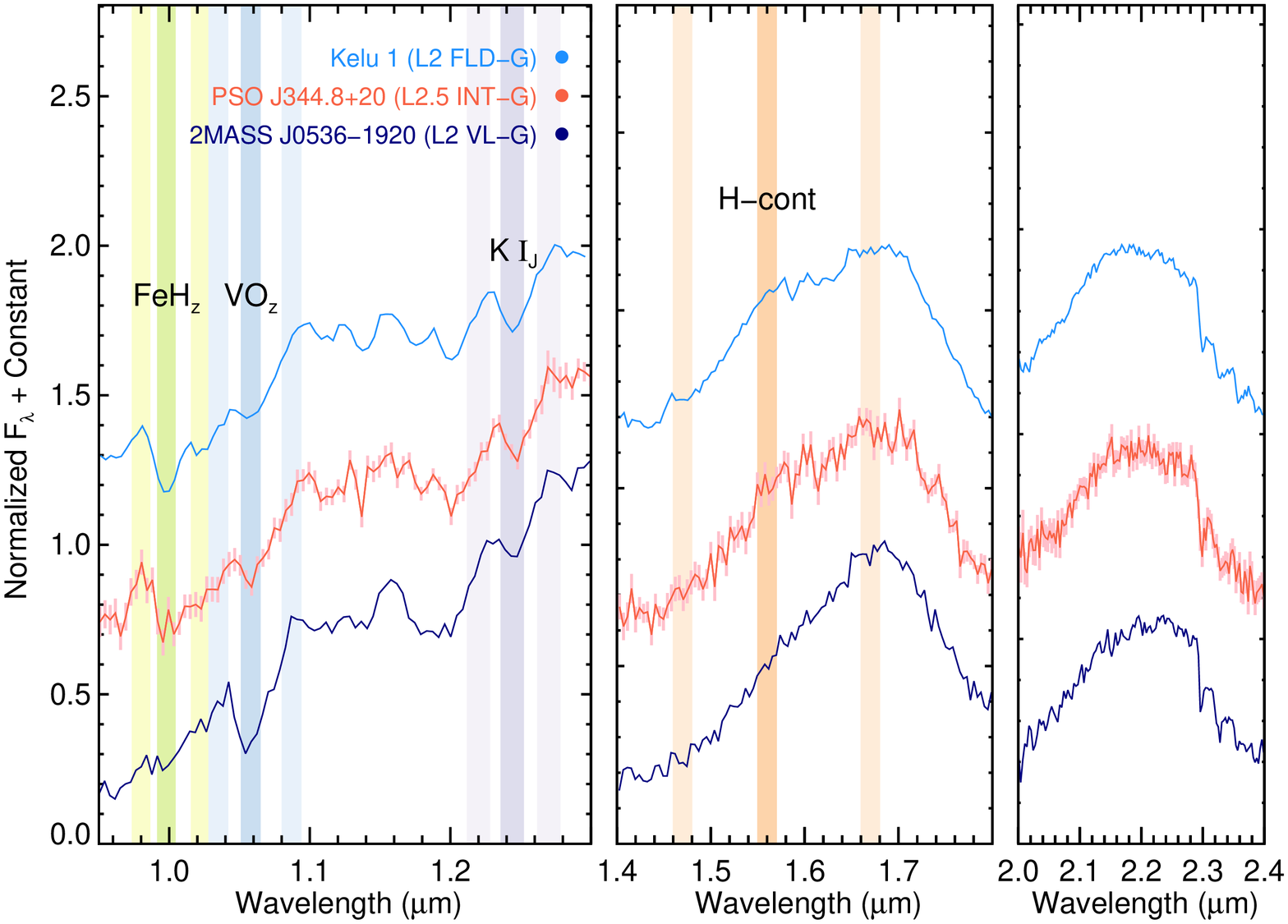}{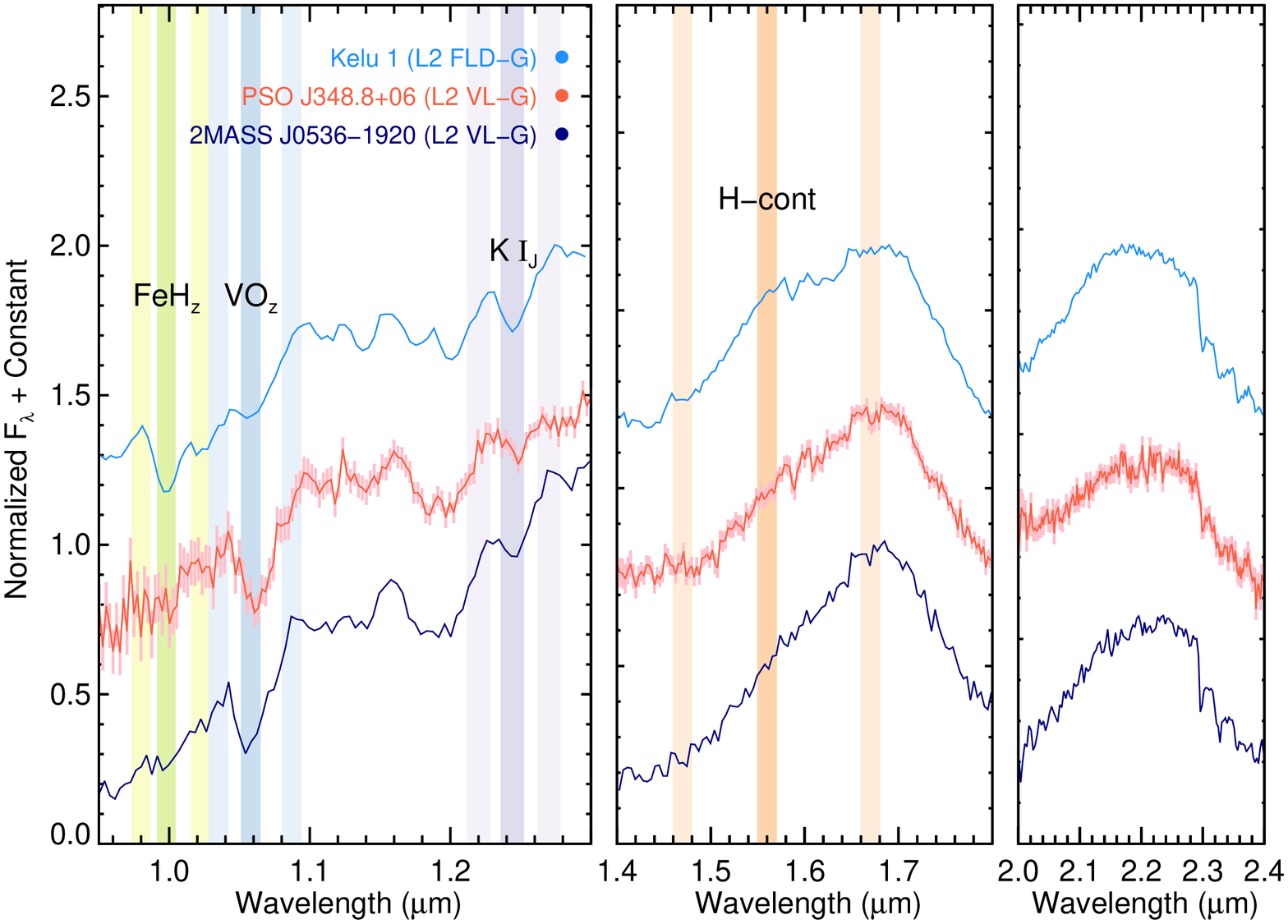}
  \caption{continued.}
  \label{fig.gravplots.def}
\end{center}
\end{figure}

Altogether, we classify 10 objects having low gravity (9 as
  \vlg, 1 as \intg) and 9 more as \fldg.  Figure~\ref{fig.lego.al13.def} plots
the gravity classes derived from the four spectral indices for these objects
against their spectral types.  Our final gravity classifications are listed in
Table~\ref{tbl.gravity.allers}, excluding six candidate members of the
  Scorpius-Centaurus Association and the Taurus star-forming region that will be
  presented in a future paper (Best et al., in prep).  The remaining four,
  PSO~J078.9+31 (L1.5~\vlg), PSO~J336.9$-$18 (L6::~red~\vlg), PSO~J344.8+20
(L2.5~\intg), and PSO~J348.8+06 (L2~\vlg), appear to be young field objects, and
their spectra are shown in Figure~\ref{fig.gravplots.def} along with field
standards from \citet{Kirkpatrick:2010dc} and \vlg\ standards from AL13 for
comparison.  Three of the objects (excluding PSO~J336.9$-$18) show
weak 0.99~\um\ \fehz\ and strong 1.06~\um\ \voz\ absorption features and a
triangular H band shape, all signs of youth.  PSO~J336.9$-$18 is an L6~dwarf,
too late-type for the \fehz\ and \voz\ features to yield reliable information
about gravity (AL13), but featuring a triangular H-band shape and very red
colors.  While these are both recognized signatures of youth, AL13 caution that
the triangular H-band shape can also appear in spectra of objects that have
evidence of old age (based on kinematics).  Therefore, while our classification
of \vlg\ is formally correct for PSO~J336.9$-$18, further evidence is needed to
support the conclusion that the object is young.

\begin{figure}
\begin{center}   
  \includegraphics[width=0.99\columnwidth, trim = 0 95mm 0 0]{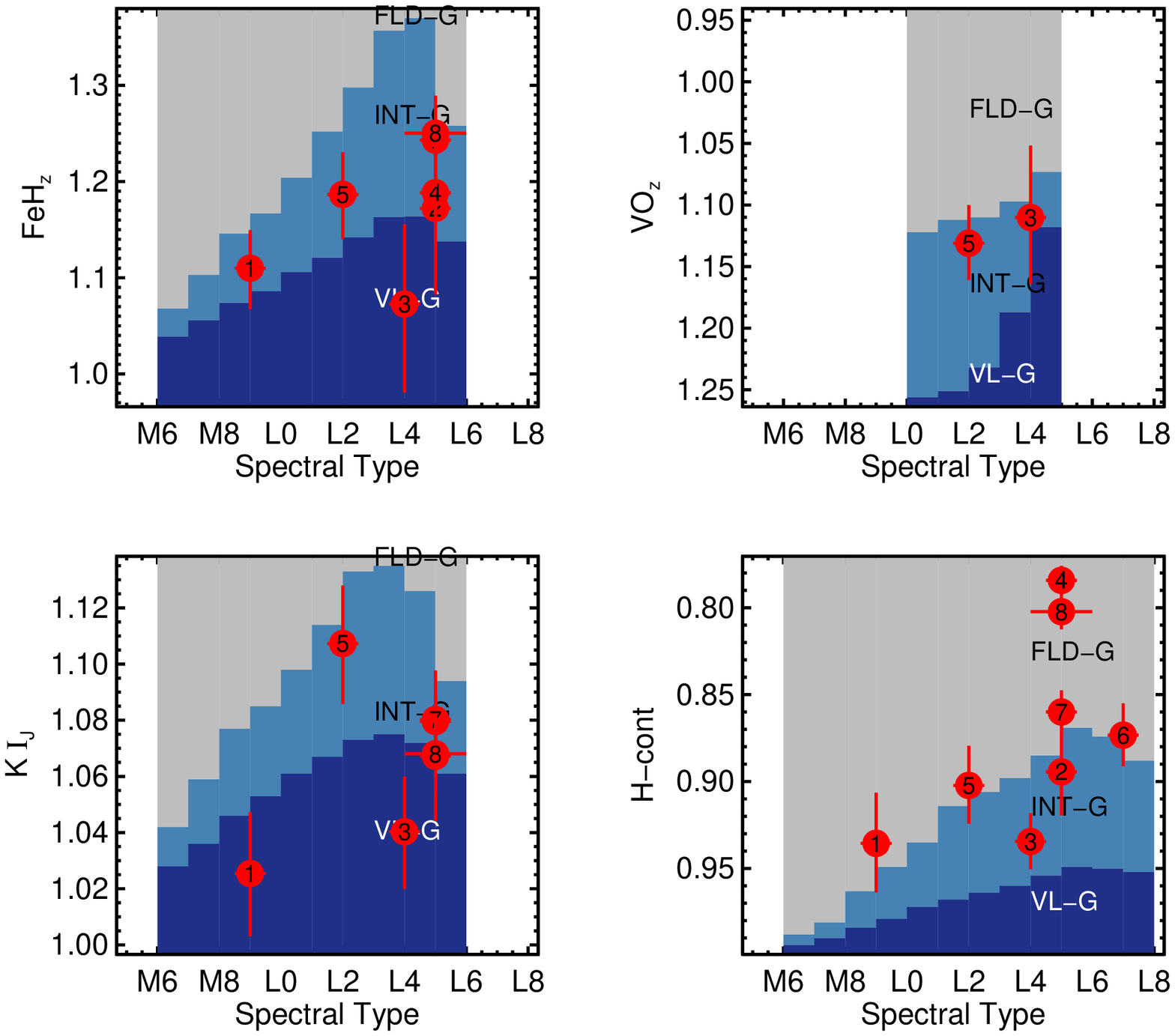}
  \caption{Same as Figure~\ref{fig.lego.al13.def}, but for objects whose
    index-based gravity classes we determine only tentatively due to modest S/N
    in the spectra.}
\label{fig.lego.al13.maybe}
\end{center}
\end{figure}

We identify another 7 objects as potentially low-gravity based on their
indices, but higher S/N spectra are needed to securely classify them.  Among
these is PSO~J068.9+13 (L6~red, candidate~\intg), identified by
\citet{Lodieu:2014jo} as a candidate member of the Hyades (see discussion in
Section~\ref{results.indices.allers}).  Figure~\ref{fig.lego.al13.maybe} shows
the gravity classes vs. spectral types for these 7 potentially
low-gravity objects, and Figure~\ref{fig.gravplots.maybe} compares their spectra
to the field and \vlg\ standards.

\begin{figure}
\begin{center}   
  \epsscale{1.8}
  \plottwo{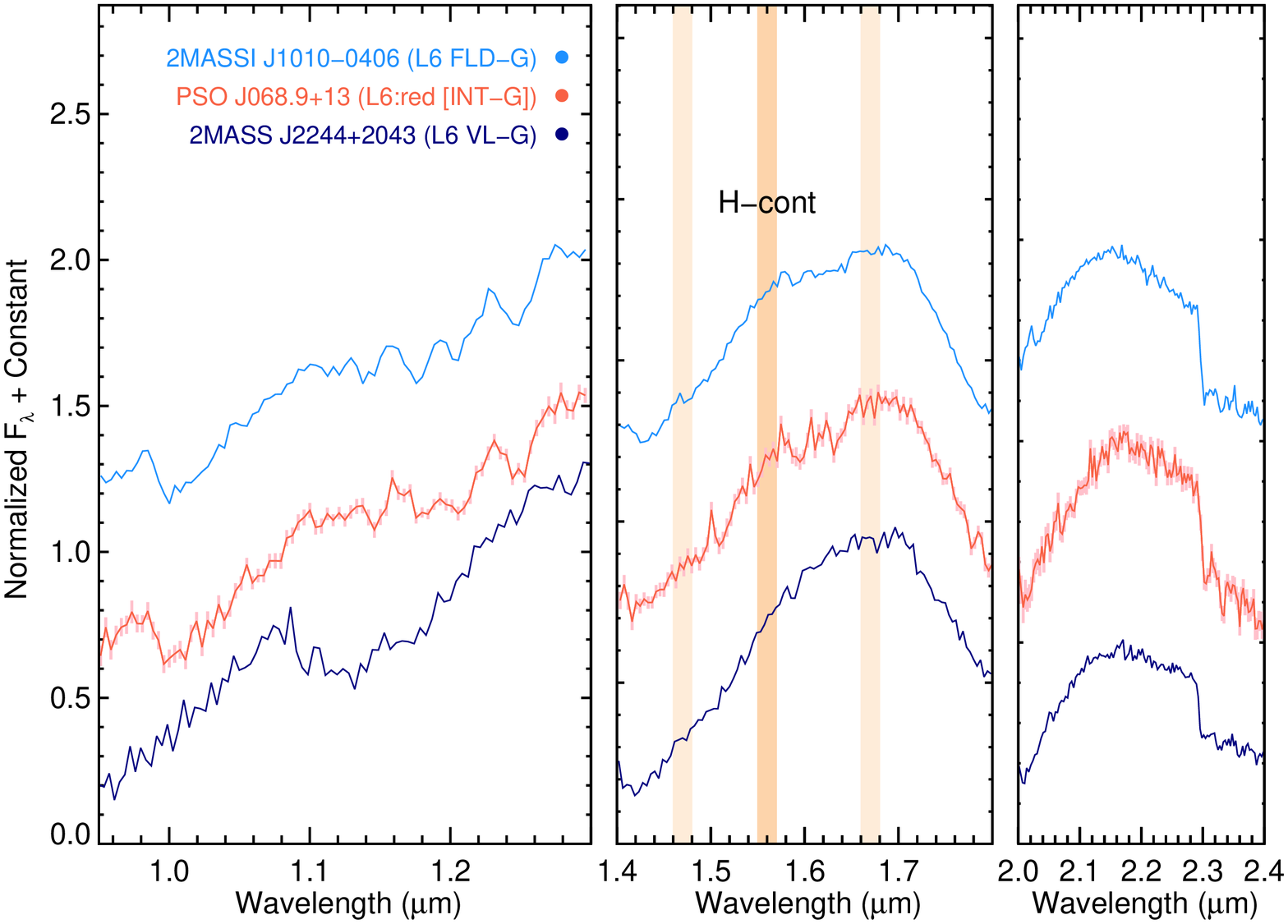}{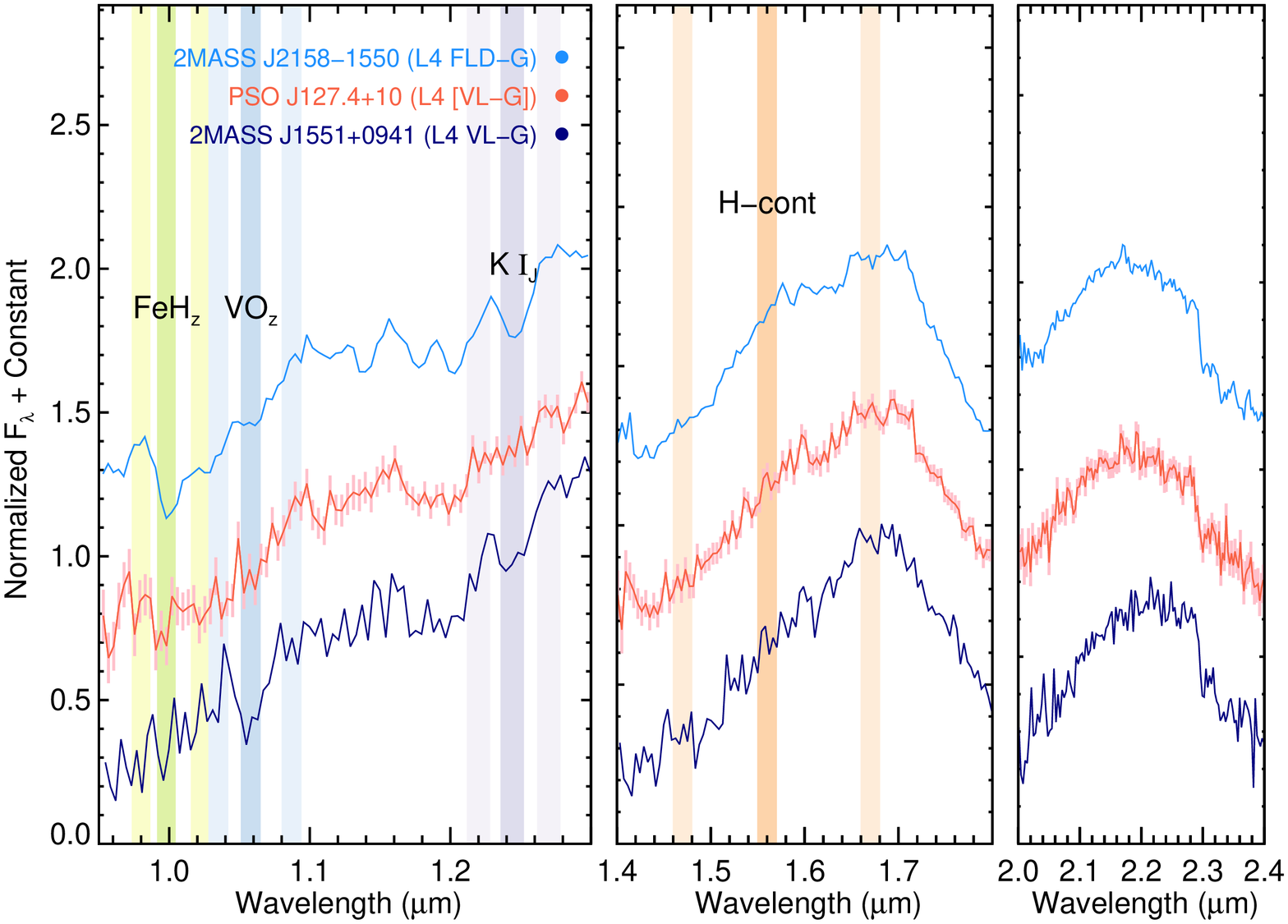}
  \caption{Same as Figure~\ref{fig.gravplots.def}, but for objects whose
    index-based gravity classes we determine only tentatively (classes indicated
    in brackets as in Table~\ref{tbl.gravity.allers}) due to modest S/N in
    the spectra.}
  \figurenum{fig.gravplots.maybe.1}
\end{center}
\end{figure}

\begin{figure}
\begin{center}
  \epsscale{1.8}
  \plottwo{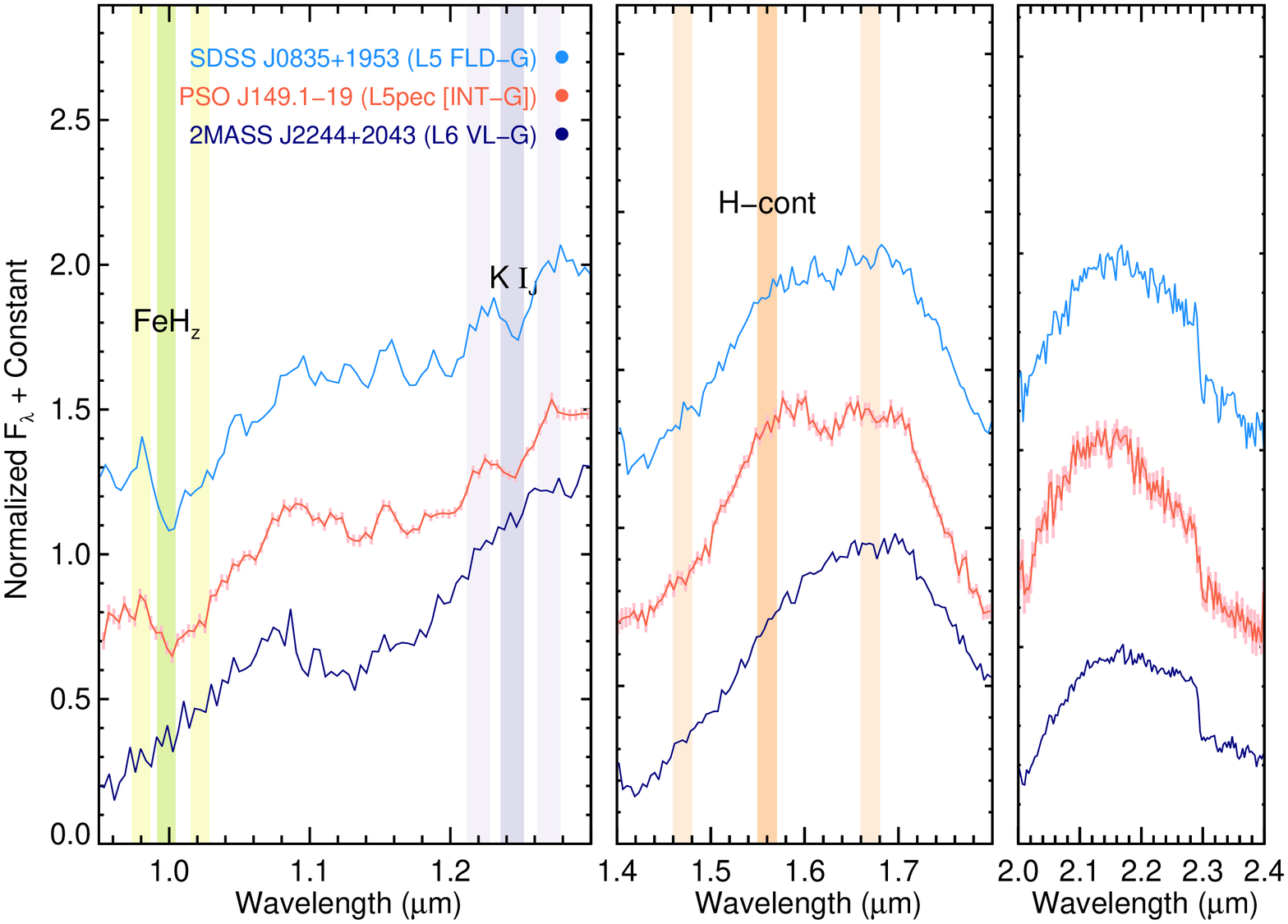}{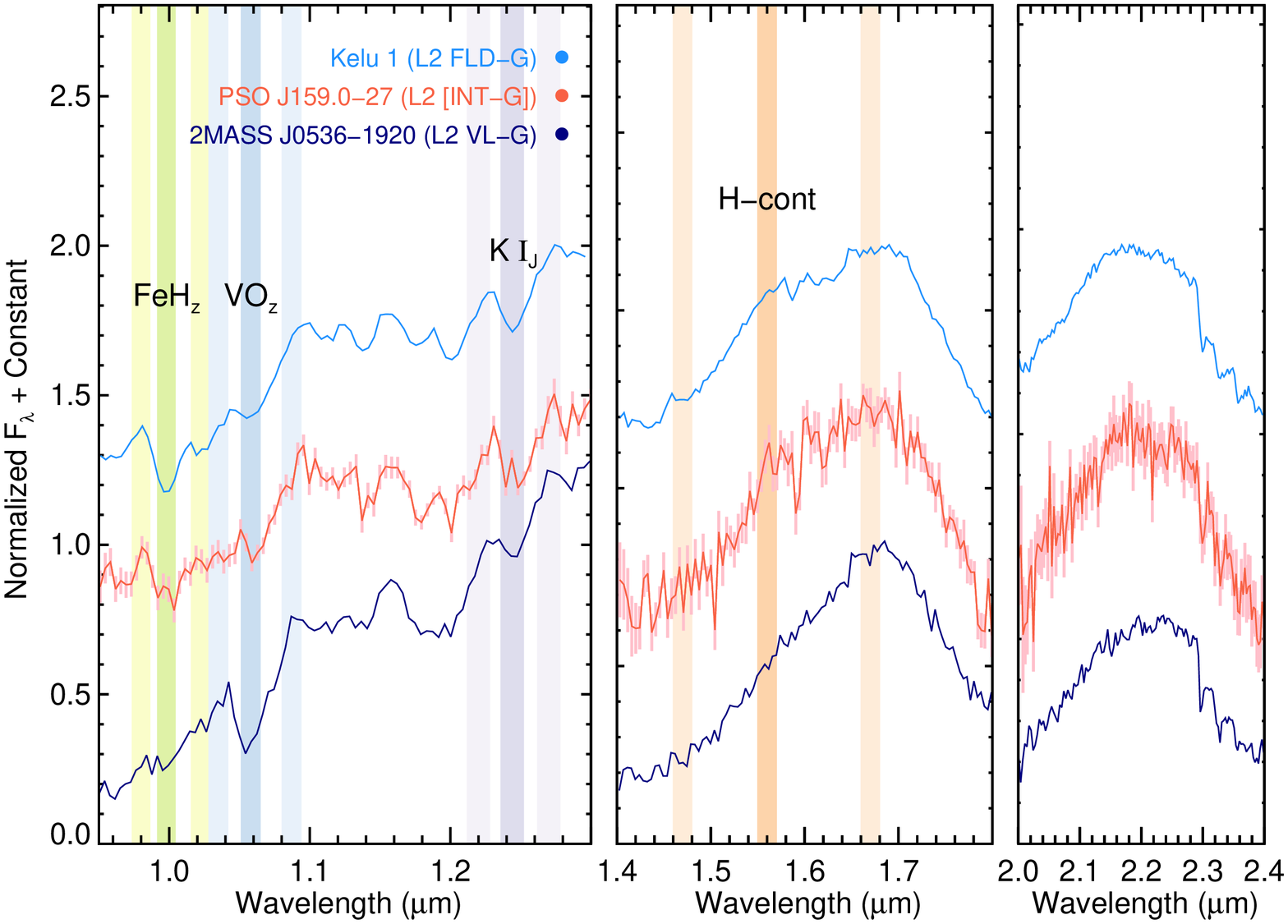}
  \caption{continued.}
  \figurenum{fig.gravplots.maybe.2}
\end{center}
\end{figure}

\begin{figure}
\begin{center}
  \epsscale{1.8}
  \plottwo{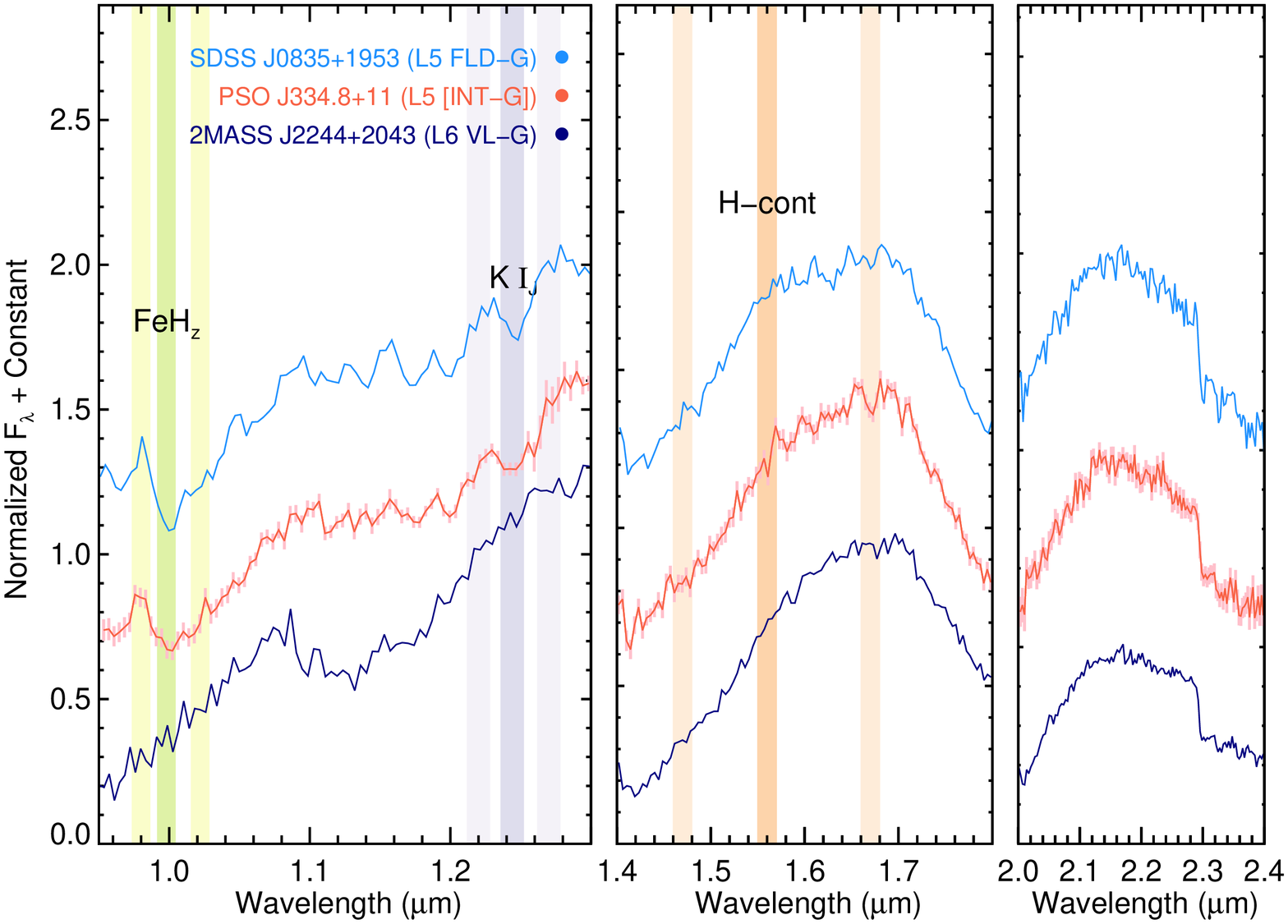}{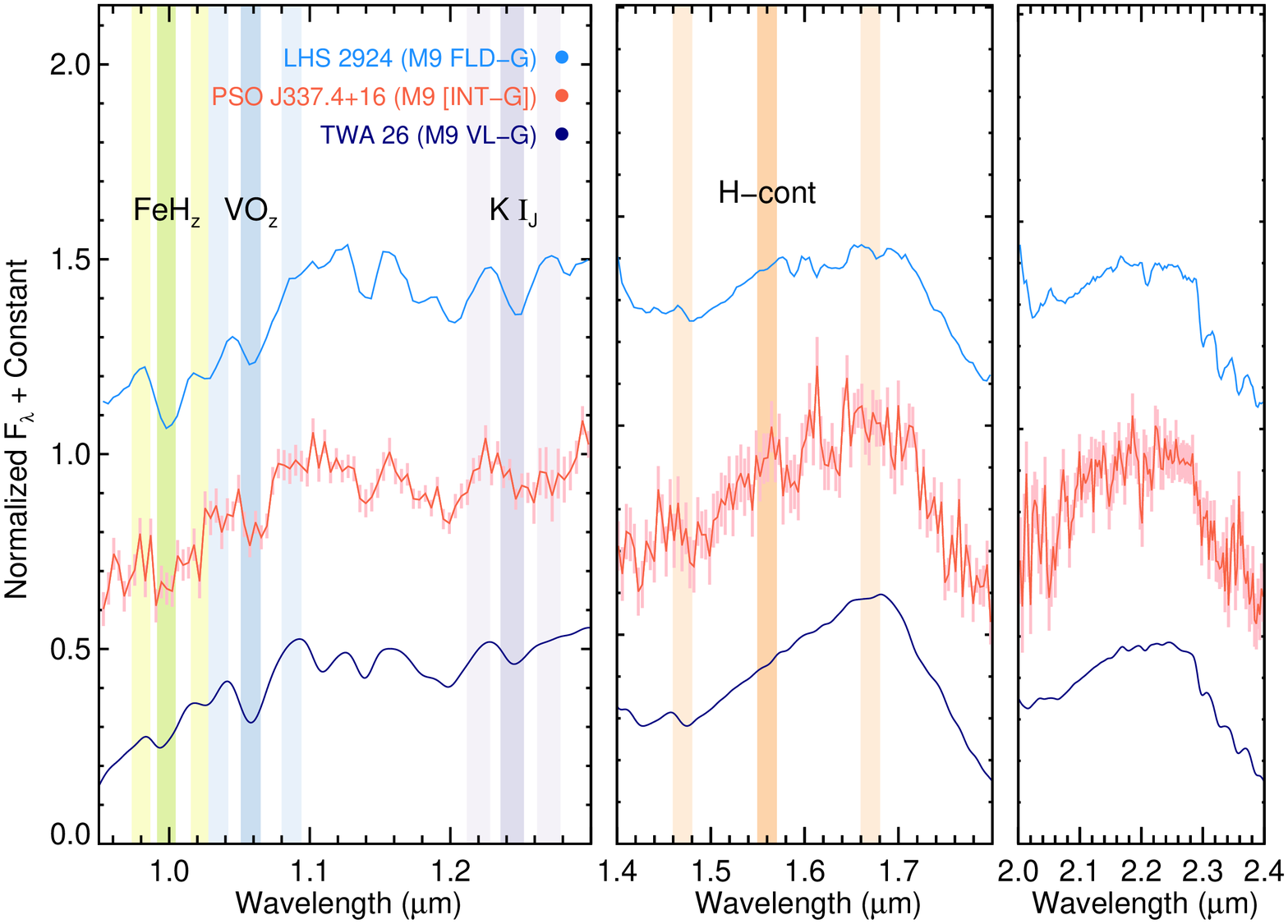}
  \caption{continued.}
  \figurenum{fig.gravplots.maybe.3}
\end{center}
\end{figure}

\begin{figure}
\begin{center}
  \epsscale{0.9}
  \plotone{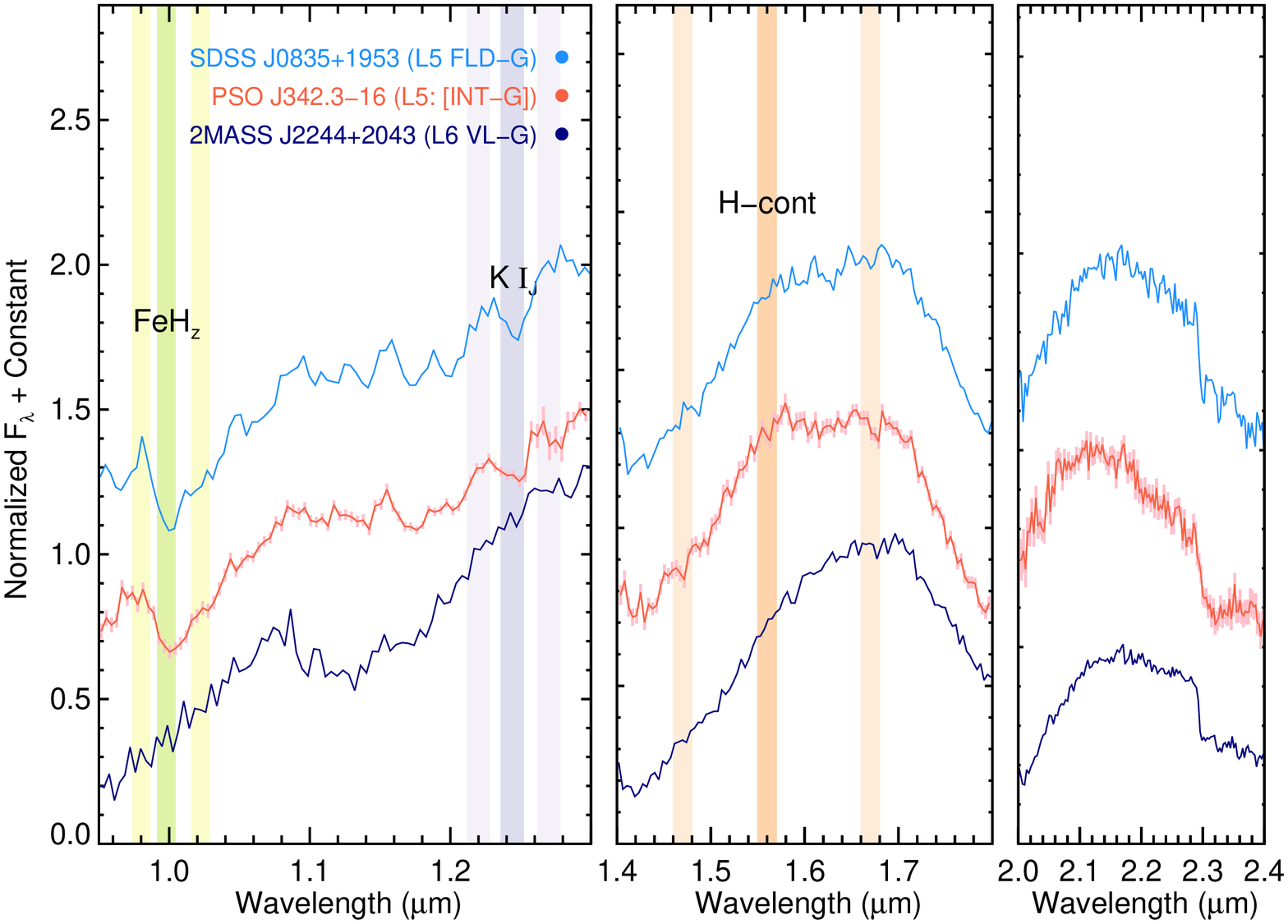}
  \caption{continued.}
  \label{fig.gravplots.maybe}
\end{center}
\end{figure}

In Table~\ref{tbl.other.young}, we list six more objects whose
spectra show indications of youth, but for which the AL13 indices were not
useful because of the spectral type of the object or the low S/N ($<30$) of our
spectrum.  These indications include the redder-than-normal colors and
triangular H-band shape described above.  Three are candidate members of
young moving groups (Section~\ref{ymg}), and three are field objects.

\subsection{Candidate Binaries}
\label{results.binaries}
Roughly $15-30$\% of ultracool dwarfs are binaries
\citep[e.g.,][]{Basri:2006eh,Liu:2006ce,Burgasser:2007fl}.  Binary systems are
important benchmarks, as the binary components are equidistant, coeval, and have
common metallicities.  If resolved with high-resolution imaging, these systems
can be monitored to determine their orbits and dynamical masses
\citep[e.g.,][]{Liu:2008ib,Konopacky:2010kr,Dupuy:2010ch}, breaking the mass/age
degeneracy and providing stringent tests for atmospheric and evolutionary models
\citep[e.g.,][]{Dupuy:2014iz,Dupuy:2015gl}.

We have examined our discoveries for unusual spectral features that suggest
unresolved binarity, using the spectral index criteria of \citet[][hereinafter
BG14]{BardalezGagliuffi:2014fl} for our M7--L7.5 dwarfs and \citet[][hereinafter
B10]{Burgasser:2010df} for our L8 and later dwarfs.  We first ranked our
discoveries by the number of index criteria satisfied, and then visually
reviewed all spectra for peculiar features indicating blends (see descriptions
of individual objects below).  We rejected objects with J band ($1.20-1.31$~\um)
S/N $<25$, as several objects with S/N below this limit satisfied many index
criteria but revealed no signs of spectral blends on visual inspection.  We used
the following scheme to identify strong, medium, and weak binary candidates.  We
ranked objects meeting at least 8 BG14 criteria or 4 B10 criteria as strong
candidates.  We ranked objects meeting at least 4 BG14 criteria or 3 B10
criteria, as well as objects having clear visual indications of blends plus at
least 2 BG14 criteria or 1 B10 criterion, as medium candidates.  We labeled
other objects showing clear visual indications as weak candidates.  This scheme
is similar to those of BG14 and B10, but here we use three categories instead of
two and we incorporate the results of visual inspection.

Overall, we identify 31 binary candidates (Table~\ref{tbl.binaries}).  We
compare the spectra of our strong, medium, and weak binary candidates with those
of field standards in Figures~\ref{fig.strong.binary}, \ref{fig.medium.binary},
and \ref{fig.weak.binary}, respectively.  About $2/3$ of these have spectral
types L9--T2.5, broadly consistent with previous studies that suggested a higher
observed frequency of binaries in the L/T transition
\citep[e.g.,][]{Liu:2006ce,Burgasser:2007fl}.  \citet{Allers:2013vv}
demonstrated that the AL13 indices' ability to identify low-gravity features is
not affected by spectral blends.  We find only one binary candidate
(PSO~J146.0+05) with mild hints of low gravity.

Below we briefly discuss individual binary candidates with notable spectral
features.

\begin{figure}
\begin{center}
  \includegraphics[width=0.8\columnwidth, trim = 10mm 0 5mm 0]{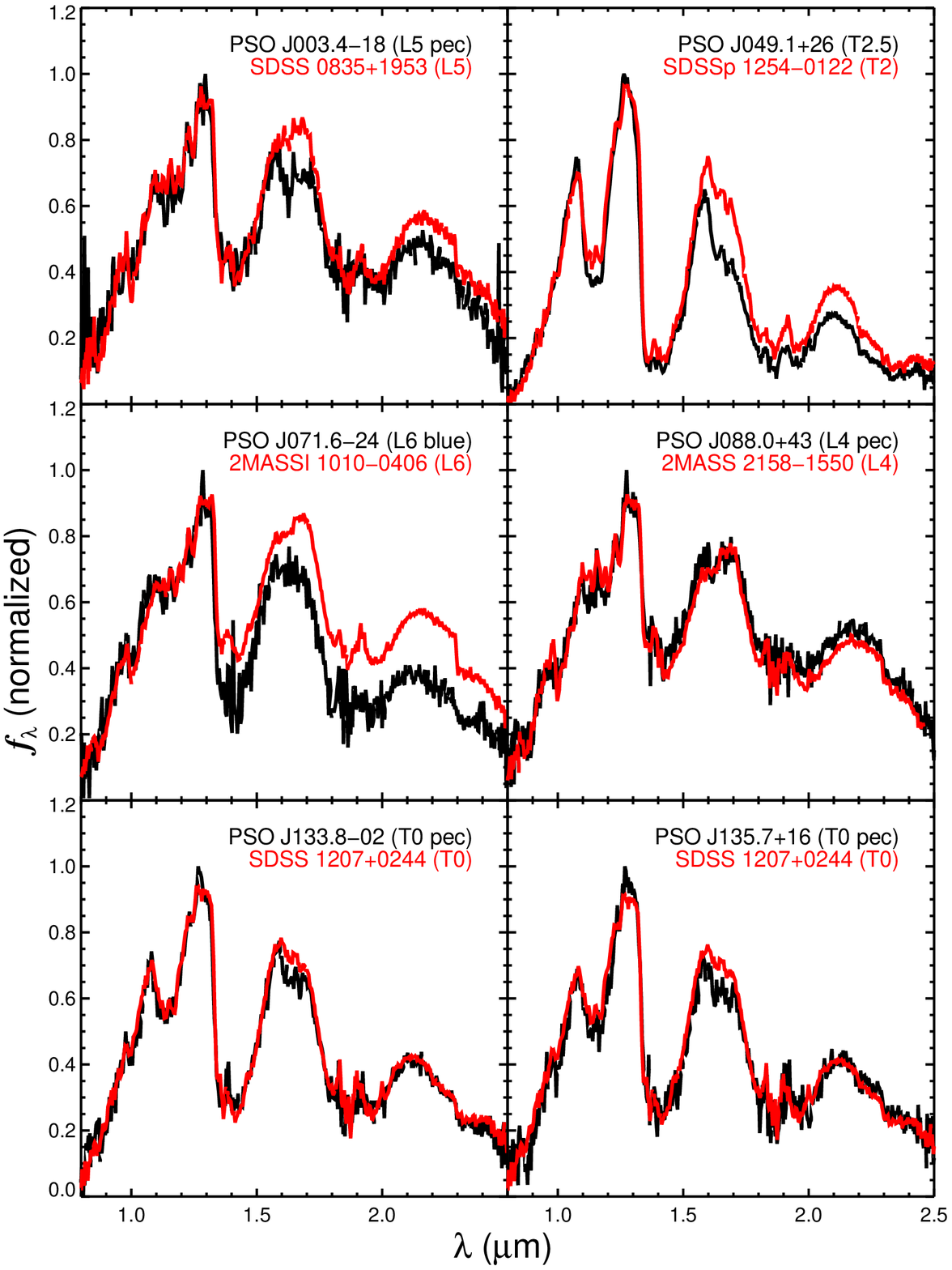}
  \caption{Plots comparing the spectra of our strong binary candidates (black)
    to the field standards of \citet{Kirkpatrick:2010dc} and
    \citet{Burgasser:2006cf} (red). Distinctive features of these spectra are
    discussed in Section~\ref{results.binaries.strong}.}
  \figurenum{fig.strong.binary.1}
\end{center}
\end{figure}

\begin{figure}
\begin{center}
  \includegraphics[width=0.8\columnwidth, trim = 10mm 0 5mm 0]{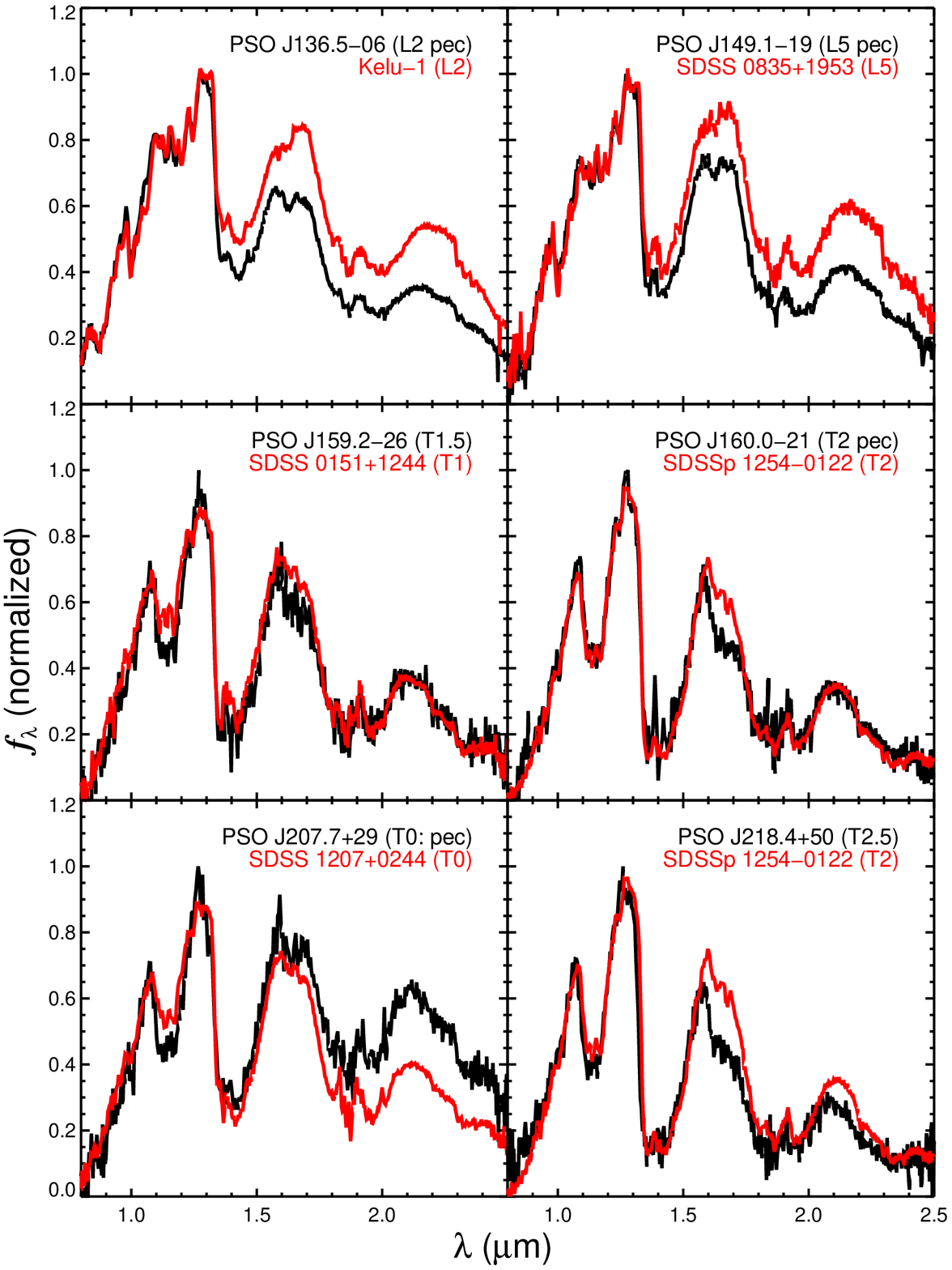}
  \caption{continued.}
  \figurenum{fig.strong.binary.2}
\end{center}
\end{figure}

\begin{figure}
\begin{center}
  \includegraphics[width=0.8\columnwidth, trim = 10mm 70mm 5mm 0]{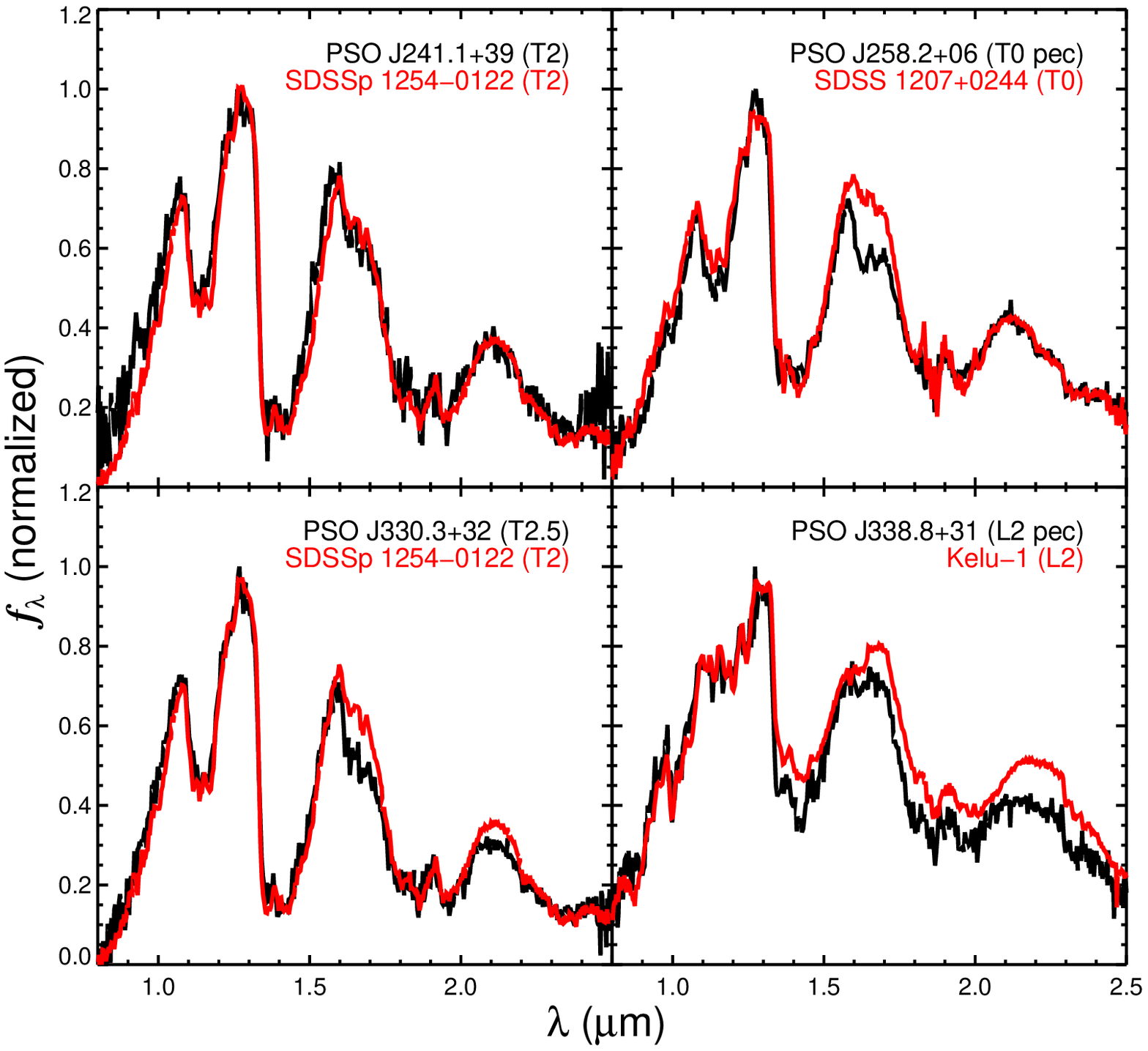}
  \caption{continued.}
  \label{fig.strong.binary}
\end{center}
\end{figure}

\subsubsection{Strong Binary Candidates}
\label{results.binaries.strong}

{\it PSO~J003.4$-$18 (2MASS~J0013$-$1816)} (L5 pec) --- This object was
independently discovered and typed by \citet{Baron:2015fn} as an L1 dwarf and a
common proper motion companion to the M3 dwarf NLTT~687. It satisfies 10 of the
12 BG14 criteria. The J band morphology of PSO~J003.4$-$18 is closest to that of
an L5 dwarf, but the deeply notched H band peak and a more subtle notch at
$\approx$2.2~\um\ are both clear indications of methane.  The peak in the J band
at 1.3~\um\ and the overall blue color are further evidence of the presence of a
T dwarf.  \citet{Baron:2015fn} used optical spectral indices to determine a
spectral type, and their optical spectrum would be dominated by the primary and
have very little flux from a T-type companion.  PSO~J003.4$-$18 is therefore
very likely to be an early-L + early-T binary.  As a companion to NLTT~687, it
would also be a rare benchmark ultracool binary (Section~\ref{comoving}).

{\it PSO~J049.1+26} (T2.5) --- This object is a near-perfect spectral match to
the T2+T7.5 binary 2MASS J12095613$-$1004008
\citep{Burgasser:2004hg,Liu:2010cw}.  The J band shape fits the T2 standard
best, but the H and K bands have the morphology of later-T dwarfs.  This object
satisfies 4 of the 6 B10 criteria.

{\it PSO~J071.6$-$24 (WISE~J0446$-$2429)} (L6 blue) --- The J band morphology
matches an L6 dwarf, but the peak in the J band at $\approx$1.3~\um\ suggests a
later T dwarf, and the overall color and H and K band shapes match a T0 dwarf.
\citet{Thompson:2013kv} independently discovered this object and typed it L5 pec
(blue), ascribing the unusual spectral features to thin large-grained clouds
rather than a L+T blend.  This object satisfies 8 of the 12 BG14 criteria.

{\it PSO~J088.0+43} (L4 pec) --- The J band peak at $\approx$1.28~\um\ and the
notched H band suggest a mid-T dwarf blended with a normal L4 dwarf. This object
satisfies 9 of the 12 BG14 criteria.

{\it PSO~J133.8$-$02} (T0 pec) --- The spectrum fits the overall shape of the T0
standard quite well, but the J and H band peaks ($\approx$1.28~\um\ and
$\approx$1.58~\um, respectively) suggest the additional presence of a later-T
dwarf. This object satisfies 5 of the 6 B10 criteria.

{\it PSO~J135.7+16} (T0 pec) --- The overall morphology is closest to that of a
T0 dwarf, but the J and H bands have the shapes of a T2 dwarf. This object
satisfies all 6 of the B10 criteria.

{\it PSO~J136.5$-$06} (L2 pec) --- The J band shape matches the L2 spectral
standard fairly well, but the deeper water absorption band at $\approx$1.4~\um\
and the blue color suggest a later-type object, and the notched H band peak and
depression at at $\approx$2.2~\um\ both indicate the presence of methane. This
object satisfies 9 of the 12 BG14 criteria.

{\it PSO~J149.1$-$19} (L5 pec) --- The J band morphology is a clear match to L5,
but the deeper water absorption band at $\approx$1.4~\um\ and the blue color
indicate a later-type object.  The notched H band peak and depression at
$\approx$2.2~\um\ both point to methane and a T-dwarf companion. This object
satisfies 10 of the 12 BG14 criteria.

{\it PSO~J159.2$-$26} (T1.5) --- The K band shape is an excellent match to the
T1 standard, but the J and H bands fit a T2 better. This object satisfies 4 of
the 6 B10 criteria.

{\it PSO~J160.0$-$21} (T2 pec) --- The overall slope of this spectrum matches
that of the T2 standard, but the J band peak at $\approx$1.28~\um\ and the blue
H band peak strongly suggest the presence of a late-T companion.  This object
satisfies all 6 of the B10 criteria.

{\it PSO~J207.7+29} (T0: pec) --- This object has no good spectral matches among
the L- and T-dwarf standards. The overall color is similar to an L9 dwarf, but
the lower flux at $\approx$1.65~\um\ and $\approx$2.2~\um\ reveal the presence
of methane, and the J band peak resembles a mid-T dwarf. This object satisfies
all 6 of the B10 criteria.

{\it PSO~J218.4+50} (T2.5) --- Similar to PSO~J049.1+26, this object is a good
spectral match to the known T2+T7.5 binary 2MASS~J12095613$-$1004008
\citep{Burgasser:2004hg,Liu:2010cw}.  The J band shape fits the T2 standard best
but not well, and the H and K bands have the morphology of later-T dwarfs.  This
object satisfies 5 of the 6 B10 criteria.

{\it PSO~J241.1+39} (T2) --- Overall and in the J band, this is a good match to
the T2 standard, but the Y and H band peaks are bluer.  This object satisfies 5
of the 6 B10 criteria.

{\it PSO~J258.2+06} (T0 pec) --- The spectrum fits the overall color and K band
shape of the T0 standard quite well, but the J and H band peaks resemble a
later-T dwarf. This object satisfies all 6 of the B10 criteria.

{\it PSO~J330.3+32} (T2.5) --- This object has unusually deep water absorption
bands at $\approx$1.15~\um\ and $\approx$1.4~\um\ for a T2.5 dwarf, and
satisfies 5 of the 6 B10 criteria.  It is a common proper motion companion to
the star Wolf~1154 (Section~\ref{comoving}), and therefore would be a rare
ultracool benchmark if confirmed as a binary binary.

{\it PSO~J338.8+31} (L2 pec) --- The spectrum is a good match to the L2 standard
in the J band, but the overall slope and K band shape are more like those of a
T0, and H band notch indicates methane.  This object satisfies 8 of the 12 BG14
criteria.

\begin{figure}
\begin{center}
  \includegraphics[width=0.8\columnwidth, trim = 10mm 0 5mm 0]{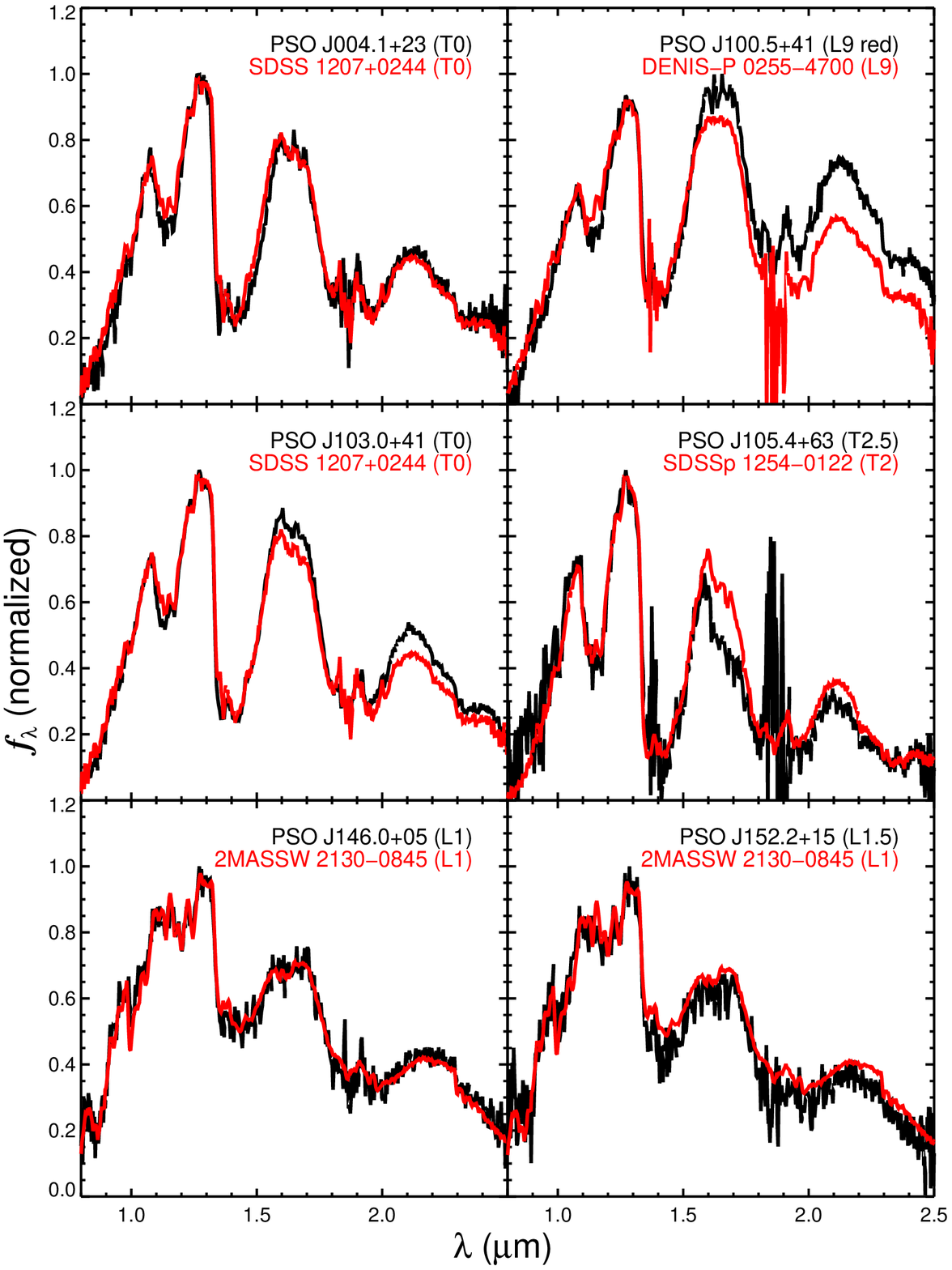}
  \caption{Same as Figure~\ref{fig.strong.binary}, but for our medium-ranked
    binary candidates. Distinctive features of these spectra are
    discussed in Section~\ref{results.binaries.medium}.}
  \figurenum{fig.medium.binary.1}
\end{center}
\end{figure}

\begin{figure}
\begin{center}
  \includegraphics[width=0.8\columnwidth, trim = 10mm 0 5mm 0]{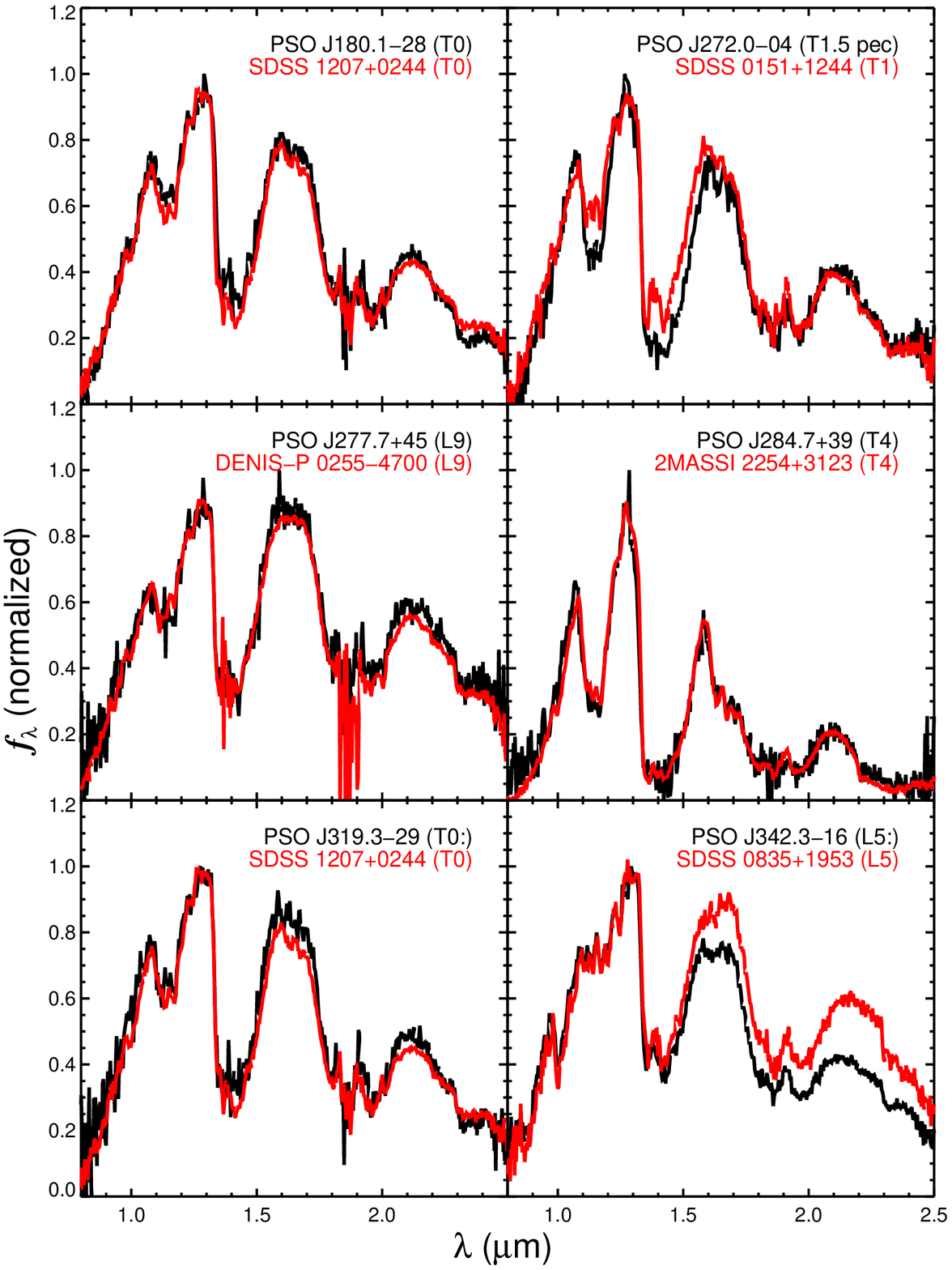}
  \caption{continued.}
  \label{fig.medium.binary}
\end{center}
\end{figure}

\begin{figure}
\begin{center}
  \includegraphics[width=0.8\columnwidth, trim = 10mm 70mm 5mm 0]{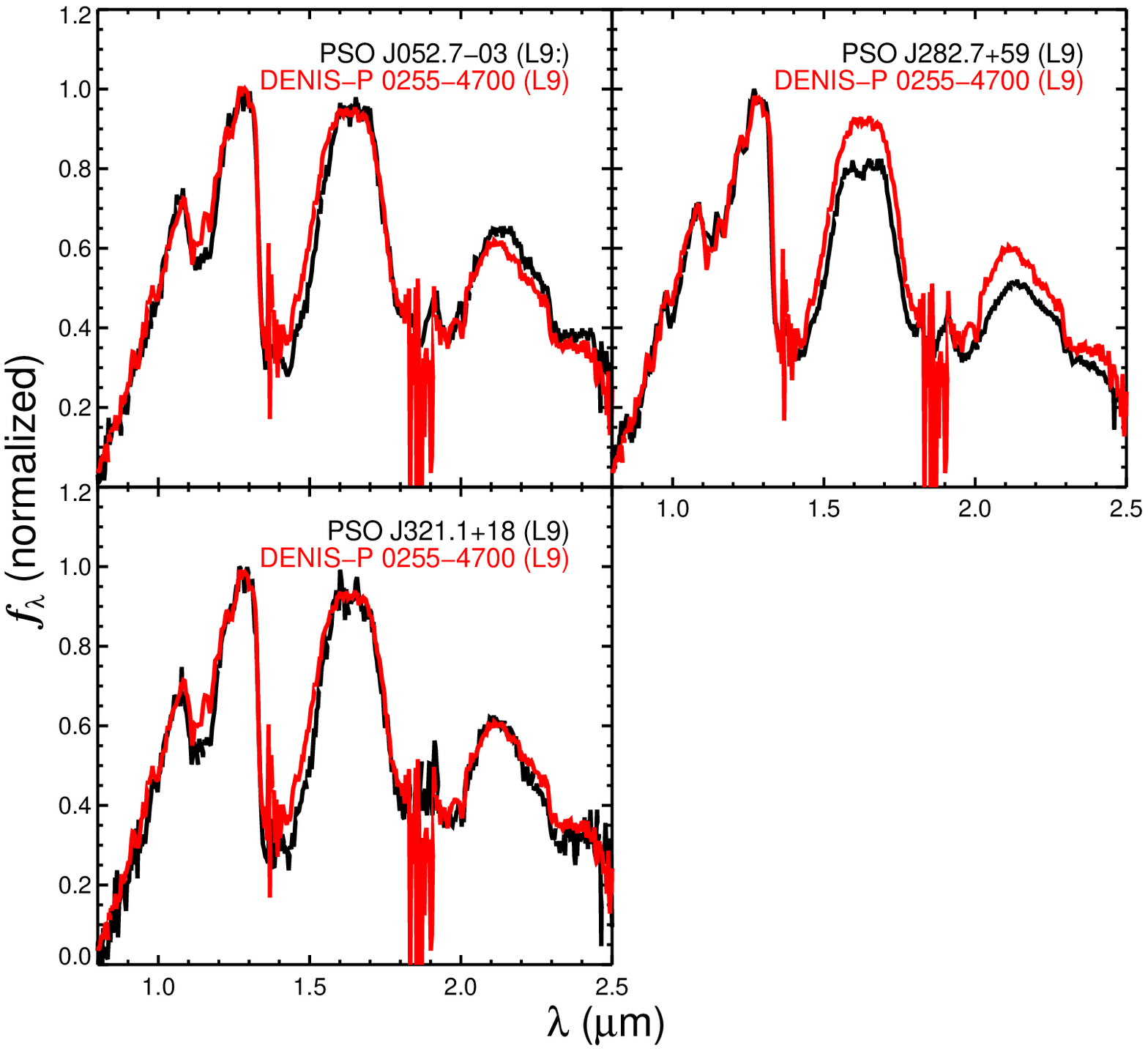}
  \caption{Same as Figure~\ref{fig.strong.binary}, but for our weak
    binary candidates. Distinctive features of these spectra are
    discussed in Section~\ref{results.binaries.weak}.}
  \label{fig.weak.binary}
\end{center}
\end{figure}

\subsubsection{Medium Binary Candidates}
\label{results.binaries.medium}

{\it PSO~J004.1+23} (T0) --- The overall morphology closely resembles a T0
dwarf, but the H band shows no clear sign of methane while the J band peak
resembles that of a T2 dwarf.  The spectrum is a good match to the L6+T2 binary
SDSSp~J042348.57-041403.5 \citep{Geballe:2002kw,Burgasser:2005gj}. It meets 2 of
the 6 B10 criteria.

{\it PSO~J100.5+41 (WISE~J0642+4101)} (L9 red) --- This unusual object was
independently identified by \citet{Mace:2013jh}, who classify it as ``extremely
red'' without assigning a spectral type.  We type it as L9 based on its
$1.2-1.3$~\um\ J band profile and the depth of its $\approx$1.4~\um\ water
absorption band, and we concur with the very red color. The redness is most
easily explained by large amounts of dusty condensates in the photosphere, but
the object also satisfies 3 of the 6 B10 criteria.

{\it PSO~J103.0+41} (T0) --- This object was identified by us as a candidate binary in
Paper I, where it is discussed in detail.  It is also a good match to the known
L6+T2 binary SDSSp~J042348.57-041403.5, and satisfies 2 of the 6 B10 criteria.

{\it PSO~J180.1$-$28} (T0) --- This object matches the overall shape of the T0
standard fairly well but shows subtle signs of a companion later-T dwarf: the
peak in the J band at 1.28~\um\ and the brighter peak in the K band.  The J and
H bands are also good matches to the L6+T2 binary
SDSSp~J042348.57-041403.5. PSO~J180.1$-$28 meets 1 of the 6 B10 criteria.

{\it PSO~J272.0$-$04} (T1.5 pec) --- The slope of this spectrum and the K band
shape fall in between the T1 and T2 standards, but the depth of the
$\approx$1.15~\um\ and $\approx$1.4~\um\ water absorption bands and the pointy J
band peak suggest a later-type companion. This object satisfies 3 of the 6 B10
criteria.

{\it PSO~J277.7+45 (WISE~J1830+4542)} (L9) --- This object, first identified by
\citet{Kirkpatrick:2011ey}, fits the L9 standard in terms of overall color and
morphology and J band shape, but there are signs of methane absorption at
$\approx$1.65~\um\ and $\approx$2.2~\um. It meets 3 of the 6 B10 criteria.

{\it PSO~J284.7+39} (T4) --- This spectrum is a good match to the T4 standard,
except for the narrow profile of the J band, which suggests the additional
presence of a later T-dwarf.  (The spike at $\approx$1.29~\um\ is likely a noise
artifact.) This object satisfies 2 of the 6 B10 criteria.

{\it PSO~J319.3$-$29} (T0:) --- Clear indications of methane absorption at
$\approx$1.65~\um\ and $\approx$2.2~\um\ point to a T dwarf, while the J band
shape and $\approx$1.15~\um\ and $\approx$1.4~\um\ water absorption band depths
are more like the L9 standard. This object meets 2 of the 6 B10 criteria.

{\it PSO~J342.3$-$16} (L5:) --- The J band morphology matches L5, but the H and
K band shapes and the bluer color indicate the additional presence of a T
dwarf. This object satisfies 5 of the 12 BG14 criteria.

\subsubsection{Weak Binary Candidates}
\label{results.binaries.weak}

{\it PSO~J052.7$-$035} (L9:) --- The best match for the J band profile and
adjacent water absorption bands is the T0 standard, but the H band shows no sign
of methane absorption and the overall slope fits the L9 standard.

{\it PSO~J282.7+59 (WISE~J1851+5935)} (L9) --- This object was identified as a candidate binary in
Paper I, where it is discussed in detail. \citet{Thompson:2013kv} type the
object as L9 pec, and also describe it as a candidate late-L + early-T
binary. Surprisingly, our spectrum meets none of the B06 criteria.

{\it PSO~J321.1+18} (L9) --- The overall slope clearly fits L9, but there is
methane absorption at $\approx$2.2~\um\ and the water absorption bands at
$\approx$1.15~\um\ and $\approx$1.4~\um\ have early-T dwarf depth.

\subsection{Proper Motions and Kinematics}
\label{kinematics}
The motion through space of (sub)stellar objects represents their kinematic
histories, as younger objects tend to have smaller tangential velocities
\citep[e.g.,][]{Wielen:1977va}.  We calculated the proper motions for our
discoveries using the individual PS1 epochs ($\approx$25--30 epochs per object,
mostly in \zps\ and \yps), along with their AllWISE reported positions.  For the
$\approx$70\% of our sample also detected in the 2MASS Point Source Catalog, we
included those positions as well (these objects have 2MASS photometry listed in
Table~\ref{tbl.nir}).  The inclusion of the 2MASS astrometry improved the
precision of our proper motions in many cases despite the fact that the
per-epoch precision for the 2MASS positions is larger ($\approx$70~mas) than for
the PS1 positions ($\approx$25~mas), as 2MASS increased the time baseline for
our calculations from $2-4$ years to $\gtrsim10$ years.

\begin{figure}    
\begin{center}
  \includegraphics[width=1.00\columnwidth, trim = 20mm 0 9mm 0]{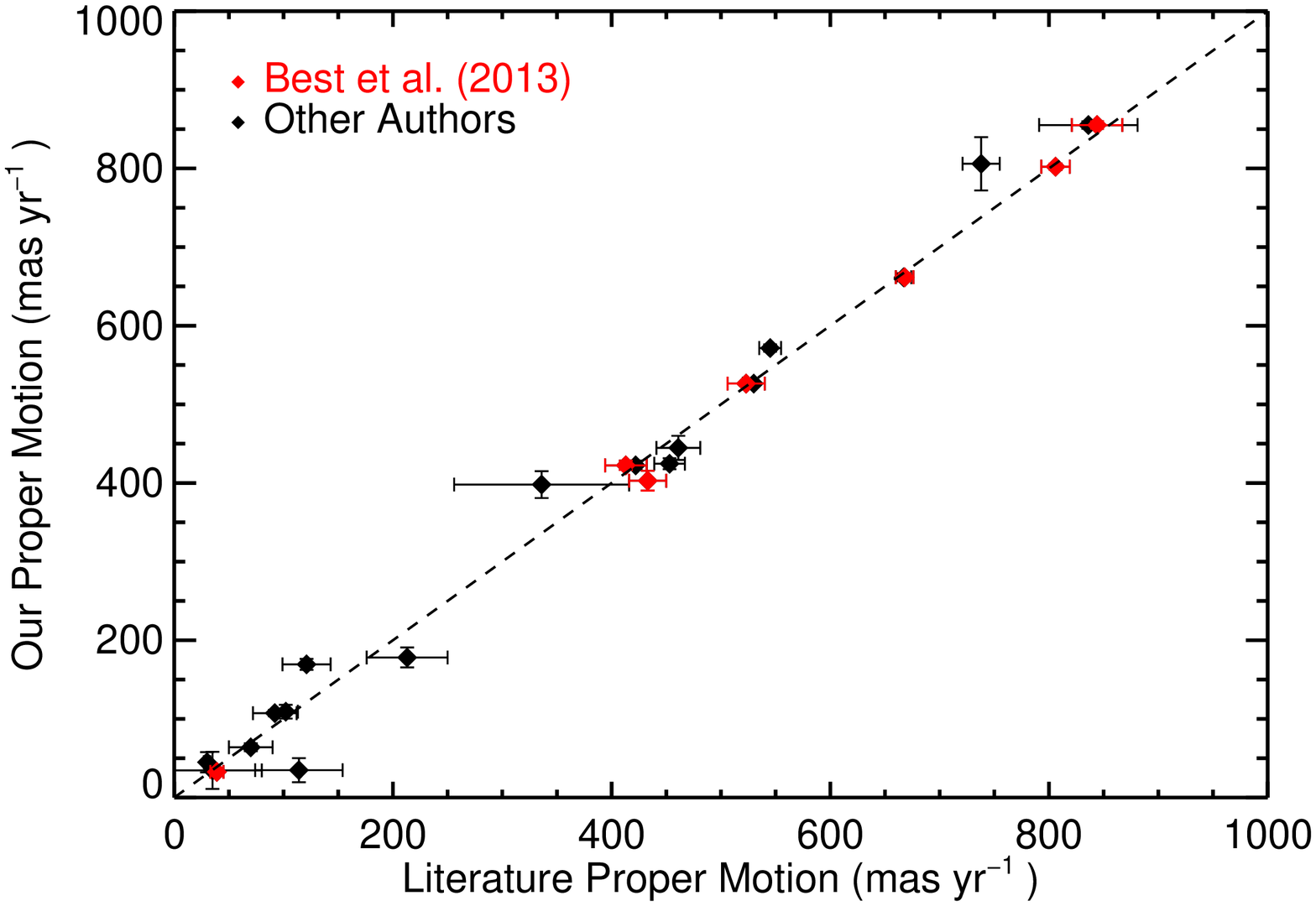}
  \caption{Comparison of our proper motions with previously published
    values from the literature.  Objects plotted in red have proper motions in
    our Paper I, which we refine in this paper.  Four objects have proper motions
    from Paper I as well as elsewhere in the literature, and we plot these as
    separate points.  Our new proper motions are consistent with previous values
    and improve on the precision by a typical factor of $2-3$.}
\label{fig.pm.compare}
\end{center}
\end{figure}

Our proper motions are presented in Table~\ref{tbl.spt.kin}.  Proper motions for
17 of our discoveries were previously published by other authors
(Table~\ref{tbl.scooped}), in addition to 7 by us in Paper I.
Figure~\ref{fig.pm.compare} demonstrates the consistency of our proper motions
with those in the literature as well as our improved precision (typically by a
factor of $2-3$).

\begin{figure}    
\begin{center}
  \includegraphics[width=1.00\columnwidth, trim = 20mm 0 9mm 0]{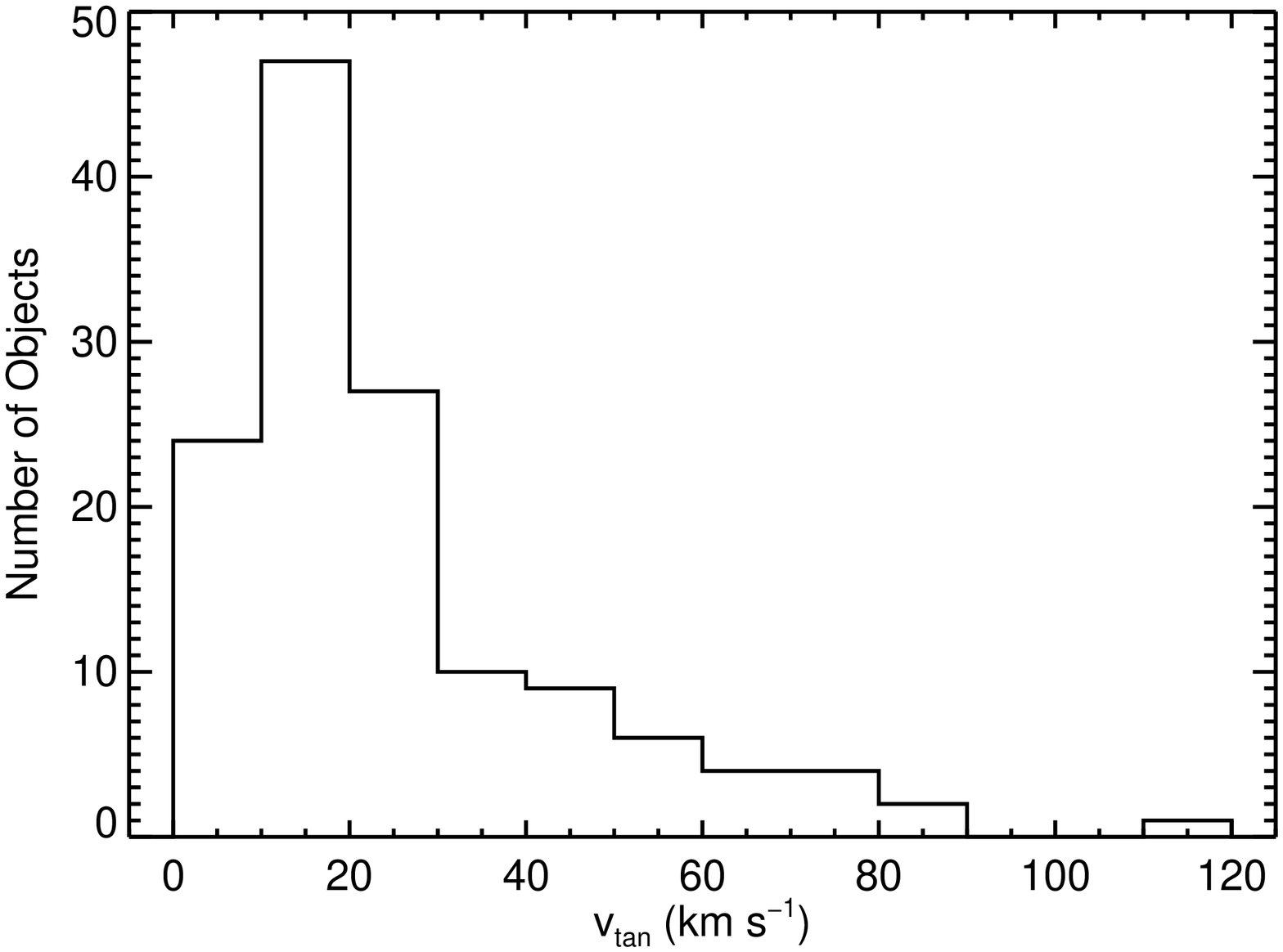}
  \caption{The distribution of tangential velocities for our discoveries.  These
    $v_{\rm tan}$ indicate that our discoveries are all very likely to be
    members of the younger thin disk population.}
\label{fig.vtan.hist}
\end{center}
\end{figure}

We calculated photometric distances for our discoveries, using $W2$ magnitudes
and the spectral type polynomial from \citet{Dupuy:2012bp}.  We used these
photometric distances along with our proper motions to determine tangential
velocities ($v_{\rm tan}$) for our discoveries.  These are also presented in
Table~\ref{tbl.spt.kin}, and we show the distribution of $v_{\rm tan}$ in
Figure~\ref{fig.vtan.hist}.  The $v_{\rm tan}$ of our discoveries are overall
$\approx$25\% lower than those of the 20~pc volume-limited sample presented in
\citet{Faherty:2009kg}, making them fully consistent with the younger thin disk
population.  One object in our sample, PSO~J329.8+03, has a notably larger
velocity ($v_{\rm tan} = 111\pm12$ km\,s$^{-1}$).  We applied the analysis of
\citet[][see their Figure 31]{Dupuy:2012bp} and found this $v_{\rm tan}$ gives
PSO~J329.8+03 a $\approx$10\% chance of being a member of the thick disk.  Older
L dwarfs typically have bluer near-IR colors \citep{Faherty:2009kg}, and while
this age--color relationship has not been clearly established for early-T dwarfs,
we note that PSO~J329.8+03 has $(J-K)_{\rm MKO}=1.26\pm0.03$~mag which is in
fact redder than the mean $(J-K)_{\rm MKO}=0.75\pm0.17$~mag for T1 dwarfs
\citep{Dupuy:2012bp}.  We consider PSO~J329.8+03 to be a thin disk object along
with the rest of our discoveries.

\subsection{Comoving Companions}
\label{comoving}
\begin{figure}[htbp]
\begin{center}\includegraphics[scale=0.8]{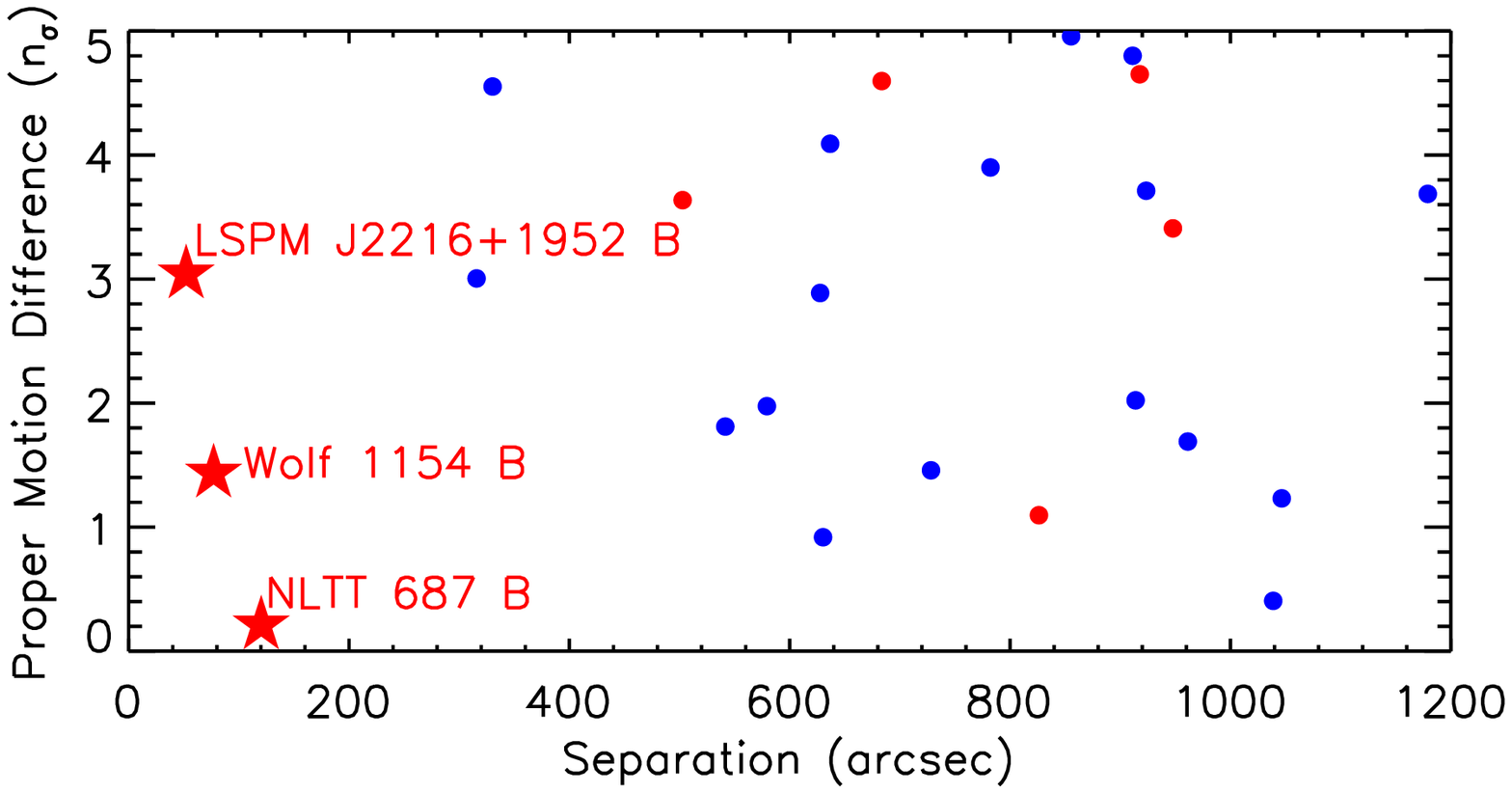}
  \caption{Our common proper motion systems (marked as red stars). The offset
    coincident pairings generated using the method of \citet{Lepine:2007cf} are
    shown as blue dots. The remaining pairings (which are likely to be
    coincident) are shown as red dots.}
\label{cpm_plot}
\end{center}
\end{figure}

To identify if any of our discoveries were members of common proper motion
systems, we cross-matched our discoveries with a large list of nearby stars from
\citet{Lepine:2005jx}, \citet{Salim:2003gv}, \citet{Lepine:2011gl},
\citet{Limoges:2013fu}, and \citet{Deacon:2007jl}. We searched for matches
within 5~arcmin and identified eight possible pairs with proper motions
differing by less than $5\sigma$ (where $\sigma$ is the quadrature sum of the
proper motion differences in each axis divided by the combined uncertainties in
that axis).  To test how many of these pairs were chance alignments of unrelated
stars, we used the method of \citet[see also
  \citealt{Deacon:2014ey}]{Lepine:2007cf}. We offset the positions in our input
catalog by $2^{\circ}$ and repeated our matching criteria, generating entirely
coincident pairings. The results are shown in Figure~\ref{cpm_plot}. Three of
our prospective pairs lie outside the area dominated by coincident pairs.

Our three pairings are described in Table~\ref{cpm_tab}. One of the pairings,
NLTT~687 and PSO~J003.4$-$18, was previously discovered by
\citet{Baron:2015fn}. Two of our secondaries, PSO~J003.4$-$18 and PSO~J334.1+19,
are identified as candidate binaries (Section~\ref{results.binaries.strong}). If
these are indeed binaries then these systems will be hierarchical triples. Such
systems are useful benchmarks as the primaries can be used to constrain their
ages and metallicities, allowing evolutionary models to estimate the masses,
radii, and effective temperatures of the binary components.  If the secondary
can be resolved with high-resolution imaging into two components, their masses
can be measured dynamically, providing a rigorous test of the evolutionary
models.  We also identify PSO~J334.1+19 as a possible $\beta$ Pictoris Moving
Group member ($p=77.8\%$, Section~\ref{ymg.banyan}). Using the BANYAN~II online
tool \citep{Gagne:2014gp} we found that its primary LSPM~J2216+1952 is also a
possible ($p=58.2\%$) member of this moving group.

\section{The Atmospheres of L/T Transition Dwarfs}
\label{lttrans.atmos}
The significant changes in the spectra and blueward shift in near-IR colors of
brown dwarfs cooling through the L/T transition arise from the formation of
methane and the depletion of photospheric condensate clouds.
\citep[e.g.,][]{Allard:2001fh,Burrows:2006ia,Saumon:2008im}.  The process by
which the clouds deplete is not well understood, and proposed scenarios involve
the clouds gradually thinning, raining out suddenly, or breaking up
\citep[e.g.,][]{Ackerman:2001gk,Knapp:2004ji,Tsuji:2005cd,Burrows:2006ia,Marley:2010kx}.
The manner in which clouds disappear from the photosphere may impact the cooling
rate, and therefore the luminosities, of the brown dwarfs
\citep{Saumon:2008im,Dupuy:2015dz}.  The colors of L/T transition objects can
therefore shed light on the cloud dispersal process(es).

An accumulation of objects at a given color on the cooling sequence would
indicate a long-live phase of evolution, with objects spending a longer time at
the temperature corresponding to that color.  The ``hybrid'' evolutionary models
of \citet{Saumon:2008im} predict a pile-up of objects in the L/T transition at
$(J-K)_{\rm MKO}\approx0.9-1.0$, as cloud clearing removes opacity from the
photospheres of brown dwarfs and the cooling slows as entropy is released from
deeper atmospheric layers.  \citet{Dupuy:2012bp} found evidence of this type of
pile-up and a subsequent gap (i.e., a short-lived evolutionary phase) in the
distribution of near-IR colors of 36 L/T transition dwarfs (selected by absolute
$H_{\rm MKO}$ magnitudes).

\begin{figure}
\begin{center}
  \includegraphics[width=0.99\columnwidth]{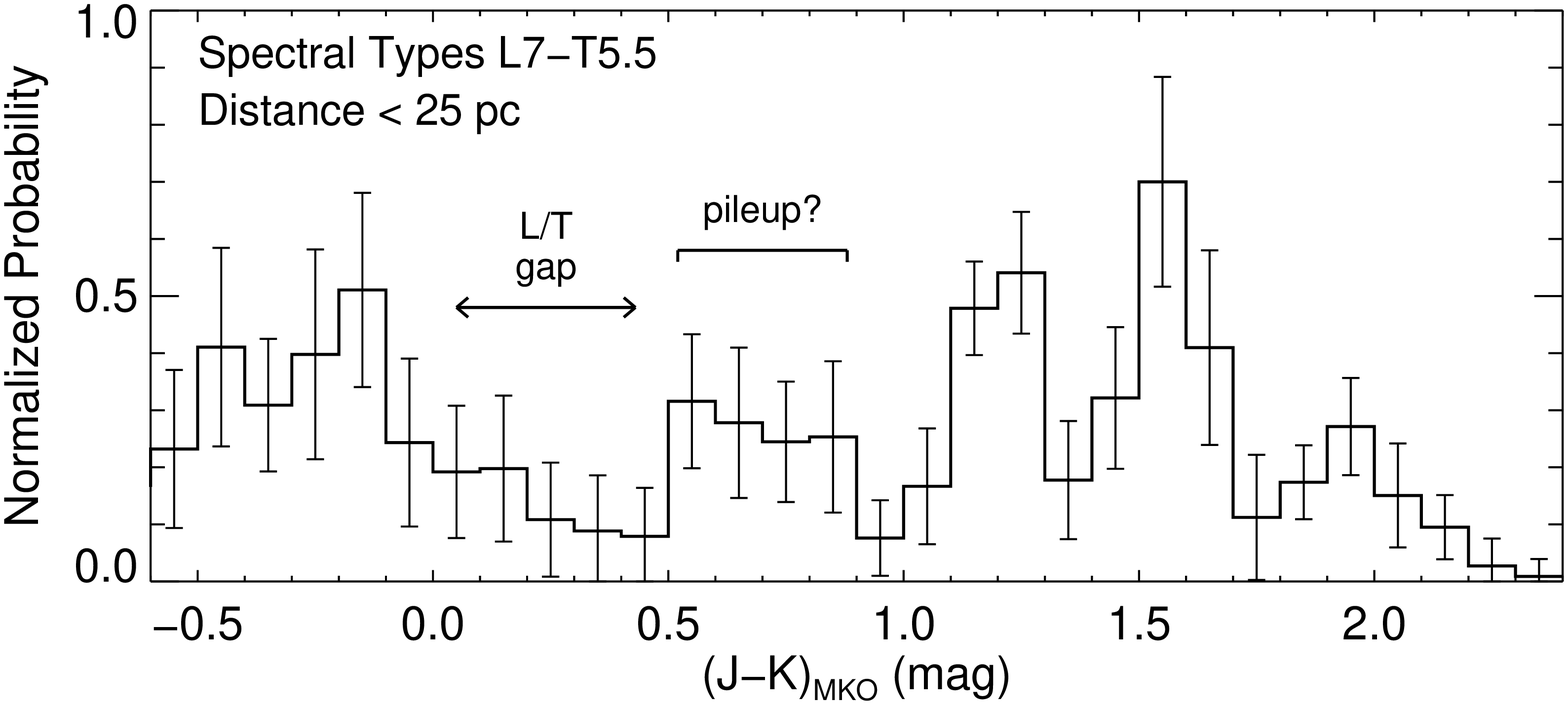}
  \caption{Distribution of $(J-K)_{\rm MKO}$ colors for 70 objects with spectral
    types L7--T5.5 and distances within 25~pc, including our discoveries and
    objects from the literature. The histogram was computed in a Monte Carlo
    fashion, accounting for errors in the photometry. The plotted uncertainties
    are the standard deviations for each color bin derived from the Monte Carlo
    simulations. The color distribution reveals signs of structure in the L/T
    transition, in particular the gap at $(J-K)_{\rm MKO}=0.0-0.5$~mag first
    detected by \citet{Dupuy:2012bp}, although the shape seen here is somewhat
    broader and shallower.  We also detect a less prominent pileup just redward
    of the gap than \citet{Dupuy:2012bp}, but see larger pileups at redder
    colors.}
\label{fig.JK.hist}
\end{center}
\end{figure}

By combining our new discoveries with objects from the literature, we have built
a larger sample of L/T transition dwarfs.  We used parallaxes when available and
photometric distances otherwise to construct a sample of 70 objects with
spectral types L7--T5.5, volume-limited at 25~pc.  In Figure~\ref{fig.JK.hist},
we show the distribution of $(J-K)_{\rm MKO}$ colors for this sample, computed
in a Monte Carlo fashion accounting for errors in the photometry.  This color
distribution suggests pile-ups and gaps across the L/T transition.  The most
prominent gap is at $(J-K)_{\rm MKO}\approx-0.1-0.5$~mag, somewhat broader and
shallower than the gap at $(J-K)_{\rm MKO}\approx0.0-0.4$~mag detected by
\citet{Dupuy:2012bp}.  We also find a less prominent pileup just redward of the
gap than \citet{Dupuy:2012bp}, but there may also be larger pileups at
$(J-K)_{\rm MKO}\approx1.2$ and 1.6~mag.

 Our larger sample supports the existence of the ``L/T gap'', but also makes
clear that a larger sample, ideally volume-limited and defined entirely by
trigonometric distances, is needed to fully delineate the color evolution in the
L/T transition.

\section{Young Discoveries}
\label{young}

\subsection{Field Objects}
\label{young.lowg}
Stars with ages $\lesssim200$~Myr are expected to be rare within 100~pc of the
Sun, at most a few percent of the population for a uniform star-forming history.
Our search was designed to identify field L/T transition dwarfs and generally
avoided known star-forming regions, so we were surprised to find 23 of our 59
M7--L7 discoveries showing confirmed or possible spectral signatures of low gravity,
i.e., youth (Section~\ref{results.gravity}), and we explored why this happened.

\begin{figure}    
\begin{center}
  \includegraphics[width=0.99\columnwidth]{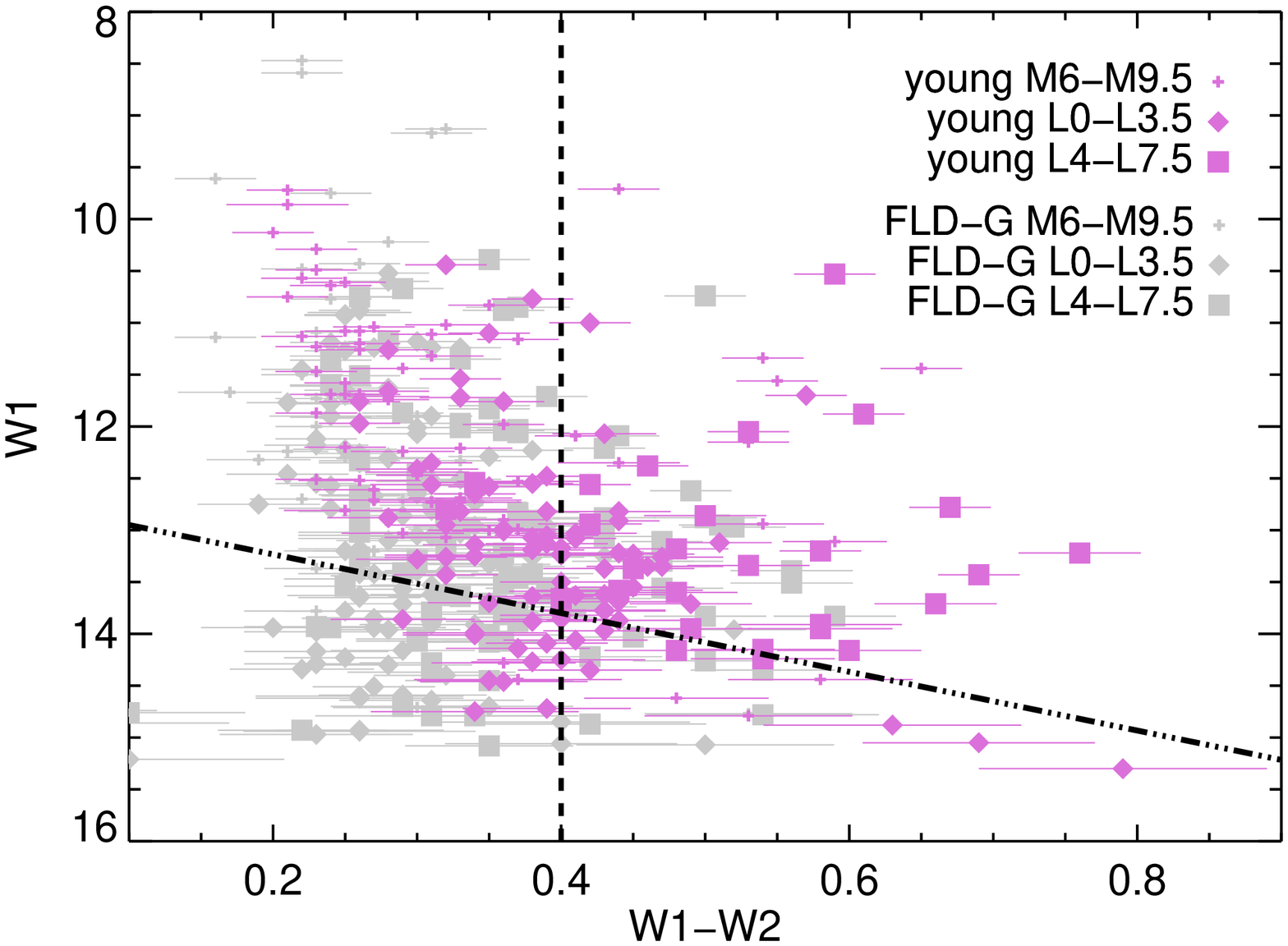}
  \caption{$W1$ vs. $W1-W2$ photometry for confirmed young objects (magenta,
    with symbols according to spectral type --- see legend at upper right) and
    \fldg\ objects (light gray, same symbols) from AL13, this paper, and
    the SpeX Prism Library. The vertical dashed line marks the \wawb\
    $\ge0.4$~mag selection criterion for our sample, while the diagonal dash-dot
    line shows the $W1$ vs. $W1-W2$ line from Figure~\ref{fig.w1.w1w2.25pc} that
    we used to identify candidates likely to be within 25~pc.  The samples here
    are drawn from different searches and likely influenced by multiple biases,
    but there is clear indication that the two criteria select (above and to the
    right of the lines) a disproportionately large number of young M and L dwarfs
    compared to the field population.}
\label{fig.w1.w1w2.young}
\end{center}
\end{figure}

Typically, young ultracool dwarfs are redder than older objects with the same
spectral types in the photometric bands we used to select candidates
\citep[e.g.,][]{Gizis:2012kv}.  They are also expected to be more luminous at
longer wavelengths (i.e., in the mid-infrared WISE bands) due to both enhanced
clouds and larger radii at younger ages.  It is therefore natural to assume that
our selection criteria, which screened out bluer and fainter objects, biased our
candidates toward young brown dwarfs.  To test this assumption, we assembled a
set of \fldg\ objects from our discoveries, AL13, and objects in the
SpeX Prism Library\footnote{http://pono.ucsd.edu/\mytilde
  adam/browndwarfs/spexprism}.  We also gathered published objects with optical
\citep[$\beta$ or $\gamma$;][]{Cruz:2009gs} or near-infrared
\citep[\intg\ or \vlg;][]{Allers:2013hk} classifications of low
gravity.  Figure~\ref{fig.w1.w1w2.young} compares the $W1$ magnitudes vs. \wawb\
colors for these sets of older and young objects.  The two sets are drawn from
multiple searches and sources, and we do not attempt to untangle the biases and
selection effects.  Nevertheless, Figure~\ref{fig.w1.w1w2.young} suggests that
our search criteria are indeed prone to selecting a disproportionately large
number of young M and L dwarfs compared to the field population.

\subsection{Young Moving Groups}
\label{ymg}
Young moving groups (YMG) are associations of young stars ($\approx$10--100~Myr)
and brown dwarfs whose similar trajectories through space imply that the members
originated in a common star-forming region \citep[e.g.,][]{Zuckerman:2004ex}.
YMG members are coeval, and therefore serve as both benchmarks for stellar and
substellar atmospheres and as empirical laboratories for testing models of star
formation.  In addition, these young stars are prime targets for direct imaging
searches for nearby exoplanets.  Our search targeted field brown dwarfs without
regard to age or space motion, but we investigated the possibility that we had
serendipitously stumbled upon members of YMGs.

\subsubsection{Candidates Selected With BANYAN II}
\label{ymg.banyan}
We used the BANYAN~II online tool \citep{Malo:2013gn,Gagne:2014gp} to calculate
probabilities of membership in nearby YMGs for our discoveries.  BANYAN~II
determines membership probabilities in a Bayesian fashion using sky position and
proper motion, as well as radial velocity and distance when available.  We
computed photometric distances using $K_{\rm MKO}$ magnitudes (and the
appropriate polynomial from \citealt{Dupuy:2012bp}) because the absolute
magnitudes of young objects and field objects are most similar in this bandpass
\citep[M. C. Liu et al., in preparation]{Gagne:2015dc}.  (We caution that
photometric distances will not be accurate for objects that are unresolved
equal-luminosity binaries.)  Based on our sky positions, proper motions, and
$K_{\rm MKO}$ photometric parallaxes, BANYAN~II found that 10 of our discoveries
have a $\gtrsim$70\% probability of membership in a YMG (Table~\ref{tbl.ymg})
and a corresponding false alarm rate of $\lesssim$10\% \citep{Gagne:2014gp}.

Interestingly, our 10 candidates all have spectral types L7--T4.5, which would
place any of them among the lowest-mass and coolest YMG members discovered to
date.  We estimated their masses assuming membership in their
  respective candidate YMGs, which have ages $149^{+51}_{-19}$~Myr for AB
  Doradus and $24\pm3$~Myr for $\beta$~Pictoris \citep{Bell:2015gw}, and
  $40\pm10$~Myr for Argus \citep{Makarov:2000dt,Torres:2008vq}.  We note that
  the Argus association lacks consensus in the literature about whether it is a
  real YMG, and if real its membership list is not yet well-defined
  \citep[e.g.,][]{Bell:2015gw}.  To estimate masses for our YMG candidates, we
  first calculated the $L_{\rm bol}$ for each object using our spectral types,
  the $K_{\rm MKO}$ bolometric corrections of \citet[their Table 6]{Liu:2010cw},
  and the $K_{\rm MKO}$-band photometric distance for each object.  We then used
  the ``hybrid'' evolutionary models of \citet{Saumon:2008im} and our
  $L_{\rm bol}$ values to determine masses at the age of each candidate's YMG.
  Our final mass estimates are included in Table~\ref{tbl.ymg}.  We propagated
  the uncertainties on our spectral types, $K_{\rm MKO}$ magnitudes, bolometric
  corrections \citep{Liu:2010cw}, distances, and ages into our mass
  determinations using Monte Carlo simulations and normal distributions for each
  uncertainty, and we quote 68th percentile confidence limits.  Mass estimates
  for these objects, assuming they are YMG members, are $\approx$6--15~\mjup,
  spanning the deuterium-burning limit and comparable to the lowest-mass
  free-floating objects ever discovered \citep{Liu:2013gy,Gagne:2015kf}.

We also repeat the warning of \citet{Shkolnik:2012cs} and others that the
spatial and kinematic locations of YMGs can be contaminated by unrelated field
objects, so other indications of youth in a candidate are helpful for confirming
membership.  Unfortunately for our candidates, the AL13 gravity indices apply
only to objects with spectral types $\le$L7, and the spectra for our two L7 YMG
candidates have S/N $<30$ so we do not regard their indices as reliable.  More
generally, low-gravity spectral signatures in the L/T transition are not as well
established as for earlier-type objects.  The young ($100\pm30$~Myr) T3.5 dwarf
GU Psc b \citep{Naud:2014jx} has an unusually red $J-K_s$ color for its spectral
type, but it is not known whether this is true for other young early-T dwarfs.
We do not see unusually red near-IR colors in our T dwarf YMG candidates.

The most promising of our candidate YMG members is PSO~J057.2+15.2 (L7), whose
spectrum reveals the triangular H band profile typical of youth, and whose
$(J-K)_{\rm 2MASS}=2.28\pm0.25$~mag color is significantly redder than the
average $(J-K)_{\rm 2MASS}=1.77\pm0.22$~mag for L7 dwarfs
\citep{Schmidt:2010ex}.  The BANYAN~II online tool gives a 91.9\% probability of
membership in the $\beta$ Pictoris Moving Group
\citep[$\beta$PMG;][]{Zuckerman:2001go} based on proper motion and photometric
distance.  If confirmed, this object would provide a nearby ($32\pm4$~pc) target
for atmospheric studies with a well-constrained age.  We estimate this object would have a mass
of $8.1^{+1.8}_{-1.5}$~\mjup, firmly in the planetary regime, and comparable to the latest
known $\beta$PMG member PSO~J318.5338$-$22.8603 \citep[spectral type
L7][]{Liu:2013gy}.

Two other L dwarf candidates have unusually red near-IR colors for their
spectral type, consistent with being low-gravity and thus young:

{\it PSO~J004.7+51} --- The BANYAN~II online tool gives this L7 dwarf a 79.9\%
probability of membership in the Argus Moving Group
\citep[ARG;][]{Zuckerman:2001go}.  We estimate it would have a mass of
$10.3^{+1.4}_{-1.2}$~\mjup.

{\it PSO~J100.5+41 (WISE~0642+4101)} --- The BANYAN~II online tool gives this
red L9 dwarf a 78.6\% probability of membership in the AB Doradus Moving Group
\citep[ABDMG;][]{Zuckerman:2004ds}.  We estimate it would have a mass of
$15^{+4}_{-3}$~\mjup.

\subsubsection{BASS Catalog}
We cross-matched our discoveries with the BASS catalog presented in
\citet{Gagne:2015ij}.  The BASS catalog contains 252 ultracool candidate YMG
members with spectral types $\ge$M5 selected in a Bayesian fashion by the full
BANYAN~II methodology \citep{Malo:2013gn,Gagne:2014gp}, which incorporates 2MASS
and \WISE\ photometry in addition to the sky position, proper motion, radial
velocity, and parallax used by the online tool.  We found only one of our
discoveries in BASS: the unusually red L dwarf PSO~J100.5+41 \citep[first
identified as WISE~0642+4101 by][]{Mace:2013jh}.  \citet{Gagne:2015ij} give this
object a 38.4\% probability of membership in ABDMG, more pessimistic than the
78.6\% probability based on our data and the online tool.  \citet{Gagne:2015ij}
also present an LP-BASS catalog with 249 ``low-priority'' candidates; none of
these are among our discoveries.  We note that our search for L/T transition
dwarfs targeted a somewhat different parameter space.  The majority of our
discoveries are near the Galactic plane ($|b|<15^\circ$), too faint
(poor-quality or non-existent 2MASS photometry), or too blue (L/T transition
objects have bluer $J-H$ colors than earlier-L dwarfs) to satisfy the criteria
used to construct the BASS sample.

\section{Summary}
\label{summary}
We have conducted a successful search for nearby L/T transition dwarfs using a
merged \PS\ $3\pi$ + \WISE\ database as our primary resource, supplemented by
near-infrared photometry from 2MASS, UKIDSS, and our own observations.  Our
search has yielded 130 ultracool dwarfs over $\approx$28,000~deg$^2$ of sky.  Of
these, 79 objects have spectral types L6--T4.5, the largest number of L/T
transition dwarfs discovered in any single search to date.  Thirty of the L/T
transition dwarfs have photometric distances less than 25~pc, and for spectral
types L9--T1.5 we have increased the number of known objects within 25~pc by
over 50\%.  We have analyzed the near-infrared colors of our L/T transition
discoveries, and we find further evidence for the pile-up in the L/T transition
first predicted by the ``hybrid'' evolutionary models of \citet{Saumon:2008im}
as well as a subsequent L/T gap first seen by \citet{Dupuy:2012bp}.

We assigned spectral types to our discoveries by visual comparison with field
spectral standards, and we compare these to types assigned using the index-based
methods of \citet[M4--L7 dwarfs]{Allers:2013hk} and \citet[L0--T8
dwarfs]{Burgasser:2006cf}.  We find that the \citet{Allers:2013hk} method
assigns spectral types generally in agreement with visually assigned types for
most objects, but earlier (by $\approx$0.5--1 subtypes) for unusually red M and
L dwarfs.  The spectral types assigned by the indices of
\citet{Burgasser:2006cf} are in good agreement with visual types for T dwarfs
but may be different by $\approx$0.5--1.0 subtypes for L dwarfs.

Among the late-M to mid-L dwarfs in our sample, we found a total of 23 objects
with spectral signatures of low gravity, indicating youth.  Using the
gravity-sensitive indices of \citet{Allers:2013hk}, we classify nine of these
discoveries as \vlg\ and one as \intg.  We assign provisional \vlg\ and \intg\
classifications to seven more objects based on spectra with modest S/N; higher
S/N spectra are needed to clarify their gravity classes.  These include
  the red L dwarf PSO~J068.3126+52.4546 (Hya12), identified by
  \citet{Lodieu:2014jo} as a candidate member of the Hyades. We identify a
further 6 objects whose spectra have clear visual suggestions of young
age but no index classification due to low S/N or spectral types outside the
applicable range of the indices.  We conclude that our candidate selection
criteria, designed to identify field L/T transition dwarfs, also favored the
discovery of young M and L dwarfs because of their redder \ywa\ and \wawb\
colors.

Thirty-one of our discoveries are candidate binaries based on their low-resolution
spectral features, making them prime targets for high-resolution imaging.  Two
of the candidate binaries are common proper motion companions to main sequence
stars: PSO~J003.4950$-$18.2802 (previously identified by \citealt{Baron:2015fn})
and PSO~J330.3214+32.3686.  If confirmed as binaries, these objects would be
ultracool binaries with ages and metallicities determined from their primaries,
making them rare empirical test cases for evolutionary models.

We also identify 11 kinematic candidates for nearby young moving groups with
spectral types L7--T4.5 using the BANYAN~II online tool, including three that
show possible spectral indications of youth.  Eight of these have spectral types
L9 or later, and if confirmed as YMG members they would provide an unprecedented
opportunity to determine the effective temperatures and test evolutionary models
of young L/T transition objects.
 
In conclusion, our discoveries include a large new set of L/T transition dwarfs
that contribute significantly to the nearby census and shed light on the
evolution of brown dwarf atmospheres in the L/T transition.  They also include
young late-M and L dwarfs, several of which are candidate very low mass brown
dwarfs in nearby star-forming regions and young moving groups.  If confirmed,
these would be exceptional age-constrained benchmarks for understanding the
properties of young cool atmospheres.


We thank the anonymous referee for a prompt and positive report.  We thank
Katelyn Allers, Michael Kotson, Brian Cabreira, Bill Golisch, Dave Griep, and
Eric Volqardsen for assisting with IRTF observations.  The \PS\ Surveys (PS1)
have been made possible through contributions of the Institute for Astronomy,
the University of Hawaii, the Pan-STARRS Project Office, the Max-Planck Society
and its participating institutes, the Max Planck Institute for Astronomy,
Heidelberg and the Max Planck Institute for Extraterrestrial Physics, Garching,
The Johns Hopkins University, Durham University, the University of Edinburgh,
Queen's University Belfast, the Harvard-Smithsonian Center for Astrophysics, the
Las Cumbres Observatory Global Telescope Network Incorporated, the National
Central University of Taiwan, the Space Telescope Science Institute, the
National Aeronautics and Space Administration under Grant No. NNX08AR22G issued
through the Planetary Science Division of the NASA Science Mission Directorate,
the National Science Foundation under Grant No. AST-1238877, the University of
Maryland, Eotvos Lorand University (ELTE), and the Los Alamos National
Laboratory.  The United Kingdom Infrared Telescope (UKIRT) is supported by NASA
and operated under an agreement among the University of Hawaii, the University
of Arizona, and Lockheed Martin Advanced Technology Center; operations are
enabled through the cooperation of the East Asian Observatory.  When the data
reported here were acquired, UKIRT was operated by the Joint Astronomy Centre on
behalf of the Science and Technology Facilities Council of the U.K.  This paper
makes use of observations processed by the Cambridge Astronomy Survey Unit
(CASU) at the Institute of Astronomy, University of Cambridge.  This project
makes use of data products from the Wide-field Infrared Survey Explorer, which
is a joint project of the University of California, Los Angeles, and the Jet
Propulsion Laboratory/California Institute of Technology, funded by the National
Aeronautics and Space Administration.  This research has made use of the 2MASS
data products; the UKIDSS data products; the VISTA data products; NASA's
Astrophysical Data System; the SIMBAD database operated at CDS, Strasbourg,
France, the SpeX Prism Spectral Libraries, maintained by Adam Burgasser at
http://pono.ucsd.edu/\mytilde adam/browndwarfs/spexprism, and the Database of
Ultracool Parallaxes, maintained by Trent Dupuy at
https://www.cfa.harvard.edu/\mytilde tdupuy/plx.  WMJB received support from NSF
grant AST09-09222.  WMBJ, MCL, and EAM received support from NSF grant
AST-1313455.  Finally, the authors wish to recognize and acknowledge the very
significant cultural role and reverence that the summit of Mauna Kea has always
held within the indigenous Hawaiian community. We are most fortunate to have the
opportunity to conduct observations from this mountain.

{\it Facilities:} \facility{IRTF (SpeX)}, \facility{PS1}, \facility{UKIRT (WFCAM)}

\begin{deluxetable}{lcclccc}
\tablecolumns{7}
\tablewidth{0pc}
\tabletypesize{\tiny}
\tablecaption{\PS\ and \WISE\ All-sky Photometry \label{tbl.ps1.wise}}  
\tablehead{   
  \colhead{\PS\ Name} &
  \colhead{\zps} &
  \colhead{\yps} &
  \colhead{\WISE\ Name} &
  \colhead{$W1$} &
  \colhead{$W2$} &
  \colhead{$W3$} \\
  \colhead{} &
  \colhead{(mag)} &
  \colhead{(mag)} &
  \colhead{} &
  \colhead{(mag)} &
  \colhead{(mag)} &
  \colhead{(mag)}
}
\startdata

PSO~J003.4950$-$18.2802 & $20.08\pm0.04$ & $18.94\pm0.07$ & J001358.81$-$181648.1 & $14.60\pm0.04$ & $14.17\pm0.05$ & $12.20\pm0.35$ \\
PSO~J004.1834+23.0741 & $20.27\pm0.06$ & $18.90\pm0.03$ & J001643.96+230426.7 & $14.29\pm0.03$ & $13.65\pm0.04$ & $12.20\pm0.35$ \\
PSO~J004.7148+51.8918 & $20.06\pm0.06$ & $19.08\pm0.07$ & J001851.51+515330.6 & $13.58\pm0.04$ & $13.04\pm0.04$ & $12.55\pm0.34$ \\
PSO~J007.7921+57.8267 & $18.18\pm0.01$ & $17.02\pm0.01$ & J003110.04+574936.3 & $12.41\pm0.02$ & $11.84\pm0.02$ & $11.30\pm0.10$ \\
PSO~J007.9194+33.5961 & $19.62\pm0.03$ & $18.60\pm0.02$ & J003140.64+333545.9 & $13.69\pm0.03$ & $13.18\pm0.03$ & $12.03\pm0.23$ \\
PSO~J010.2132+41.6091 & $19.99\pm0.06$ & $19.37\pm0.04$ & J004051.14+413631.4 & $16.11\pm0.07$ & $15.31\pm0.09$ & $>\!\!12.97$ \\
PSO~J023.8557+02.0884 & $19.62\pm0.02$ & $18.76\pm0.02$ & J013525.37+020518.4 & $14.31\pm0.03$ & $13.90\pm0.05$ & $>\!\!12.04$ \\
PSO~J024.1519+37.6443 & $20.46\pm0.11$ & $19.84\pm0.08$ & J013636.31+373840.6 & $16.80\pm0.09$ & $14.76\pm0.06$ & $>\!\!12.72$ \\
\enddata
\tablecomments{\PS\ photometry is quoted as of March 2015. The photometric
  selections described in this paper were done using \PS\ photometry from January
  2012. \WISE\ photometry is from the \WISE\ All-sky Release.}
\end{deluxetable}

\begin{deluxetable}{lcccrcccc}
\tablecolumns{9}
\tablewidth{0pc}
\tabletypesize{\tiny}
\tablecaption{Near-infrared Photometry\label{tbl.nir}}  
\tablehead{   
  \colhead{} &
  \multicolumn{3}{c}{2MASS Photometry} &
  \colhead{} &
  \multicolumn{4}{c}{MKO Photometry} \\
  \cline{2-4} \cline{6-9} \\
  \colhead{Name} &
  \colhead{$J_{\rm 2MASS}$} &
  \colhead{$H_{\rm 2MASS}$} &
  \colhead{$K_{\rm 2MASS}$} &
  \colhead{} &
  \colhead{$J_{\rm MKO}$} &
  \colhead{$H_{\rm MKO}$} &
  \colhead{$K_{\rm MKO}$} &
  \colhead{Ref.\tablenotemark{a}} \\
  \colhead{} &
  \colhead{(mag)} &
  \colhead{(mag)} &
  \colhead{(mag)} &
  \colhead{} &
  \colhead{(mag)} &
  \colhead{(mag)} &
  \colhead{(mag)} &
  \colhead{}
}
\startdata

PSO~J003.4$-$18 & $16.54\pm0.14$ & $15.89\pm0.18$ & $15.04\pm0.13$ & & $16.68\pm0.27$ & $15.87\pm0.30$ & $[15.18\pm0.28]$ & 1 \\
PSO~J004.1+23 & $>\!\!16.41$ & $15.70\pm0.14$ & $>\!\!14.97$ & & $16.58\pm0.03$ & $15.72\pm0.02$ & $[15.24\pm0.05]$ & 1 \\
PSO~J004.7+51 & $16.82\pm0.15$ & $15.29\pm0.11$ & $14.64\pm0.07$ & & $16.70\pm0.05$ & $15.40\pm0.11$ & $[14.51\pm0.07]$ & 1 \\
PSO~J007.7+57 & $14.95\pm0.04$ & $13.78\pm0.04$ & $13.22\pm0.03$ & & $14.80\pm0.01$ & $13.86\pm0.01$ & $[13.21\pm0.03]$ & 1 \\
PSO~J007.9+33 & $16.45\pm0.13$ & $15.43\pm0.12$ & $14.49\pm0.08$ & & $16.40\pm0.03$ & $[15.47\pm0.08]$ & $[14.69\pm0.08]$ & 1 \\
PSO~J010.2+41 & \nodata & \nodata & \nodata & & $18.27\pm0.14$ & $[17.34\pm0.20]$ & $[16.78\pm0.22]$ & 1 \\
PSO~J023.8+02 & $16.62\pm0.13$ & $15.48\pm0.10$ & $15.12\pm0.12$ & & $16.48\pm0.15$ & $15.66\pm0.14$ & $14.99\pm0.12$ & 2 \\
PSO~J024.1+37 & \nodata & \nodata & \nodata & & $18.23\pm0.11$ & $17.46\pm0.10$ & $[17.14\pm0.16]$ & 1 \\
\enddata
\tablecomments{All 2MASS photometry is from the 2MASS Point Source Catalog
  \citep{Cutri:2003vr} except where noted. For MKO photometry, H and K band
  magnitudes enclosed in single brackets were synthesized using observed J band
  magnitudes and our low-resolution spectra. MKO magnitudes enclosed in double
  brackets were synthesized using the 2MASS magnitudes for the corresponding
  filters and our low-resolution spectra.}
\tablenotetext{a}{References for MKO photometry: (1) this work, (2) UKIDSS DR9
  \citep{Lawrence:2013wf}, (3) VISTA Hemisphere Survey \citep{Cross:2012jz}.}
\tablenotetext{b}{2MASS photometry from 2MASS Point Source Reject Table
  \citep{Skrutskie:2006hl}.}
\tablenotetext{c}{Calibrated using the 2MASS $H$ magnitude because the 2MASS
  $K$ magnitude is an upper limit.}
\end{deluxetable}

\begin{deluxetable}{lcccc}
\tablecolumns{5}
\tablewidth{0pc}
\tabletypesize{\tiny}
\tablecaption{IRTF/SpeX Observations \label{tbl.obslog}}  
\tablehead{   
  \colhead{Object} &
  \colhead{Date} &
  \colhead{Slit} &
  \colhead{$t_\mathrm{int}$} & 
  \colhead{A0 V Standard} \\
  \colhead{} &
  \colhead{(UT)} &
  \colhead{('')} &
  \colhead{(s)} &
  \colhead{} 
}
\startdata
PSO~J003.4$-$18 & 2012 Dec 1 & 0.8 & 720 & HD 3604 \\
PSO~J004.1+23 & 2012 Dec 1 & 0.8 & 720 & HD 9711 \\
PSO~J004.7+51 & 2013 Sep 22 & 0.8 & 540 & HD 19844 \\
PSO~J007.7+57 & 2012 Sep 24 & 0.8 & 200 & HD 240290 \\
PSO~J007.9+33 & 2013 Sep 22 & 0.8 & 540 & HD 15240 \\
PSO~J010.2+41 & 2012 Oct 28 & 0.8 & 2880 & HD 219290 \\
PSO~J023.8+02 & 2012 Nov 8 & 0.8 & 720 & HD 18571 \\
PSO~J024.1+37 & 2012 Oct 25 & 0.8 & 2880 & HD 12381 \\
\enddata
\tablecomments{All observations performed in prism mode with the $0.8\times15$
  arcsec ($R\approx75$) or $0.5\times15$ arcsec ($R\approx120$)
  slit. Observations for PSO~J007.7+57, PSO~J103.0+41, PSO~J140.2+45,
  PSO~J272.4-04, PSO~J282.7+59, PSO~J307.6+07, and PSO~J339.0+51 were originally
  presented in \citet{Best:2013bp}. }
\end{deluxetable}

\begin{deluxetable}{lcccccc}
\tablecaption{Visual Spectral Types and Kinematics\label{tbl.spt.kin}}  
\tablecolumns{7}
\tabletypesize{\scriptsize}
\tablewidth{0pt}
\tablehead{
  \colhead{Name} &
  \colhead{SpT} &
  \colhead{$d_{\rm phot}$\tablenotemark{a}} &
  \colhead{$\mu_\alpha$\,cos\,$\delta$\tablenotemark{b}} & 
  \colhead{$\mu_\delta$\tablenotemark{b}} &
  \colhead{$v_\mathrm{tan}$} &
  \colhead{Discovery} \\
  \colhead{ } &
  \colhead{(visual)} &
  \colhead{(pc)} &
  \colhead{(mas\,yr$^{-1}$)} &
  \colhead{(mas\,yr$^{-1}$)} &
  \colhead{(km\,s$^{-1}$)} &
  \colhead{References} \\
}
\startdata

PSO~J003.4$-$18 & L5 pec & $46.1\pm5.6$ & $-31\pm12$ & $-175\pm13$ & $39\pm6$ & 1 \\
PSO~J004.1+23 & T0 & $24.7\pm2.5$ & $398\pm13$ & $36\pm15$ & $47\pm5$ & 2 \\
PSO~J004.7+51 & L7 & $23.4\pm2.8$ & $293\pm4$ & $-12\pm7$ & $33\pm4$ & 2 \\
PSO~J007.7+57 & L9 & $11.6\pm1.4$ & $526\pm2$ & $-12\pm4$ & $29\pm3$ & 3,4 \\
PSO~J007.9+33 & L9 & $21.4\pm2.6$ & $-19\pm5$ & $-10\pm6$ & $2\pm1$ & 2 \\
PSO~J023.8+02 & L9.5 & $28.8\pm3.0$ & $87\pm11$ & $-470\pm12$ & $65\pm7$ & 2 \\
PSO~J024.1+37 & M7 & $134.8\pm22.6$ & $-6\pm10$ & $-7\pm15$ & $6\pm8$ & 2 \\
PSO~J031.5+20 & T5.5 & $24.7\pm2.3$ & $286\pm34$ & $88\pm28$ & $35\pm5$ & 2 \\
\enddata
\tablecomments{Uncertainties for these visual spectral types are $\pm0.5$
  subtypes, except for those listed with : ($\pm1.0$ subtype) or ::
  ($\ge\pm1.5$ subtypes).}
\tablenotetext{a}{Photometric distances calculated using $W2$ magnitudes and the
  polynomial from \citet{Dupuy:2012bp}.}
\tablenotetext{b}{Proper motions calculated using PS1 astrometry, as well as
  2MASS astrometry when available.}
\tablenotetext{c}{Photometric candidate.}
\tablerefs{Discovery References: (1) \citet{Baron:2015fn}, (2) this paper, (3)
  \citet{Thompson:2013kv}, (4) \citet{Best:2013bp} (Paper I), 
  (5) \citet{Hogan:2008ha}, (6) \citet{Lodieu:2014jo}, (7) \citet{Mace:2013jh},
  (8) \citet{Zhang:2009kw}, (9) \citet{BardalezGagliuffi:2014fl}, (10)
  \citet{Marocco:2015iz}, (11) \citet{Aberasturi:2011bk}, (12)
  \citet{Zhang:2010bq}, (13) \citet{Kirkpatrick:2011ey}, (14)
  \citet{Bihain:2013gw}.  Objects first published by other authors (and
  independently discovered by us) are detailed in Table~\ref{tbl.scooped}.}
\end{deluxetable}

\begin{deluxetable}{lccccrlccc}
\tablecaption{Spectral Types and Proper Motions of Previously Discovered Objects\label{tbl.scooped}}  
\tablecolumns{10}
\tabletypesize{\scriptsize}
\tablewidth{0pt}
\rotate
\tablehead{
  \multicolumn{5}{c}{Previous Observations} &
  \colhead{} &
  \multicolumn{4}{c}{This Paper} \\
  \cline{1-5} \cline{7-10} \\
  \colhead{Name} &
  \colhead{References} &
  \colhead{SpT} &
  \colhead{$\mu_\alpha$\,cos\,$\delta$} & 
  \colhead{$\mu_\delta$} &
  \colhead{} &
  \colhead{Name} &
  \colhead{SpT} &
  \colhead{$\mu_\alpha$\,cos\,$\delta$} & 
  \colhead{$\mu_\delta$} \\
  \colhead{} &
  \colhead{} &
  \colhead{} &
  \colhead{(mas\,yr$^{-1}$)} &
  \colhead{(mas\,yr$^{-1}$)} &
  \colhead{} &
  \colhead{} &
  \colhead{(visual)} &
  \colhead{(mas\,yr$^{-1}$)} &
  \colhead{(mas\,yr$^{-1}$)}
}
\startdata

\cutinhead{Spectroscopically Confirmed Objects}

2MASS J00135882$-$1816462 & \citet{Baron:2015fn} & L1 & $-20\pm41$ & $-212\pm37$ & & PSO~J003.4$-$18 & L5 pec & $-31\pm12$ & $-175\pm13$ \\[6pt]
WISE J003110.04+574936.3 & \citet{Thompson:2013kv} & L8 & $530\pm10$ & $-10\pm10$ & & PSO~J007.7+57 & L9 & $526\pm2$ & $-12\pm4$ \\
 & \citet{Best:2013bp} (Paper I) & L9 & $523\pm17$ & $-1\pm16$ & & & & & \\[6pt]
Hya12 & \citet{Hogan:2008ha} & L\tablenotemark{a} & $112.04\pm7$\tablenotemark{b} & $-17.86\pm7$\tablenotemark{b} & & PSO~J068.9+13 & L6 (red) & $108\pm9$ & $-14\pm10$ \\
 & \citet{Lodieu:2014jo} & L3.5 & $101\pm11$ & $-15\pm11$ & & & & & \\[6pt]
WISE J044633.45$-$242956.8 & \citet{Thompson:2013kv} & L5 pec (blue) & $-210\pm20$ & $410\pm20$ & & PSO~J071.6$-$24 & L6 (blue) & $-199\pm13$ & $398\pm16$ \\[6pt]
WISE J064205.58+410155.5 & \citet{Mace:2013jh} & extremely red & \nodata & \nodata & & PSO~J100.5+41 & L9 (red) & $12\pm4$ & $-372\pm6$ \\[6pt]
PSO J103.0927+41.4601 & \citet{Best:2013bp} (Paper I) & T0 & $-8\pm6$ & $-38\pm6$ & & PSO~J103.0+41 & T0 & $3\pm3$ & $-32\pm5$ \\
\enddata
\vspace{-12pt}
\tablecomments{Uncertainties for our visual spectral types are $\pm0.5$
  subtypes, except for those listed with : ($\pm1.0$ subtype) or ::
  ($\ge\pm1.5$ subtypes). }
\tablenotetext{a}{Based on photometry only.}
\tablenotetext{b}{\citet{Hogan:2008ha} report typical proper motion errors of
  $\pm7$~mas\,yr$^{-1}$.}
\tablenotetext{c}{Derived from CH$_4$s--CH$_4$l photometry.}
\end{deluxetable}

\begin{deluxetable}{lcl}
\rotate
\tablecolumns{3}
\tablewidth{0pc}
\tabletypesize{\scriptsize}
\tablecaption{Non-ultracool Interlopers \label{tbl.interlopers}}
\tablehead{   
  \colhead{Name} &
  \colhead{Probable Type\tablenotemark{a}} &
  \colhead{Description\tablenotemark{b}}
}
\startdata
PSO~J010.2+41 & carbon star & Carbon star in the direction of NGC 205 \\
PSO~J249.4$-$10 & RBO & Nondescript continuum broadly peaked in J band \\
PSO~J255.6+10 & galaxy &  Flat continuum with strong emission features
at $\approx$1.05~\um, $\approx$1.75~\um, and $\approx$2.05~\um \\
PSO~J342.9$-$09 & unknown & Nondescript continuum, flux decreases with wavelength (low S/N) \\
PSO~J276.0$-$01 & RBO & Nondescript continuum broadly peaked in H band, wide
absorption feature at $\approx$1.8~\um. $b=5.1^\circ$\\
PSO~J202.5$-$26 & K2 V & Good match to the SpeX Spectral Library K2 V standard
HD 3765 \citep{Rayner:2009ki} \\
PSO~J053.3+30 & RBO & Nondescript continuum sharply peaked in H band \\
PSO~J068.3+52 & RBO & Sharp peaks in J, H, and K bands. $b=3.1^\circ$ \\
PSO~J076.1+25 & RBO & Nondescript continuum sharply peaked in H band. $b=-9.7^\circ$ \\
PSO~J085.3+36 & RBO & Nondescript continuum sharply peaked in H band. $b=3.1^\circ$ \\
PSO~J303.7+31 & RBO & Nondescript continuum broadly peaked in J band. $b=-1.6^\circ$ \\
PSO~J296.0+35 & K5 V & Good match to the SpeX Spectral Library K5 V standard
HD 36003 \citep{Rayner:2009ki} \\
\enddata
\tablenotetext{a}{RBO = Reddened background object (likely a giant or supergiant
star).}
\tablenotetext{b}{Five of these objects lie within $10^\circ$ of the galactic
  plane. Their galactic latitudes ($b$) are indicated.}
\end{deluxetable}

\begin{deluxetable}{lccc}
\tablecolumns{4}
\tablewidth{0pc}
\tabletypesize{\scriptsize}
\tablecaption{\label{tbl.new.nir.spt} New Near-infrared Spectral Types}
\tablehead{
\colhead{Object Name} &
\colhead{Opt SpT} &
\colhead{Opt Ref} &
\colhead{NIR SpT}
}
\startdata
2MASS J09153413+0422045 & L6\tablenotemark{a} & \citet{Reid:2008fz} & L5 \\[3pt]
2MASSW J1239272+551537 & L5 & \citet{Kirkpatrick:2000gi} & L5 \\[3pt]
\enddata
\tablenotetext{a}{Binary brown dwarf \citep{Reid:2006dy}; both components have
  spectral type L6.}
\end{deluxetable}

\begin{deluxetable}{lcccccc}
\tablecaption{Index-based Spectral Types from \citet{Allers:2013hk} for M7--L8 Objects
  \label{tbl.indices.allers}}   
\tablecolumns{7}
\tabletypesize{\scriptsize}
\tablewidth{0pt}
\tablehead{   
  \colhead{} &
  \multicolumn{5}{c}{Index-derived Spectral Types} &
  \colhead{} \\
  \cline{2-6} &
  \multicolumn{4}{c}{} &
  \colhead{} &
  \colhead{Adopted} \\
  \colhead{Name} &
  \colhead{\hto} &
  \colhead{\htod} &
  \colhead{\htoa} &
  \colhead{\htob} &
  \colhead{SpT\tablenotemark{a}} &
  \colhead{SpT} \\
  \multicolumn{5}{c}{} &
  \colhead{(avg.)} &
  \colhead{(visual)\tablenotemark{b}}
}
\startdata

\cutinhead{M Dwarfs}
PSO~J024.1+37\tablenotemark{c} & \nodata & \nodata & \nodata & \nodata & \nodata & M7 \\
PSO~J101.8+39 & M7.6$_{-1.9}^{+2.1}$ & \nodata & M6.6$_{-1.5}^{+1.5}$ & M9.3$_{-1.6}^{+1.6}$ & M8.2$_{-1.2}^{+1.4}$ & M9.5 \\
PSO~J115.0+59 & M9.0$_{-2.3}^{+2.2}$ & \nodata & M7.9$_{-1.4}^{+1.4}$ & M7.3$_{-1.8}^{+1.9}$ & M8.3$_{-1.5}^{+1.5}$ & M9.5 \\
PSO~J133.8+06 & M8.0$_{-0.7}^{+0.7}$ & \nodata & M6.5$_{-1.2}^{+1.2}$ & M8.2$_{-0.7}^{+0.7}$ & M8.0$_{-0.5}^{+0.5}$ & M9 \\
PSO~J174.6$-$18 & L0.1$_{-1.4}^{+1.4}$ & \nodata & M8.3$_{-1.4}^{+1.3}$ & M9.5$_{-1.8}^{+1.6}$ & M9.7$_{-1.0}^{+1.0}$ & M9 \\
PSO~J186.5+21 & L0.8$_{-2.0}^{+1.9}$ & \nodata & M6.9$_{-1.5}^{+1.5}$ & M9.4$_{-1.9}^{+1.7}$ & L0.0$_{-1.4}^{+1.3}$ & M9 \\
PSO~J304.7$-$07 & \nodata & \nodata & L0.4$_{-1.3}^{+1.3}$ & M8.4$_{-1.7}^{+1.7}$ & M8.8$_{-1.4}^{+1.4}$ & M9 \\
PSO~J337.4+16 & L1.1$_{-1.8}^{+1.7}$ & \nodata & M8.5$_{-1.3}^{+1.3}$ & M8.1$_{-1.5}^{+1.5}$ & M9.8$_{-1.2}^{+1.1}$ & M9 \\
\enddata
\tablecomments{None of these indices are valid for spectral types later than L8,
  so objects with those spectral types are not listed here.}
\tablenotetext{a}{Calculated as the weighted average of the spectral types from
  Monte Carlo simulations for all indices, excluding those that fell
  outside the valid range for each index.}
\tablenotetext{b}{Spectral types determined by visual comparison with spectral
  standards, which we adopt as the final spectral types for our
  discoveries. Uncertainties for these visual spectral types are $\pm0.5$
  subtypes, except for those listed with : ($\pm1.0$ subtype) or ::
  ($\ge\pm1.5$ subtypes).}
\tablenotetext{c}{No indices yielded spectral types within the valid range of
any index for this object, so no average spectral type was derived.}
\end{deluxetable}

\begin{deluxetable}{lccccccc}
\tablecolumns{8}
\tablewidth{0pc}
\tabletypesize{\scriptsize}
\tablecaption{Index-based Spectral Types from \citet{Burgasser:2006cf} for L0--T7 Objects\label{tbl.indices.burg}}  
\tablehead{   
  \colhead{} &
  \multicolumn{6}{c}{Index Values (Derived Spectral Types)\tablenotemark{a}} &
  \colhead{} \\
  \cline{2-7} &
  \multicolumn{6}{c}{} &
  \colhead{Adopted} \\
  \colhead{Name} &
  \colhead{H$_2$O-$J$} &
  \colhead{CH$_4$-$J$} &
  \colhead{H$_2$O-$H$} &
  \colhead{CH$_4$-$H$} &
  \colhead{CH$_4$-$K$} &
  \colhead{SpT} &
  \colhead{SpT} \\
  \multicolumn{6}{c}{} &
  \colhead{(avg.)} &
  \colhead{(visual)\tablenotemark{b}}
}
\startdata

\cutinhead{L Dwarfs}
PSO~J003.4$-$18 & 0.693 (L7.5) & 0.780 ($<$T0) & 0.703 (L6.4) & 0.934 (T1.3) & 0.924 (L6.8) & L8.0$\pm$2.3 & L5 pec \\
PSO~J004.7+51 & 0.667 (L8.3) & 0.783 ($<$T0) & 0.721 (L5.7) & 1.136 ($<$T1) & 1.007 (L4.2) & L6.1$\pm$2.1 & L7 \\
PSO~J007.7+57 & 0.659 (L8.5) & 0.834 ($<$T0) & 0.665 (L7.8) & 1.047 ($<$T1) & 0.844 (L8.7) & L8.4$\pm$0.5 & L9 \\
PSO~J007.9+33 & 0.697 (L7.4) & 0.747 ($<$T0) & 0.658 (L8.1) & 1.009 ($<$T1) & 0.899 (L7.5) & L7.6$\pm$0.4 & L9 \\
PSO~J023.8+02 & 0.662 (L8.4) & 0.742 ($<$T0) & 0.625 (L9.2) & 1.069 ($<$T1) & 0.870 (L8.2) & L8.6$\pm$0.5 & L9.5 \\
PSO~J041.5+01 & 0.680 (L7.9) & 0.866 ($<$T0) & 0.646 (L8.5) & 0.998 (T1.0) & 0.875 (L8.0) & L8.1$\pm$0.3 & L8.5 \\
PSO~J048.9+07 & 0.794 (L4.4) & 0.905 ($<$T0) & 0.674 (L7.5) & 1.029 ($<$T1) & 0.914 (L7.1) & L6.3$\pm$1.7 & L6: (blue) \\
PSO~J049.1+17 & 0.635 (L9.2) & 0.751 ($<$T0) & 0.611 (L9.6) & 0.996 (T1.0) & 0.869 (L8.2) & L9.0$\pm$0.7 & L9.5 \\
\enddata
\tablenotetext{a}{Spectral types were calculated using the polynomials defined
  in \citet{Burgasser:2007fl}.}
\tablenotetext{b}{Spectral types determined by visual comparison with spectral
  standards, which we adopt as the final spectral types for our
  discoveries. Uncertainties for these visual spectral types are $\pm0.5$
  subtypes, except for those listed with : ($\pm1.0$ subtype) or ::
  ($\ge\pm1.5$ subtypes).}
\end{deluxetable}

\begin{deluxetable}{lcccccccc}
\tablecaption{Low-Resolution Gravity Indices from
  \citet{Allers:2013hk} \label{tbl.gravity.allers}}  
\tablecolumns{9}
\tabletypesize{\scriptsize}
\tablewidth{0pt}
\tablehead{
  \colhead{Name} &
  \colhead{FeH$_{z}$} &
  \colhead{VO$_{z}$} &
  \colhead{KI$_{J}$} & 
  \colhead{H-cont} &
  \colhead{Index} &
  \colhead{Gravity} &
  \colhead{Gravity} &
  \colhead{SpT} \\
  \colhead{ } &
  \colhead{ } &
  \colhead{ } &
  \colhead{ } &
  \colhead{ } &
  \colhead{Scores\tablenotemark{a}} &
  \colhead{Score\tablenotemark{b}} &
  \colhead{Class\tablenotemark{c}} &
  \colhead{(visual)}
}
\startdata

\cutinhead{M dwarfs}
PSO~J133.8+06 & 1.170$_{-0.021}^{+0.021}$ & 1.080$_{-0.014}^{+0.014}$ & 1.078$_{-0.010}^{+0.010}$ & 0.926$_{-0.011}^{+0.011}$ & 0n10 (0n?0) & 0.0$_{-0.0}^{+1.0}$ & \fldg\tablenotemark{d} & M9 \\
PSO~J337.4+16 & 1.110$_{-0.042}^{+0.040}$ & 1.103$_{-0.028}^{+0.027}$ & 1.026$_{-0.022}^{+0.022}$ & 0.936$_{-0.029}^{+0.028}$ & 1n20 (1n20) & 1.0$_{-0.0}^{+1.0}$ & [\intg] & M9 \\

\cutinhead{L dwarfs}
PSO~J068.9+13 & 1.355$_{-0.045}^{+0.043}$ & 1.037$_{-0.019}^{+0.019}$ & 1.120$_{-0.015}^{+0.015}$ & 0.884$_{-0.013}^{+0.012}$ & nnn1 (nnn?) & 1.0$_{-1.0}^{+0.0}$ & [\intg] & L6 (red) \\
\enddata
\tablecomments{This table includes M7--L7 discoveries for which our spectrum has
  high enough S/N to extract useful measurements of the AL13 gravity indices,
  corroborated by visual inspection. No AL13 index is valid for spectral types
  later than L7, so objects with those spectral types are not included.}
\tablenotetext{a}{Scores in parentheses were determined using the original AL13
  classification scheme, in which objects with index values corresponding to
  \intg\ but within $1\sigma$ of the \fldg\ value are classified
  with a score of ``?''.}
\tablenotetext{b}{The overall gravity classification value and the 68\%
  confidence limits calculated as described in \citet{Aller:2015ti}.}
\tablenotetext{c}{Gravity classes in brackets are based on lower-S/N spectra and
  could not be confirmed visually, and therefore should be considered
  tentative. Higher-S/N spectra are needed to clarify the gravity.}
\tablenotetext{d}{Although this object is classified as \fldg\ under the
  AL13 system, we note that within the uncertainties in our gravity score, this
  object shows signs of intermediate gravity. A higher resolution
  spectrum is needed to more accurately classify the gravity of this object.}
\tablenotetext{e}{Strong or medium candidate binary (see text). The spectral
  type may therefore be based on a composite spectrum.}
\end{deluxetable}

\clearpage

\begin{deluxetable}{lccc}
\tablecaption{Other Objects Showing Low Gravity Spectral Features \label{tbl.other.young}}  
\tablecolumns{4}
\tabletypesize{\scriptsize}
\tablewidth{0pt}
\tablehead{
  \colhead{Name} &
  \colhead{SpT} &
  \colhead{Reason for No Gravity Class\tablenotemark{a}} &
  \colhead{Youth\tablenotemark{b}} \\
  \colhead{ } &
  \colhead{(visual)} &
  \colhead{ } &
  \colhead{ }
}
\startdata

PSO~J004.7+51 & L7 & Low S/N & Red \\
PSO~J054.8$-$11 & L3: & Low S/N & H band; Red \\
PSO~J057.2+15 & L7 (red) & Low S/N & H band; Red \\
PSO~J100.5+41\tablenotemark{c} & L9 (red) & SpT $>$ L7 & Red \\
PSO~J244.1+06 & L9 (red) & SpT $>$ L7; Low S/N & Red \\
PSO~J353.8+45 & L7.5 & SpT $>$ L7 & Red \\

\enddata
\tablenotetext{a}{``SpT $>$ L7'': None of the AL13 spectral indices are defined
  for spectral types later than L7.  ``Low S/N'': Spectra with J band S/N $<30$
  produced AL13 gravity scores with uncertainties too large to yield useful
  results.}
\tablenotetext{b}{``Red'': Redder-than-normal near-IR colors for the spectral
  type.  ``H band'': Triangular H band profile.}
\tablenotetext{c}{Previously identified by \citet{Mace:2013jh} as an ``extremely
  red'' L dwarf, WISE~0642+4101, and by \citet{Gagne:2014gp} as a candidate
  member of the AB Dor young moving group (Section~\ref{ymg}).}
\end{deluxetable}

\clearpage

\begin{deluxetable}{lcccc}
\tablecaption{Candidate Binaries\label{tbl.binaries}}  
\tablecolumns{5}
\tabletypesize{\scriptsize}
\tablewidth{0pt}
\tablehead{
  \colhead{Name} &
  \colhead{SpT} &
  \colhead{BG14\tablenotemark{a}} &
  \colhead{B10\tablenotemark{b}} & 
  \colhead{Visual Signs\tablenotemark{c}} \\
  \colhead{ } &
  \colhead{(Visual)} &
  \colhead{Criteria} &
  \colhead{Criteria} &
  \colhead{ }
}
\startdata

\cutinhead{Strong Candidates}
PSO~J003.4$-$18 & L5 pec & 10 & \nodata & Y \\
PSO~J049.1+26 & T2.5 & \nodata & 4 & Y \\
PSO~J071.6$-$24 & L6 (blue) & 8 & \nodata & Y \\
PSO~J088.0+43 & L4 pec & 9 & \nodata & Y \\
PSO~J133.8$-$02 & T0 pec & \nodata & 5 & Y \\
PSO~J135.7+16 & T0 pec & \nodata & 6 & Y \\
PSO~J136.5$-$06 & L2 pec & 9 & \nodata & Y \\
PSO~J149.1$-$19 & L5 pec & 10 & \nodata & Y \\
\enddata
\tablecomments{Our spectra for all binary candidates listed above have
  S/N~$\ge25$ averaged over the J band ($1.20-1.31$~\um).}
\tablenotetext{a}{Number of index based-criteria (out of 12) for binarity from
  \citet{BardalezGagliuffi:2014fl} satisfied by this object's spectrum. These
  indices apply only to M7--L7 dwarfs (we included our L7.5 discoveries).}
\tablenotetext{b}{Number of index based-criteria (out of 6) for binarity from
  \citet{Burgasser:2010df} satisfied by this object's spectrum. These indices
  apply only to L8, L9, and T dwarfs.}
\tablenotetext{c}{See text (Section~\ref{results.binaries}) for descriptions.}
\end{deluxetable}

\begin{deluxetable}{lcccccc}
\tablecolumns{7}
\tablewidth{0pc}
\tabletypesize{\scriptsize}
\tablecaption{\label{cpm_tab} Common Proper Motion Pairings}
\tablehead{
\colhead{Name} &
\colhead{$\mu_{\alpha}\cos\delta$} &
\colhead{$\mu_{\delta}$} &
\colhead{Dist.} &
\colhead{SpT} &
\colhead{$r$} &
\colhead{$r$} \\
&
\colhead{(mas\,yr$^{-1}$)} &
\colhead{(mas\,yr$^{-1}$)} &
\colhead{(pc)} &
\colhead{} &
\colhead{(arcsec)} &
\colhead{(AU)}
}
\startdata
NLTT 687\tablenotemark{a} & $-33\pm5$ & $-173\pm5$\tablenotemark{b} & $42^{+26}_{-16}$\tablenotemark{c} & M3\tablenotemark{c} & 120.3 & 5053 \\[3pt]
PSO J003.4950$-$18.2802 & $-31\pm12$ & $-175\pm13$ & $46.1\pm5.6$ & L5 pec \\[3pt]
\hline\\[-6pt]
Wolf 1154 & $121.0\pm4.1$ & $62.0\pm4.0$\tablenotemark{d} & $27.5^{+11.9}_{-6.3}$\tablenotemark{e} & M1\tablenotemark{f} & 77.1 & 2313 \\[3pt]
PSO J330.3214+32.3686 & $105\pm8$ & $65\pm9$ & $20.1\pm2.1$ & T2.5 \\[3pt]
\hline\\[-6pt]
LSPM J2216+1952 & $146\pm8$ & $-96\pm8$\tablenotemark{g} & $30.0^{+11.2}_{-8.2}$\tablenotemark{h} & M4 & 52.2 & 1566 \\[3pt]
PSO J334.1193+19.8800 & $120\pm8$ & $-72\pm9$ & $30.7\pm3.2$ & T3 \\[3pt]
\hline
\enddata
\tablenotetext{a}{This pairing was previously discovered by \citet{Baron:2015fn}.}
\tablenotetext{b}{\citet{Salim:2003gv}}
\tablenotetext{c}{\citet{Baron:2015fn}}
\tablenotetext{d}{\cite{Hog:2000wk}}
\tablenotetext{e}{\cite{Lepine:2011gl}}
\tablenotetext{f}{Estimated from $V-J$ and the color-SpT relation of \cite{Lepine:2011gl}.}
\tablenotetext{g}{\cite{Lepine:2005jx}}
\tablenotetext{h}{Estimated using the distance relations of \cite{Lepine:2005cd}.}
\end{deluxetable}

\begin{deluxetable}{lccccccccc}
\rotate
\tablecaption{Candidate Members of Young Moving Groups \label{tbl.ymg}}  
\tablecolumns{10}
\tabletypesize{\scriptsize}
\tablewidth{0pt}
\tablehead{
  \colhead{Name} &
  \colhead{SpT} &
  \colhead{$d_{\rm phot}$\tablenotemark{a}} &
  \colhead{$\mu_\alpha$\,cos\,$\delta$} &
  \colhead{$\mu_\delta$} &
  \colhead{Youth\tablenotemark{b}} &
  \multicolumn{2}{c}{BANYAN II} &
  \colhead{log($L_{\rm bol}$/\lsun)} &
  \colhead{Mass\tablenotemark{c}}\\
  \colhead{ } &
  \colhead{(visual)} &
  \colhead{(pc)} &
  \colhead{(mas\,yr$^{-1}$)} &
  \colhead{(mas\,yr$^{-1}$)} &
  \colhead{} &
  \colhead{YMG} &
  \colhead{Prob. (\%)} &
  \colhead{(dex)} &
  \colhead{(\mjup)}
}
\startdata

PSO~J004.7148+51.8918 & L7 & $26.4\pm3.7$ & $293\pm4$ & $-12\pm7$ & Red & Argus & 79.9 & $-4.36^{+0.12}_{-0.14}$ & $10.3^{+1.4}_{-1.2}$ \\
PSO~J007.7921+57.8267 & L9 & $11.2\pm1.5$ & $526\pm2$ & $-12\pm4$ & \nodata & Argus & 98.7 & $-4.54^{+0.12}_{-0.13}$ & $8.8^{+1.3}_{-1.1}$ \\
PSO~J057.2893+15.2433 & L7 (red) & $31.5\pm4.4$ & $68\pm11$ & $-127\pm12$ & H band; Red & $\beta$ Pictoris & 92.0 & $-4.36^{+0.12}_{-0.13}$ & $8.1^{+1.8}_{-1.5}$ \\
PSO~J071.8769$-$12.2713 & T2: & $25.9\pm3.1$ & $20\pm19$ & $-89\pm19$ & \nodata & $\beta$ Pictoris & 86.4 & $-4.69^{+0.13}_{-0.14}$ & $6.1\pm0.7$ \\
PSO~J076.7092+52.6087 & T4.5 & $18.1\pm4.4$ & $57\pm4$ & $-196\pm7$ & \nodata & Argus & 71.4 & $-4.8\pm0.2$ & $7.0^{+1.5}_{-1.2}$ \\
PSO~J100.5233+41.0320\tablenotemark{d} & L9 (red) & $18.5\pm2.6$ & $12\pm4$ & $-372\pm6$ & Red & AB Doradus & 78.6 & $-4.54^{+0.12}_{-0.13}$ & $15^{+4}_{-3}$ \\
PSO~J272.4689$-$04.8036 & T1 & $13.1\pm1.3$ & $-46\pm4$ & $-400\pm13$ & \nodata & AB Doradus & 93.8 & $-4.65\pm0.09$ & $14\pm2$ \\
PSO~J319.3102$-$29.6682 & T0: & $15.6\pm1.6$ & $148\pm4$ & $-162\pm4$ & \nodata & $\beta$ Pictoris & 97.1 & $-4.60^{+0.10}_{-0.11}$ & $6.5^{+0.7}_{-0.6}$ \\
\enddata
\tablenotetext{a}{Photometric distances calculated using $K_{\rm MKO}$
  magnitudes and the polynomial from \citet{Dupuy:2012bp}.}
\tablenotetext{b}{``Red'': Redder-than-normal near-IR
  colors for the spectral type.  ``H band'': Triangular H band profile.}
\tablenotetext{c}{Mass estimates derived assuming the objects are members of the
  given YMGs.  We used $L_{\rm bol}$ and the ``hybrid'' evolutionary models of
  \citet{Saumon:2008im}, following the method described in
  Section~\ref{ymg.banyan}.}
\tablenotetext{d}{Previously identified by \citet{Gagne:2014gp,Gagne:2015ij} as a candidate
  member of the AB Doradus Moving Group.  First discovered as WISE~0642+4101
  \citep{Mace:2013jh}.}
\end{deluxetable}

\end{document}